\documentclass[11pt]{aastex}
\usepackage{emulateapj5,apjfonts}
\usepackage{epsf}
\slugcomment{{\sc Accepted to ApJ:} July 28, 2001}
\received{XXth XXXXXXXX 2001}
\newcommand{\begit}{\begin{itemize}}
\newcommand{\enit}{\end{itemize}}
\newcommand{\begen}{\begin{enumerate}}
\newcommand{\enen}{\end{enumerate}}
\newcommand{\beq}{\begin{equation}} 
\newcommand{\eeq}{\end{equation}} 
\newcommand{\beqa}{\begin{eqnarray}} 
\newcommand{\eeqa}{\end{eqnarray}} 
\newcommand{\p}{\partial}    
\newcommand{\pr}{^\prime}

\newcommand{\lumnue}{L_{\nu_e}}
\newcommand{\lumunu}{L_{\nu_\mu}}
\newcommand{\lumanue}{L_{\bar{\nu}_e}}

\newcommand{\avenue}{\langle\varepsilon_{\nu_e}\rangle}
\newcommand{\aveanue}{\langle\varepsilon_{\bar{\nu}_e}\rangle}
\newcommand{\aveunu}{\langle\varepsilon_{\nu_\mu}\rangle}

\begin{document} 

\title{The Physics of Protoneutron Star Winds: 
Implications for $r$-Process Nucleosynthesis}
\author{Todd A. Thompson}
\affil{Department of Physics, 
The University of Arizona, Tucson, AZ 85721\\
thomp@physics.arizona.edu}
\author{Adam Burrows}
\affil{Steward Observatory, 
The University of Arizona, Tucson, AZ 85721\\
burrows@jupiter.as.arizona.edu}
\author{Bradley S. Meyer}
\affil{Department of Physics and Astronomy, Clemson University, Clemson, SC
29634-0978\\
mbradle@ces.clemson.edu}

\vspace{.5cm}

\begin{abstract}

We solve the general-relativistic steady-state eigenvalue problem of neutrino-driven protoneutron
star winds, which immediately follow core-collapse supernova explosions. 
We provide velocity, density, temperature, and composition profiles and explore the
systematics and structures generic to such a wind for a variety of protoneutron 
star characteristics. Furthermore, we derive the entropy,
dynamical timescale, and neutron-to-seed ratio in the general relativistic framework 
essential in assessing this site as a candidate for $r$-process nucleosynthesis. 
Generally, we find  that for a given mass outflow rate ($\dot{\,M}$), the dynamical timescale of the
wind is significantly shorter than previously thought. 

We argue against the existence or viability of a high entropy ($\gtrsim300$ per k$_{\rm B}$ per baryon), 
long dynamical timescale $r$-process epoch.  In support
of this conclusion, we model the protoneutron star cooling phase, calculate nucleosynthetic yields 
in our steady-state profiles, and estimate the integrated mass loss.
We find that transonic winds enter a high entropy phase only with very low $\dot{M}$ 
($\lesssim1\times10^{-9}$ M$_\odot$ s$^{-1}$) and extremely long dynamical timescale 
($\tau_\rho\gtrsim0.5$ seconds). Our results support the possible existence of an early $r$-process 
epoch at modest entropy ($\sim150$) and very short dynamical timescale, consistent in our calculations 
with a very massive or very compact protoneutron star that contracts rapidly after the preceding supernova.
We explore possible modifications to our models, which might yield significant $r$-process nucleosynthesis generically.  

Finally, we speculate on the 
effect of fallback and shocks on both the wind physics and nucleosynthesis.
We find that a termination or reverse shock in the wind, but exterior to the wind sonic point, may 
have important nucleosynthetic consequences.
The potential for the $r$-process in protoneutron star winds remains an open question.

\end{abstract}

\keywords{Winds, neutrinos, supernovae, neutron stars, nucleosynthesis}


\section{INTRODUCTION}

The successful two-dimensional Type-II supernova simulation of Burrows, Hayes, and Fryxell (1995)
shows clearly a post-explosion neutrino-driven wind, emerging approximately half a second after
bounce.  The convective plumes and fingers due to Rayleigh-Taylor instabilities that accompany shock 
re-ignition in the gain region are pushed out and cleared from the area closest to the neutron star 
by the pressure of the neutrino-driven wind.  The last 50 milliseconds (ms) of the simulation show that
a nearly spherically symmetric wind has established itself as the protoneutron star, newly born,
begins its Kelvin-Helmholtz cooling phase.\footnote{See these 2D simulations at 
http://www.astrophysics.arizona.edu/movies.html.}  Although these simulations employed only crude
neutrino transport, did not address the issue of fallback, and did not study the wind as a function
of progenitor mass and structure, they are suggestive of a general phenomenon that might naturally accompany
many core-collapse supernovae.  

Some multiple of 10$^{53}$ erg will be lost via neutrino 
radiation by the protoneutron star as it cools.  A small fraction of that energy will be deposited in the surface layers of the
nascent neutron star, heating and driving material from its surface.
Although the wind is interesting in its own right, hydrodynamically and as a phenomenon that attends both the supernova
and the Kelvin-Helmholtz cooling phase,
perhaps its most important ramification is
the potential production of $\sim$50\% of all the nuclides above the iron group in rapid($r$) neutron-capture 
nucleosynthesis.  

In the $r$-process, rapid interaction of neutrons with heavy, neutron-rich, seed nuclei allows
a neutron capture-disintegration equilibrium to establish itself among the isotopes of each element.
Beta decays occur on a longer timescale and increase the nuclear charge.  For sufficiently large
neutron-to-seed ratio, the `nuclear flow' proceeds to the heaviest nuclei, forming abundance 
peaks at $A\sim80$, 130, and 195 (Burbidge et al.~1957; Wallerstein et al.~1997).
The neutron-to-seed ratio itself is largely set by the dynamical timescale ($\tau_\rho$; see eq.~\ref{taurho}), 
entropy ($s$), and neutron richness in the earlier phase of the expansion (Hoffman, Woosley, and Qian 1997; Meyer and Brown 1997; 
Freiburghaus et al. $\!$1999).
While the nuclear physics is fairly well understood, the astrophysical site has not been unambiguously 
established.  Currently, neutron star mergers 
(Freiburghaus, Rosswog, and Thielemann 1999, and references therein; Rosswog et al.$\!$ $\!$1999) 
and protoneutron star winds (Meyer et al.~1992; Woosley and Hoffman 1992; Woosley et al.~1994; 
Qian and Woosley 1996; Hoffman, Woosley, 
and Qian 1997; Otsuki et al.~2000) are considered the most viable candidates.

In addition to attaining the requisite neutron-to-seed ratios, dynamical timescales, and temperatures, 
the astrophysical site must consistently reproduce the observed solar abundance pattern of $r$-process elements with $A\gtrsim135$.
Recent observations of neutron-capture elements in ultra-metal-poor halo stars 
(Burris et al.~2000; McWilliam et al.~1995a,b; Sneden et al.~1996; Cowan et al.~1996; Westin et al.~2000;
Hill et al.~2001) show remarkable agreement with the solar $r$-process abundance pattern in this mass range.  
This suggests a universal mechanism for producing the second and third abundance peaks, which acts early in the chemical enrichment 
history of the galaxy.  In this paper, beyond addressing the physical nature and systematics of the wind, we investigate
its potential as a site for $r$-process nucleosynthesis up to and beyond the third abundance peak.

\cite{dsw} were the first to address the physics of steady-state neutrino-driven neutron star winds.
Although interested in the relative importance of the neutrino and photon luminosity in determining the wind dynamics,
they also identified some of the basic systematics and scaling relations generic to 
the problem.   More recent investigations have focused less on the general physics of the wind and
more on its potential nucleosynthetic yield.  \cite{woosley1994} conducted the first such detailed study.
They followed the nucleosynthesis in a protoneutron star wind that emerged in a one-dimensional post-supernova environment.
Approximately 18 seconds after collapse and explosion,
their model attained entropies of
$\sim$400 (throughout, we quote entropy per k$_{\rm B}$ per baryon), 
long dynamical timescales, and electron fraction ($Y_e$) in the range $0.36-0.44$.  
However, in their model the supernova
shock reached only 50,000 km at these late times.  In turn, this external boundary caused the wind material 
to move slowly. It remained in the heating regime for an extended period, thus raising the entropy
above what any simulation or analytical calculation has since obtained.  
Although the $r$-process proceeded to the third abundance peak in their calculation, 
nuclei in the mass range near $A\sim90$ (particularly, $^{88}$Sr, $^{89}$Y, and $^{90}$Zr)
were overproduced by more than a factor of 100.

\cite{twj1994} conducted a similar investigation, but did not attain the entropies of
\cite{woosley1994}.  In fact, they fell short by a factor of $\sim5$. 
Later, \cite{qw1996} showed that for reasonable protoneutron star characteristics, including post-Newtonian corrections,
an entropy of 400 is unrealistic.  \cite{qw1996} also provided many analytical scaling relations that have 
since framed the discussion of neutrino-driven winds.  Following this work,
\cite{hwq} conducted nucleosynthetic calculations in the wind models of \cite{qw1996}.  They concluded that
the standard wind models of \cite{qw1996} did not produce third peak $r$-process nucleosynthesis.  
They also employed an adiabatic cooling prescription to survey the parameter space relevant to 
protoneutron star winds and noted several important systematics in the nucleosynthesis.  Particularly, they identified
a low entropy ($\sim120$), fast timescale ($\tau_\rho\sim2$ ms),
and $high$ electron fraction ($Y_e\gtrsim0.48$) window where third peak $r$-process could take place (see their Fig.~10).

\cite{cardall} generalized some of the scaling relations presented in \cite{qw1996} to include general
relativity and found significant enhancements in the entropy and dynamical timescale of the wind in this framework.
Recently, \cite{otsuki} have sought to solve the general-relativistic wind equations and to conduct $r$-process calculations in the 
winds they obtained.
They concluded that $r$-process nucleosynthesis can proceed to the third abundance peak at $A\sim195$ for a 
protoneutron star with radius 10 km, a gravitational mass of 2 M$_\odot$, and total neutrino luminosity of $10^{52}$ erg s$^{-1}$.  
These conditions produce modest entropies ($s\sim140$) and fast dynamical timescales ($\tau_\rho\sim2-3$ ms).
Although a significant parameter in protoneutron star winds, $Y_e$ was fixed by hand in their calculations.
Furthermore, they employed a simple equation of state and, in the context of the transonic wind problem, an unphysical 
external boundary condition.  

These uncertainties and ambiguities in the conclusions of previous groups suggest that a re-evaluation is in order. 
Our goal in this paper is to solve the full eigenvalue problem of the steady-state transonic wind problem in general
relativity, employing physical boundary conditions.
Using this formalism, we survey the relevant parameter space, identify the major systematic trends, and explore some
of the particulars of the general-relativistic treatment. We then use these steady-state solutions to model
the whole of the Kelvin-Helmholtz cooling phase, including radial contraction of the protoneutron star,
as well as the evolution of the neutrino luminosity and average neutrino energy.  We estimate the total amount of material ejected 
during this cooling epoch, and put significant constraints on the range of entropies and dynamical timescales that 
might actually occur in Nature.  In addition, for a subset of the models generated, we calculate the
total nucleosynthetic yield as a function of atomic mass.

In \S\ref{sec:equations}, we present the fundamental equations for a time-independent neutrino-driven wind (general-relativistic and
Newtonian), including the equation for the evolution of the electron fraction.  In addition, we present the 
integrals of the motion and a discussion of the equation of state we employ.  In \S\ref{sec:numerical}, we describe our
solution to the wind problem using an iterative relaxation procedure on
an adjustable radial grid and the necessary boundary conditions. 
In \S\ref{sec:neutrino}, we present the
neutrino heating rates used in this study, review the effects of general relativity, 
discuss potential modifications to the energy deposition rates, and explore (approximately) the effects of transport.
In \S\ref{sec:fiducial}, we present our results for the wind problem itself.  The wind structures and general characteristics 
as a function of neutron star mass, radius, and neutrino spectral character are explored.  We cover the entire 
relevant parameter space and include some modifications to the power laws presented in \cite{qw1996}.  
In \S\ref{sec:evolution}, we use our steady-state wind results
to construct a sequence of such models that represent the time evolution of the wind during the
protoneutron star cooling phase.  We estimate the total mass ejected for a given evolutionary trajectory
and put useful constraints on possible epochs of $r$-process nucleosynthesis.
In \S\ref{sec:nucleosynthesis}, we present nucleosynthetic results from a subset of our wind trajectories.
In \S\ref{sec:what}, we discuss reasonable modifications to our wind models that might 
yield a successful $r$-process, including changes to the energy deposition function.
Finally, in \S\ref{sec:discussion} we review
our results, summarizing the constraints our calculations impose on the viability of the protoneutron star wind
as the astrophysical site of the $r$-process.  Furthermore, we speculate on the effects of progenitor structures and fallback,
as well as hydrodynamical and transport considerations left to be addressed in future work.


\section{THE STEADY-STATE WIND EQUATIONS}
\label{sec:equations}
The time-independent hydrodynamical equations for flow in a Schwarzschild spacetime
can be written in the form (Nobili, Turolla, and Zampieri 1991; Flammang 1982)
\beq
\frac{1}{vy}\frac{d(vy)}{dr}+\frac{1}{\rho}\frac{d\rho}{dr}+\frac{2}{r}=0,
\label{nobilimass}
\eeq
\beq
\frac{1}{y}\frac{dy}{dr}+\frac{1}{\varepsilon+P}\frac{dP}{dr}=0,
\label{nobilimom}
\eeq
and
\beq
\frac{d\varepsilon}{dr}-\frac{\varepsilon+P}{\rho}\frac{d\rho}{dr}+\rho\frac{\dot{q}}{vy}=0,
\label{nobilien}
\eeq
where $u_r(=vy)$ is the radial component of the fluid four-velocity, 
$v$ is the velocity of the matter measured by a stationary observer,
\beq
y=\left(\frac{1-2GM/rc^2}{1-v^2/c^2}\right)^{1/2},
\eeq
$\varepsilon\,(=\rho c^2+\rho\epsilon)$ is the total mass-energy density, $\rho$ is the rest-mass density, 
$P$ is the pressure, $\epsilon$ is the specific internal energy, and $\dot{q}$ is the energy deposition rate per unit mass.
These equations assume that the mass of the wind is negligible.  Although 
not readily apparent in the form above, eqs.~(\ref{nobilimass})$-$(\ref{nobilien}) exhibit a critical point when $v$ equals the local 
speed of sound.  In order to make the solution to this system tractable and 
the critical point manifest we recast the equations as
$$
\frac{\p v}{\p r}=
\frac{v}{2r}\left[\frac{v_e^2}{y^2}\left(\frac{1-c_s^2/c^2}{c_s^2-v^2}\right)
-4c_s^2\left(\frac{1-v^2/c^2}{c_s^2-v^2}\right)\right] $$
\beq
+\frac{D}{C_V T}\,\frac{\dot{q}}{y}\left(\frac{1-v^2/c^2}{c_s^2-v^2}\right),
\label{grv}
\eeq

\beq
\frac{\p\rho}{\p r}=
\frac{2\rho}{r}\left(\frac{v^2-v_e^2/4y^2}{c_s^2-v^2}\right)
-\frac{\rho}{(vy)}\frac{D}{C_V T}\frac{\dot{q}}{c_s^2-v^2},
\label{grr}
\eeq
and
$$\frac{\p T}{\p r}=
\frac{2}{r\rho}\frac{D}{C_V}\frac{(P+\varepsilon)}{c^2}\left(\frac{v^2-v_e^2/4y^2}{c_s^2-v^2}\right)\hspace*{1cm}$$
\beq
\hspace*{2cm}+\frac{\dot{q}}{C_V (vy)}\left(\frac{(1-D/c^2)c_T^2-v^2}{c_s^2-v^2}\right).
\label{grt}
\eeq
In the above expressions, $C_V$ is the specific heat at constant volume, $c_s$ is the local
adiabatic sound speed, $c_T$ is the isothermal sound speed, $v_e=(2GM/r)^{1/2}$, $M$ is the protoneutron star gravitational mass,
and 
\beq
D=c^2\frac{T}{\varepsilon+P}\left.\frac{\p P}{\p T}\right|_{\rho}.
\eeq
Taking the limits $v/c\ll1$ and $c_s/c\ll1$ we recover the Newtonian wind equations in critical form;
\beq
\frac{\p v}{\p r}=
\frac{v}{2r}\left(\frac{v_e^2-4c_s^2}{c_s^2-v^2}\right)
+\frac{D}{C_V T}\,\frac{\dot{q}}{c_s^2-v^2},
\label{nwv}
\eeq
\beq
\frac{\p\rho}{\p r}=\frac{2\rho}{r}\left(\frac{v^2-v_e^2/4}{c_s^2-v^2}\right)
-\frac{\rho}{v}\frac{D}{C_V T}\frac{\dot{q}}{c_s^2-v^2}
\label{nwr}
\eeq
and
\beq
\frac{\p T}{\p r}=\frac{2}{r}\,\frac{D}{C_V}\left(\frac{v^2-v_e^2/4}{c_s^2-v^2}\right)+
\frac{\dot{q}}{C_V v}\left(\frac{c_T^2-v^2}{c_s^2-v^2}\right).
\label{nwt}
\eeq
In this limit, $D$ becomes $(T/\rho)\left. \p P/\p T\right|_\rho$.
The differential wind  equations, both Newtonian and general-relativistic,
are solved in precisely the form above, as a two-point boundary value problem, using a relaxation algorithm 
described in \S\ref{sec:numerical}.  Note that direct integration of the continuity equation (eq.~\ref{nobilimass})
yields the eigenvalue of the steady-state wind problem, the mass outflow rate $\dot{M}=4\pi r^2 \rho v y$.
In addition, for $\dot{q}=0$, eqs.~(\ref{nobilimass})-(\ref{nobilien}) admit a second integral of the flow,
the Bernoulli integral. We impose neither as a mathematical 
constraint in solving the system. Instead, we use the degree to which each is conserved 
to gauge the accuracy of our converged models.  In the Newtonian case, the Bernoulli integral can be expressed by
\beq
\dot{M}\Delta\left(\epsilon+\frac{1}{2}v^2+\frac{P}{\rho}-\frac{GM}{r}\right)=\int_{R_\nu}^r\,d^3r^{\prime}\,\rho\,\dot{q}=Q(r),
\eeq
where $R_\nu$ is the coordinate radius of the protoneutron star surface.
In general relativity, with $\dot{q}=0$, $\gamma h \sqrt{-g_{00}}$ is a constant.  
Here, $\gamma$ is the Lorenz factor and $h$ is the specific enthalpy. 
With a source term, the differential change in neutrino luminosity is given by
\beq
e^{-2\phi}\frac{\p}{\p\mu}(L_\nu e^{2\phi})=-\dot{q},
\eeq
where $d\mu/dr=4\pi r^2\rho \,e^{\Lambda}$ and $ds^2=-e^{2\phi}dt^2+e^{2\Lambda}dr^2+r^2d\Omega$
defines the metric.  The total energy deposition rate is then,
\beq
Q=4\pi\int_{R_\nu}^\infty\,dr\,r^2\,\rho\,\dot{q}\,e^{\Lambda}\,e^{2\phi}.
\label{bigQtot}
\eeq
Using these prescriptions for the Bernoulli integral and the equation for
$\dot{M}$ and employing modest radial zoning (2000 points), 
we typically conserve both to better than 0.1\%.

In our solution to the wind problem, we couple the three wind equations in critical form
to the differential equation describing the evolution of the electron fraction, $Y_e$,
due to the charged-current electron-type neutrino interactions with free nucleons:
$\nu_e n\leftrightarrow e^-p$ and $\bar{\nu}_e p\leftrightarrow e^+n$.  This differential 
equation does not contain a critical point and can be written as
\beq
(vy)\frac{dY_e}{dr}=X_n[\Gamma_{\nu_en}+\Gamma_{e^+n}]-X_p[\Gamma_{\bar{\nu}_ep}+\Gamma_{e^-p}],
\label{yeeq}
\eeq
where $X_n$ and $X_p$ are the neutron and proton fraction, respectively. The $\Gamma$'s are the
number rates for emission and absorption, taken from the approximations of \cite{qw1996}. 
The number rate subscripts denote initial-state particles. 
The asymptotic value of the electron fraction ($Y_e^a$) is generally determined within $\sim$10 km of the protoneutron 
star surface. Ignoring the details of transport and neutrino decoupling near the 
neutrinospheres, $Y_e^a$ is determined by both the luminosity ratio $L_{\bar{\nu}_e}/L_{{\nu}_e}$ and the energy ratio 
$\aveanue/\avenue$, where $\langle\varepsilon_\nu\rangle=\langle E_\nu^2\rangle/\langle E_\nu\rangle$, and
$E_\nu$ is the neutrino energy.  To rough (but useful) approximation (Qian et al.~1993; Qian and Woosley 1996),
\beq
Y_e^a\simeq\frac{\Gamma_{\nu_en}}{\Gamma_{\nu_en}+\Gamma_{\bar{\nu}_ep}}\simeq
\left(1+\frac{\lumanue}{\lumnue}
\frac{\aveanue-2\Delta+1.2\Delta^2/\aveanue}
{\avenue+2\Delta+1.2\Delta^2/\avenue}\right)^{-1},
\label{yeqw}
\eeq
The energy threshold ($\Delta=m_n-m_p\simeq1.293$ MeV) for the $\bar{\nu}_e$ neutrino absorption process, 
$\bar{\nu}_e p \rightarrow n e^+$, is manifest in eq.~(\ref{yeqw}). 
Note the difference in sign on the $2\Delta$ term between 
numerator and denominator.  This implies that simply having $\lumanue/\lumnue>1$ and $\aveanue/\avenue>1$
is not sufficient to guarantee $Y_e^a<0.5$;  the magnitudes of $\aveanue$ and $\avenue$ are also important.
For example, taking $\lumanue/\lumnue=1.1$ and $\aveanue/\avenue=13\,\,{\rm MeV}/10\,\,{\rm MeV}$ gives $Y_e^a\simeq0.52$
Hence, if $\lumanue$ and $\lumnue$ are correlated with $\aveanue$ and $\avenue$, respectively, then even
for constant $\lumanue/\lumnue$ and $\aveanue/\avenue$, $Y_e^a$ must increase as the total neutrino luminosity
of the protoneutron star decays in time.  This phenomenon, which we refer to as the 
{\it threshold effect}, is important for the viability of an $r$-process epoch in protoneutron star winds at late
times and low $\aveanue$ and $\avenue$.  Other possible effects that might 
bear materially on $Y_e^a$ include the formation of alpha particles from free nucleons (the alpha effect) (Fuller and Meyer 1995; 
McLaughlin, Fuller, and Wilson 1996), the differential redshift of $\bar{\nu}_e$ neutrinos versus $\nu_e$ neutrinos, due to the 
physical separation of their 
respective neutrinospheres (Fuller and Qian 1996), and charge conjugation violation (Horowitz and Li 2000). 

Coupled to the wind equations and the equation for $Y_e$ evolution is a simple equation of state (EOS)
well-suited to the conditions in the neutrino-driven wind ($T\lesssim5$ MeV, $\rho\lesssim10^{13}$ g cm$^{-3}$,
and $0.0\lesssim Y_e\lesssim0.5$).  Under these conditions, to good approximation, free neutrons, protons, and alpha
particles may be treated as non-relativistic ideal gases. 
A fully general electron/positron equation of state is employed.  Photons are also included.
Past wind studies have approximated electrons and positrons as non-degenerate and relativistic.  Although
$\eta_e(=\mu_e/T)$ divided by $\pi$ may approach $\sim$10 at the protoneutron surface, it drops steeply with
the density so that in the main heating region the non-degenerate assumption is justified.
In contrast, however, it is important to include the non-relativistic character of 
the electrons and positrons.  For a broad range of protoneutron star characteristics,
the temperature drops to $\sim0.5$ MeV within $\sim50-100$ km of the neutron star surface.  The dynamics
and asymptotic character of the wind, including mass outflow rate, asymptotic velocity, and composition
can be significantly affected by assuming relativistic electrons and positrons 
throughout the wind profile.  In addition, the important range of matter temperatures for $r$-process nucleosynthesis
occurs for $T(r)\lesssim0.5$ MeV.  \cite{sumiyoshi} have found that using a general electron/positron EOS can decrease
the dynamical timescale in the nucleosynthetic region of the wind by as much as a factor of two,
a potentially important modification when considering the viability of the neutrino-driven wind as a candidate site
for the $r$-process.


\section{NUMERICAL TECHNIQUE}
\label{sec:numerical}

Past numerical studies of steady-state protoneutron star winds have at times been hampered 
by unphysical boundary conditions and ill-defined numerics not generally suited to the solution
of the protoneutron star wind problem.

The solution to the wind equations constitutes an eigenvalue problem.  For a given set of
protoneutron star characteristics and boundary conditions, there exists a
unique mass outflow rate ($\dot{M}=4\pi r^2\rho vy$) and critical radius 
($R_c$) where the matter velocity is equal to the local speed of sound ($v(R_c)=c_s$).
Although eqs.~(\ref{grv})-(\ref{grt}) are ordinary differential equations,
one cannot treat the wind as an initial value problem;  the critical point
necessitates a two-point boundary value prescription.

Even though shooting methods determine $\dot{M}$ precisely, $R_c$
can remain uncertain (London and Flannery 1982).   In addition, although \cite{dsw} made 
effective use of this method, they were forced to employ two
types of shooting: one for high $\dot{M}$ solutions and another for low $\dot{M}$ solutions.
In an effort to circumvent these problems and to construct a robust algorithm with flexible boundary conditions,
we solve both the Newtonian and general-relativistic wind equations
as a two-point boundary value problem using a relaxation algorithm 
on an adaptive radial mesh, as described in \cite{london_flannery}
(see also Kippenhahn, Weigert, and Hoffmeister 1968; Eggleton 1971; Press et al. $\!$1992).  

The relaxation algorithm involves replacing the differential equations
with algebraic difference equations at each point on the radial grid.  Then,
given an initial guess for each variable, at each point, the solution is
obtained by iteration.  However, because the condition $v(R_c)=c_s$ defines the outer boundary
and this point is not known {\it a priori} - such knowledge would effectively
constitute the solution - we follow the procedure of \cite{london_flannery}
and introduce a new independent variable $q$ that labels radial 
mesh points by integers ($q_1\leq q \leq q_N$).  The price paid is three more differential equations:

$$\frac{dr}{dq}=\frac{\psi}{\phi(r)},$$
$$\frac{dQ}{dq}=\psi,$$
and
\beq
\frac{d\psi}{dq}=0.
\label{mesheq}
\eeq
In this scheme, $r$ becomes a dependent variable, $Q(r)$ is a mesh spacing function 
(e.g., $Q(r)=\log r$), $\psi$ is an intermediate variable, and $\phi(r)$ is proportional to the 
density of mesh points.  The system is solved on the mesh of $q$ values.
Hence, the outer radial coordinate, at which the $v(R_c)=c_s$ boundary condition is to obtain, 
adjusts in a Newton-Raphson sense to simultaneously satisfy all boundary conditions. 
Typically, when we begin a calculation, we start with an initial guess that extends to a 
radius with Mach number of $\sim0.9$.  With each iteration, $R_c$ is adjusted so that the 
Mach number goes to 1.

We now have three wind equations for $\rho(q)$, $T(q)$, and $v(q)$, three extra differential equations
for the mesh algorithm, and the equation for $Y_e(q)$.  Seven boundary conditions must then
be imposed to close the system.  The boundary conditions on the first two of eqs.~(\ref{mesheq}) are 
simply $r(q_1)=R_\nu$ and $Q(q_1)=\log(R_\nu)$ (for $\log$ spacing), where $R_\nu$ is
the protoneutron star radius. Two more boundary conditions obtain at the critical point.
Taking $q_N$ as the outer mesh point, we have that
$v(q_N)=c_s$ and we have that the numerator of any of the differential equations for 
$\rho(r)$, $T(r)$, or $v(r)$ must simultaneously be zero to ensure continuity of the solution
through the sonic point.  Three additional boundary conditions are required to specify the problem completely.
Although simply setting $Y_e(q_1)$, $\rho(q_1)$, and $T(q_1)$ at $R_\nu$ is sufficient,
we do not use this prescription.
Instead, we assume that the radius of neutrino decoupling coincides with the coordinate radius of
the protoneutron star surface.  In addition, we assume that this neutrinosphere ($R_\nu$) is the same for all neutrino species:
electron ($\nu_e$), anti-electron ($\bar{\nu}_e$), and mu and tau  neutrinos (collectively, $\nu_\mu$'s).
Indeed, as the protoneutron star cools we expect these neutrinospheres to be separated by 
just tenths of kilometers.  As the $\nu_e$ neutrinos have the largest net opacity ($\kappa_{\nu_e}$) of any species,
we set an integral boundary condition on their optical depth ($\tau_{\nu_e}$):

\beq
\tau_{\nu_e}(R_\nu)\,=\,\int^\infty_{R_\nu}\,\kappa_{\nu_e}\rho\,dr\,=\,\frac{2}{3}.
\label{optical}
\eeq
Included in the $\nu_e$ opacity are contributions from scattering with free nucleons,
scattering on electron/positron pairs, $\nu_e n\rightarrow p e^-$, and alpha scattering.
As a second boundary condition, we assume that the net neutrino heating balances the net cooling
at $R_\nu$.  That is, $\dot{q}(R_\nu)=0$.
Finally, our third boundary condition assumes that the charged-current processes are in equilibrium
at the protoneutron star surface.  
Explicitly,
\beq
\left.\frac{dY_e}{dr}\right|_{R_\nu}=0.
\label{yebound}
\eeq
For a given protoneutron star mass and radius and a set of average neutrino energies and luminosities,
we start with a guess for the mass density at the surface of the protoneutron star.  We use 
a two-dimensional Newton-Raphson algorithm to simultaneously satisfy the conditions on $\dot{q}$ and
$dY_e/dr$ at $R_\nu$.  This determines $T(R_\nu)$ and $Y_e(R_\nu)$ for the first step.
At each subsequent iterative step, the relaxation algorithm attempts to satisfy the integral boundary condition
on $\tau_{\nu_e}$.  Effectively, this procedure results in a new $T(R_\nu)$ and $Y_e(R_\nu)$.  
In this way, we satisfy all boundary conditions simultaneously.
Given a good initial guess for
the solution (i.e., maximum deviations from convergence in any variable of $\lesssim20$\%) we obtain
a solution in just 5$-$10 iterations. Once the profile for $R_\nu \leq r\leq R_c$ is obtained, we use l'Hospital's rule to bridge
$R_c$ and then integrate to larger radii as an initial value problem using a fourth-order Runga-Kutta scheme.
Successfully converged models then serve as an initial guess for the next protoneutron star model with adjacent
characteristics (i.e., in mass, radius, or neutrino spectral characteristics).

\subsection{Tests of the Code}

We do not impose $d\dot{\rm M}/dr=0$ or the Bernoulli integral as mathematical constraints on
the system.  Instead, we use the degree to which these conditions obtain to gauge the precision
of our method.  Typically, both are conserved to better than 0.1\% in both general-relativistic 
and Newtonian calculations.   In addition, as we increase the number of mesh points, the error
in both of these quantities decreases significantly.

One may argue that eq.~(\ref{yebound}) need not hold generally.  
In fact, we adjusted the code to accept a fixed $Y_e(R_\nu)$
boundary condition to make sure this has no effect on the asymptotic character of the wind.
We find that the number rates are large enough at the surface that $Y_e$ is forced from
$Y_e(R_\nu)$ to a value such that $dY_e/dr\sim0$ in the first radial zone with no appreciable
effect on any aspect of the wind.  For simplicity, then, we have chosen to enforce eq.~(\ref{yebound}).

Similarly, in a fully dynamical calculation, one would not expect $\dot{q}(R_\nu)=0$,
generally.  While this may certainly be true, in our solution to the  steady-state wind
we encounter numerical instabilities that preclude solution of the equations with finite
$\dot{q}(R_\nu)$.  It is difficult to estimate the importance of such an assumption without
employing the full machinery of radiation hydrodynamics.


\section{THE NEUTRINO HEATING FUNCTION}
\label{sec:neutrino}

The neutrino energy deposition rate ($\dot{q}$) is a sum of contributions
from the charged-current $\nu_e$ and $\bar{\nu}_e$ neutrino absorption processes,
neutrino-electron/positron scattering, neutrino-nucleon scattering, and the process 
$\nu\bar{\nu}\leftrightarrow e^+ e^-$.  We describe each in turn.

\subsection{The Charged-Current Processes}
\label{sec:cc}

At the entropies encountered in supernovae ($\lesssim40$),
the charged-current or beta processes ($\nu_e n\leftrightarrow e^-p$ and $\bar{\nu}_e p\leftrightarrow e^+n$)
dominate the opacity and energy exchange 
for the electron and anti-electron neutrinos.  In
the protoneutron star wind context, at much higher entropies, we expect these
processes to compete with neutrino-electron/positron scattering in the net energy deposition. 
Ignoring final-state blocking and assuming relativistic electrons and positrons, the 
charged-current cooling function can be written as
\beq	
C_{\rm{cc}}\simeq2.0\times10^{18}\,T^6\,\left[X_p\frac{F_5(\eta_e)}{F_5(0)}+
X_n\frac{F_5(-\eta_e)}{F_5(0)}\right]\,\,\,\,\,\,\,\,\,{\rm erg\,\,g^{-1}\,\,s^{-1}},
\label{ccc}
\eeq
where
$$F_n(y)=\int_0^\infty\frac{x^n}{e^{x-y}+1}\,dx$$
are the Fermi integrals, $X_n$ and $X_p$ are the neutron and proton fractions, respectively, $T$ is in MeV,
and $\eta_e=\mu_e/T$.  
The heating rate due to neutrino captures on free nucleons can be written as
\beqa
H_{\rm{cc}}&\simeq&\,\,\,9.3\times10^{18}\,\,\,R_{\nu6}^{\,-2}\,\,\,\,\,{\rm erg\,\,g^{-1}\,\,s^{-1}}\nonumber \\
&\times&\left[\,X_n\, L^{\rm{51}}_{\nu_e}\,\langle\varepsilon^2_{\nu_e}\rangle\,+\,
X_p\, L^{\rm{51}}_{\bar{\nu}_e}\,\langle\varepsilon^2_{\bar{\nu}_e}\rangle\,\right]\,\,\,\Phi^6\,\,\Xi(r),
\label{hcc}
\eeqa
where $R_{\nu6}$ is the neutrinosphere radius in units of $10^{6}$ cm and
both $L_\nu$ and $\langle\varepsilon_\nu\rangle$ are defined at $R_\nu$.
We separate $L_\nu$ and $\langle\varepsilon_\nu\rangle$ in this heating rate and those below.
Although $L_\nu$ and $\langle\varepsilon_\nu\rangle$ are correlated, $L_\nu$ and the tail of the
neutrino energy distribution are generally not.  We retain
$L_\nu$ and $\langle\varepsilon_\nu\rangle$ separately so we have the freedom to change them independently.
With eqs.~(\ref{ccc}) and (\ref{hcc}), the net energy deposition due to the charged-current processes 
is then $\dot{q}_{cc}=H_{cc}-C_{cc}$.  
In eq.~(\ref{hcc}),
\beq
\langle\varepsilon^2_\nu\rangle=
\int_{-1}^1d\mu\int\,d\varepsilon_\nu\varepsilon_\nu^5\,f_\nu\,\cdot\,
\left[\int_{-1}^1d\mu\int\,d\varepsilon_\nu\varepsilon_\nu^3\,f_\nu\,\right]^{-1},
\eeq
where $\mu(=\cos\theta)$ is the cosine of the zenith angle and $f_\nu$ is the neutrino distribution function.
$\Phi$ accounts for the gravitational redshift of neutrinos from the protoneutron star surface
and is given by
\beq
\Phi=\left(\frac{1-2GM/R_\nu c^2}{1-2GM/rc^2}\right)^{1/2}.
\label{red}
\eeq
Also present in eq.~(\ref{hcc}) is the spherical dilution function, $\Xi(r)$, which describes the
radial dependence of the neutrino energy and number densities.  In the vacuum approximation, assuming a sharp
neutrinosphere,
\beq
\Xi(r)= 1-\sqrt{1-(R_\nu/r)^2/\Phi^2}
\label{null}
\eeq
and is related to the flux factor $\langle\mu\rangle(=F_\nu/J_\nu)$ by $\langle\mu\rangle=R_\nu^2/2\Phi^2\Xi(r)r^2$,
where $F_\nu$ and $J_\nu$ are the neutrino flux and energy density, respectively.
In eq.~(\ref{null}), the factor $\Phi^2$ accounts for the bending of null geodesics in a curved spacetime 
(Cardall and Fuller 1997; Salmonson and Wilson 1999).  This effectively increases the neutrino number density a given 
mass element sees at any radius, thus augmenting the energy deposition.  In contrast, since the heating rate for any neutrino 
interaction is proportional to positive powers of $L_\nu$ and $\langle\varepsilon_\nu\rangle$, the gravitational redshift terms
which modify these quantities can only decrease the energy deposition rate at a given radius.
As pointed out by \cite{cardall}, the augmentation of $\dot{q}$ by the bending of neutrino trajectories and
the degradation of $\dot{q}$ by the gravitational redshift compete as $M/R_\nu$ increases, but the latter dominates.

\cite{qw1996}, \cite{otsuki}, and \cite{wanajo} have employed eq.~(\ref{null}) or
its Newtonian analog in their studies of neutrino-driven winds.  In an effort to address neutrino decoupling
more fully, we compared our wind solutions using this spherical dilution factor to those
obtained with an effective $\Xi(r)$ derived from the Monte-Carlo transport results 
of \cite{janka1991} who connected $\langle\mu\rangle$ with the density gradient at 10$^{10}$ g cm$^{-3}$ 
and the curvature of the opacity profile at the radius of decoupling. The difference between the radial 
dependence of $\langle\mu\rangle$
using this approach and the effective flux factor obtained in a vacuum approximation is significant, as
would be expected, only when the atmosphere is sufficiently extended.  That is, as the density
gradient just exterior to the neutrinosphere goes to infinity, so too should the Monte-Carlo results of \cite{janka1991}
approach the vacuum approximation.  Hence, significant deviations from the vacuum approximation only present themselves
when the radius of the neutron star is large ($\sim20-40$ km) and/or the total neutrino luminosity is very high 
(i.e., the temperature at $R_\nu$ is large).  As the protoneutron star cools and the luminosity decreases for a given $R_\nu$,
the density gradient becomes steeper (see Fig.~\ref{fig:rp} below), thus making the vacuum approximation more appropriate.
Over the range of models presented here, we have examined the effects of
using the \cite{janka1991} $\langle\mu\rangle$ in order to characterize the spherical dilution of neutrinos
for all of the relevant heating processes and find them negligible, particularly for the compact protoneutron star wind models
most likely to be important for $r$-process nucleosynthesis.  

We emphasize that this simple comparison is not a complete analysis of the transport effects that might be important
in determining $\dot{q}(r)$.  While this result suggests the vacuum approximation is appropriate in our scheme
for handling the neutrinospheres and boundary conditions, it says nothing about actual transport effects in the decoupling region.

\subsection{Neutrino-Electron/Positron Scattering}
\label{sec:nuescatt}

At high entropies, electron-positron pairs are produced in abundance.
Therefore, neutrino-pair scattering is expected to contribute substantially to the total
energy deposition in the protoneutron star wind. The energy transfer associated with 
a single neutrino-electron or positron scattering event is well approximated by 
$\omega_i\simeq(\varepsilon_{\nu_i}-4T)/2$, where $\omega$ is the energy transfer,
$T$ is the matter temperature, and $\varepsilon_{\nu_i}$ is the neutrino energy of species $i$ (Bahcall 1964).  
We have confirmed this approximation using the thermalization code described in \cite{thompson},
which employs the fully relativistic structure function formalism of \cite{reddy_1998}.
The net heating rate due to the interaction of the neutrino fluid with the pair plasma can be 
approximated by $\dot{q}\simeq cn_e n_{\nu_i}\langle\sigma_{\nu_i e}\,\omega\rangle$,
where $n_e$ and $n_{\nu_i}$ are the number density of electrons and neutrinos, respectively, and $\sigma_{\nu_i e}$
is the cross section for scattering.
We obtain
\beq
\dot{q}_{\nu_i e}=\int\,\omega\, c\, \sigma_{\nu_i e}\, \frac{dn_e}{d\varepsilon_e}d\varepsilon_e\,
         \frac{dn_{\nu_i}}{d\varepsilon_{\nu_i}}d\varepsilon_{\nu_i},
\eeq
where $\sigma_{\nu_i e}(=\kappa_i\,T\,\varepsilon_{\nu_i})$ is the cross-section for neutrino scattering on relativistic, 
non-degenerate
electrons (Tubbs and Schramm 1975).  
$\kappa_i=\sigma_o\Lambda_i/2m_e^2$ is a neutrino species dependent constant, where $m_e$ is the mass of the
electron in MeV, $\sigma_o\simeq1.71\times10^{-44}$ cm$^2$, and 
\beq
\Lambda_i=(c_V+c_A)^2+\frac{1}{3}(c_V-c_A)^2.
\eeq
$c_V$ and $c_A$ are the  vector and axial-vector coupling constants for a given neutrino species.
We find that the energy deposition rate can be expressed as
\beqa
\dot{q}_{\nu_i e}&=&
\frac{c}{\rho}\left(\frac{T^3}{(\hbar c)^3}\frac{F_2(\eta_e)}{\pi^2}\right)
\frac{L_\nu}{4\pi r^2 c\langle\varepsilon_\nu\rangle\langle\mu\rangle} 
\,\,\,\Phi^2\,\,\,\,\rm{erg\,\,\,g^{-1}\,\,\,s^{-1}}\nonumber \\
&\times&\left[\frac{\kappa}{2}\langle\varepsilon_\nu\rangle\frac{F_4(\eta_\nu)}{F_3(\eta_\nu)}
T\left(\langle\varepsilon_\nu\rangle\Phi\frac{F_2(\eta_\nu)}{F_3(\eta_\nu)}-4T\frac{F_3(\eta_e)}{F_4(\eta_e)}\right)\right].
\label{qdotes}
\eeqa
We drop the subscript $i$ here for brevity.  Note that $\langle\mu\rangle(=R_\nu^2/2\Phi^2\Xi(r)r^2)$ contains
two powers of the redshift.  $\eta_\nu$ is an effective degeneracy parameter obtained from
fitting the neutrino distribution function with a Fermi-Dirac distribution with an appropriate $T_\nu$ 
and $\eta_\nu$ (Janka and Hillebrandt 1989).  Although we retain the general form here, in the wind calculations
presented below we take $\eta_\nu=0$ for all neutrino species.
In eq.~(\ref{qdotes}), the first term  in parentheses is the number density of electrons, $n_e$.  
Alternatively, with the replacement $\eta_e\rightarrow-\eta_e$, it is the number density of positrons.
The next term is the number density of neutrinos, a function of radius, which depends also on the flux factor, 
$\langle\mu\rangle$. The term in square brackets is the appropriately averaged product of the cross 
section for scattering and the energy transfer per scattering, $\langle\sigma_{\nu_i e}\,\omega\rangle$.  
We retain the form $\dot{q}=cn_e n_{\nu_i}\langle\sigma_{\nu_i e}\,\omega\rangle$ here for clarity.
Note that, for $\eta_\nu=\eta_e=0$, $\langle\varepsilon_\nu\rangle$ equals $3.15T_\nu$ and the second quantity 
in parentheses in eq.~(\ref{qdotes}) 
is simply $(T_{\nu_i}-T)$; net heating occurs if $T_\nu>T(r)$ at any $r$.  
In order to obtain the contribution to the net heating from 
neutrino-positron scattering, in addition to the change in $\eta_e$, one must also make appropriate changes to $\Lambda_i$.


\subsection{Neutrino-Nucleon Scattering}
\label{sec:nunscatt}

The energy transfer associated with a single neutrino-nucleon scattering event is 
much smaller than that for neutrino-electron scattering.  The cross section, however, is much
larger.  Using our neutrino thermalization code, which solves the 
Boltzmann equation in an isotropic, homogeneous thermal bath of scatterers, including the
full collision term, with Pauli blocking and explicit coupling between all energy bins 
(Thompson, Burrows, and Horvath $\!$2000), we confirm that the average energy transfer for neutrino-nucleon scattering,
in non-degenerate nuclear matter, for neutrino energies below $\sim$40 MeV is well approximated by
$\omega\simeq\varepsilon_\nu(\varepsilon_\nu-6T)/m_N$, where $m_N$ is the nucleon mass in MeV (Tubbs 1979).  
We derive the heating rate for
neutrino-neutron scattering as 
$$\dot{q}_{\nu_i n}=\frac{cn_n}{\rho}
\frac{L_\nu}{4\pi r^2 c\langle\varepsilon_\nu\rangle\langle\mu\rangle}\,\,\,\Phi^4\,\,\,\,\rm{erg\,\,\,g^{-1}\,\,\,s^{-1}}$$
\beq
\times\left[\kappa\left(\frac{F_2(\eta_\nu)}{F_3(\eta_\nu)}\langle\varepsilon_\nu\rangle\right)^2
\frac{\langle\varepsilon_\nu\rangle}{m_n}\frac{F_6(\eta_\nu)}{F_3(\eta_\nu)}
\left(\langle\varepsilon_\nu\rangle\Phi\frac{F_2(\eta_\nu)}{F_3(\eta_\nu)}-6T\frac{F_5(\eta_e)}{F_6(\eta_e)}\right)\right].
\label{nunscatt}
\eeq
$\kappa=\sigma_o/(16m_e^2)(1+3g_A^2)$ for neutron scattering and $\kappa=\sigma_o/(4m_e^2)
[4\sin^4\theta_W-2\sin^2\theta_W+(1+3g_A^2)/4]$ for neutrino-proton scattering, where $\sin^2\theta_W\simeq0.231$ and
$g_A(\simeq-1.26)$ is the axial-vector coupling constant.

\subsection{$\nu_i\bar{\nu_i}\leftrightarrow e^+e^-$}
\label{sec:pairpr}

Cooling due to $e^+e^-$ annihilation, assuming relativistic electrons and positrons,
and ignoring Pauli blocking in the final state, can be written as
\beq
C_{e^+e^-\rightarrow\nu\bar{\nu}}\simeq
\,1.4\times10^{17}\,
\frac{T^9}{\rho_8}\,f(\eta_e)
\,\,\,{\rm ergs\,\,\,g^{-1}\,\,\,s^{-1}},
\label{coldpr}
\eeq
where
\beq
f(\eta_e)\,=\,\frac{F_4(\eta_e)F_3(-\eta_e)+F_4(-\eta_e)F_3(\eta_e)}{2F_4(0)F_3(0)}.
\eeq
$\rho_8$ is the mass density in units of $10^8$ g cm$^{-3}$ and $T$ is in MeV.
We employed  eq.~(\ref{coldpr}) with and without the
$\eta_e$ dependence and compared it to the results using the fit given in \cite{itoh1996}.
Such a modification amounted to no more than a 1-2\% change in the entropy, dynamical timescale,
total energy deposition, or mass outflow rate.  

Much more uncertain than eq.~(\ref{coldpr}) is the heating due to $\nu\bar{\nu}\rightarrow e^+e^-$.  
The spherical dilution of this process from the neutrinospheres is complicated.
For a given flux, the local energy density depends sensitively on $\langle\mu\rangle$,
which can only be properly treated by solving the full transport problem.  
We save such an investigation for a later work.  Instead, we
compare two approaches.  The first is based on the vacuum spherical dilution approximation, written simply as
(Qian and Woosley 1996)
\beqa
H&\simeq&1.6\times10^{19}\frac{\Psi(x)}{\rho_8\,R_{\nu\,6}^4}\,\,\,\Phi^9\,\,\,{\rm erg \,\,\, g^{-1}\,\,\,s^{-1}}\nonumber \\
&\times&\left[\lumanue^{51}\lumnue^{51}(\aveanue+\avenue)+\frac{6}{7}(L^{51}_{{\nu}_\mu})^2\aveunu\right],
\label{heatpr}
\eeqa
where $\Psi(x)=(1-x)^4(x^2+4x+5)$, $x=(1-(R_\nu/r)^2/\Phi^2)^{1/2}$, $R_{\nu\,6}$ is the 
neutrinosphere radius in units of 10$^6$ cm, and $L^{51}_{\nu_i}$ is the neutrino luminosity in units of 10$^{51}$ erg s$^{-1}$.
The redshift term, $\Phi$, appearing in $x$, accounts for the amplification of this process 
due to the bending of null geodesics in general relativity as in eq.~(\ref{null}) (Salmonson and Wilson 1999).
We compared this approximation to the heating rate obtained by \cite{janka1991}.  
Because of the extreme nature of the density gradient just exterior to 
the protoneutron star, we find that the vacuum approximation adequately characterizes the energy deposition.  
In addition, the fact that there is no obvious way to include the general relativistic effects in 
the parameterization of $\langle\mu\rangle$ by \cite{janka1991} led us to employ eq.~(\ref{heatpr}) 
in the wind models we present here.

\subsection{Other Possible Neutrino Processes}

We were motivated to consider nucleon-nucleon bremsstrahlung, plasma, and 
photo-neutrino processes by the sensitivity of the dynamical timescale and asymptotic entropy
to changes in the energy deposition profile (see \S\ref{sec:energy_source} and Qian and Woosley 1996).
We found that none of these processes contributed significantly.  Qian and Fuller (1995a,b) have 
addressed the possibility that neutrino oscillations may effectively provide an extra energy source,
at larger radii, beyond the point where the mass outflow rate is determined.  In \S\ref{sec:energy_source},
we include an ad hoc energy source to test the sensitivity of our results to changes in $\dot{q}(r)$,
but do not address the issue of neutrino oscillations directly.


\section{RESULTS: FIDUCIAL MODELS}
\label{sec:fiducial}

With a robust and efficient means by which to solve the wind problem, coupled with physical boundary
conditions and a well-motivated neutrino heating algorithm we can now survey the protoneutron star wind
parameter space.

Supernova and protoneutron star calculations, coupled with observations of
neutron star binary systems and our knowledge of the high density nuclear equation of state,
place useful limits on the parameter space that protoneutron stars actually inhabit.
Particularly, we are interested in a range of protoneutron star masses from 1.2 to 2.0 M$_\odot$,
total neutrino luminosities from $4\times10^{52}$ to $1\times10^{51}$ erg s$^{-1}$,
average neutrino energies as high as 35 MeV for $\nu_\mu$ neutrinos and 15-20 MeV for $\nu_e$ and $\bar{\nu}_e$ neutrinos
with $\avenue<\aveanue<\aveunu$, and a range of neutron star radii from 20 km to perhaps 9 km.

Our main goal in this section is to map the possible protoneutron star wind parameter space, taking as input 
the physical ranges specified above.  In what follows, we present the basic structure of the wind and
identify some of the systematics we can use to assess this site as a candidate for $r$-process
nucleosynthesis. Particularly, we include a discussion of the dynamical timescale, the electron fraction, 
and the asymptotic entropy.  In identifying some of the approximate power law relations which characterize the wind,
we refer to {\it low} and {\it high} total neutrino luminosities, with {\it low} denoting $\lesssim10^{52}$ erg s$^{-1}$
and {\it high} meaning $\gtrsim10^{52}$ erg s$^{-1}$.  We write these power laws in terms of $L_\nu$ and 
$\langle\varepsilon_\nu\rangle$.  These stand for a representative luminosity and a representative average energy, respectively.
That is, we take $\lumanue/\lumnue$, $\lumanue/\lumunu$, $\aveanue/\avenue$, and $\aveanue/\aveunu$ as constant.

We employ the notation $L_{\nu}^{51}$ to refer to luminosities in units of 10$^{51}$ erg s$^{-1}$. 
All neutrino luminosities quoted  throughout this paper are local quantities at the neutron star surface.  
The luminosity at infinity can be obtained from the luminosities quoted here by multiplying the local luminosity
by two powers of the gravitational redshift  (eq.~\ref{red}).  For example, taking $M=1.4$ M$_\odot$ and $R_\nu=10$ km,
$\Phi^2\simeq0.58$.  In order to keep track of our models, we use the $\bar{\nu}_e$ neutrino luminosity ($\lumanue$) 
to label each individual model.  The total luminosity can then be obtained from the ratios $\lumanue/\lumnue$ and 
$\lumanue/\lumunu$.  Finally, in our expressions for $\dot{q}$ we included  $\eta_{{\nu}_i}$ terms 
for completeness.  In what follows, we take $\eta_{\nu_e}=\eta_{\bar{\nu}_e}=\eta_{\nu_\mu}=0$.
For an assessment of this assumption, see \cite{janka_hillebrandt} and \cite{myra}.


\subsection{The Structure of Neutrino-Driven Winds}
\label{sec:structure}

\begin{figure} 
\vspace*{6.0in}
\hbox to\hsize{\hfill\includegraphics{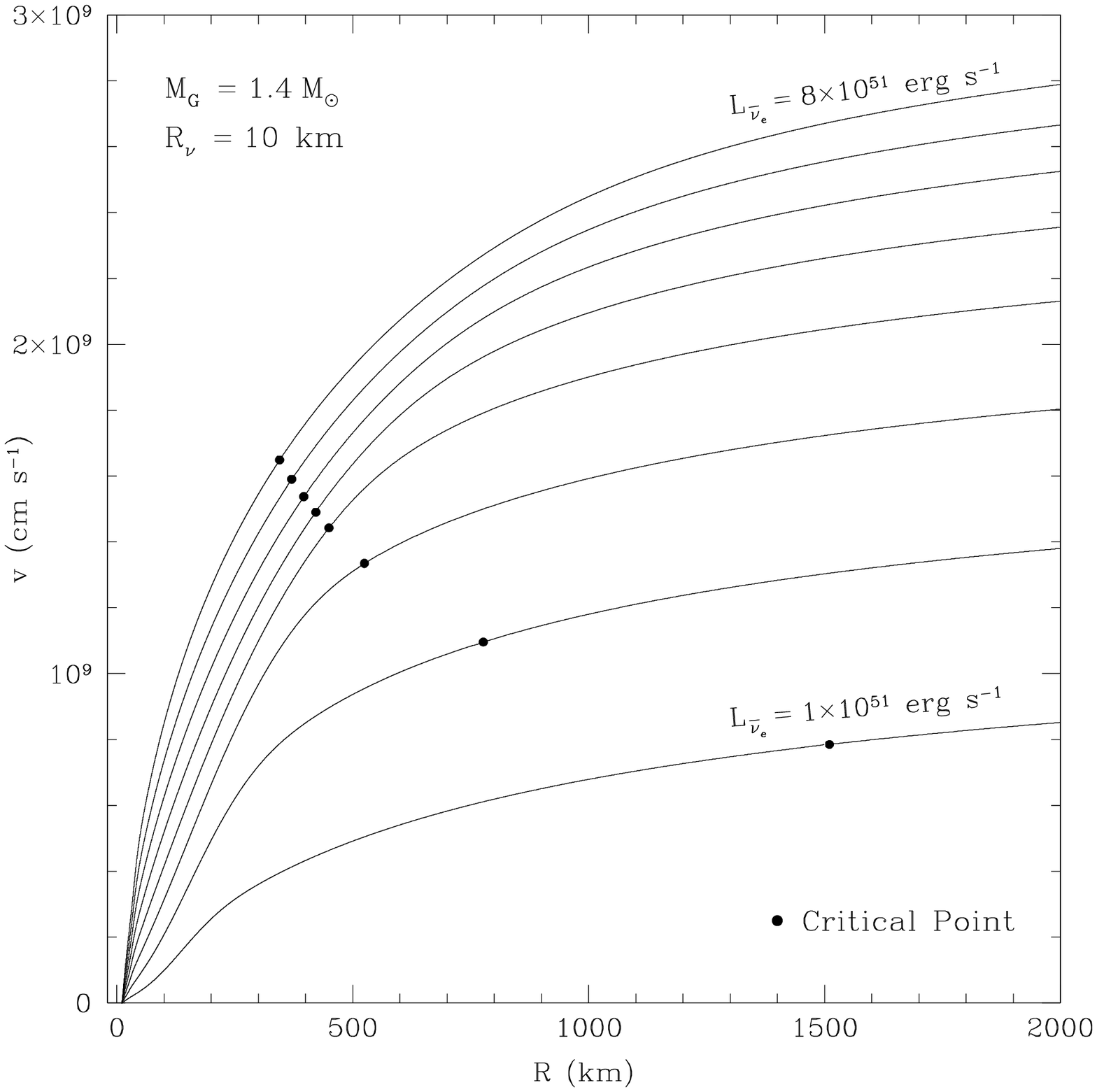}\kern+6in\hfill}
\caption{Matter velocity (${\rm v}$) in cm s$^{-1}$ as measured in the Schwarzschild frame as a function of radius ($R$) in km 
for a 1.4 $M_\odot$ 
(gravitational) protoneutron star with $L_{\bar{\nu}_e}=8$, 7, 6, 5, 4, 3, 2, and 1$\times10^{51}$ erg s$^{-1}$. 
For $L_{\bar{\nu}_e}=8\times10^{51}$  erg s$^{-1}$, we set $\avenue=11$ MeV, $\aveanue=14$ MeV, and $\aveunu=23$ MeV.   
For each subsequent luminosity, the average neutrino energy for each species  was decreased 
according to $\langle\varepsilon_\nu\rangle\propto L_\nu^{1/4}$.  The luminosities were set in the ratios
$\lumanue/\lumnue=1.3$ and $\lumanue/\lumunu=1.4$.
The dots mark the critical point for each wind profile, where the matter velocity is equal to the local speed of sound.  
The neutrinosphere radius is held fixed at 10 km. For all profiles, for $r\lesssim60$ km, the flow is nearly homologous with 
$v\propto L_\nu r$. \label{fig:vp} }
\end{figure}

\begin{figure} 
\vspace*{6.0in}
\hbox to\hsize{\hfill\includegraphics{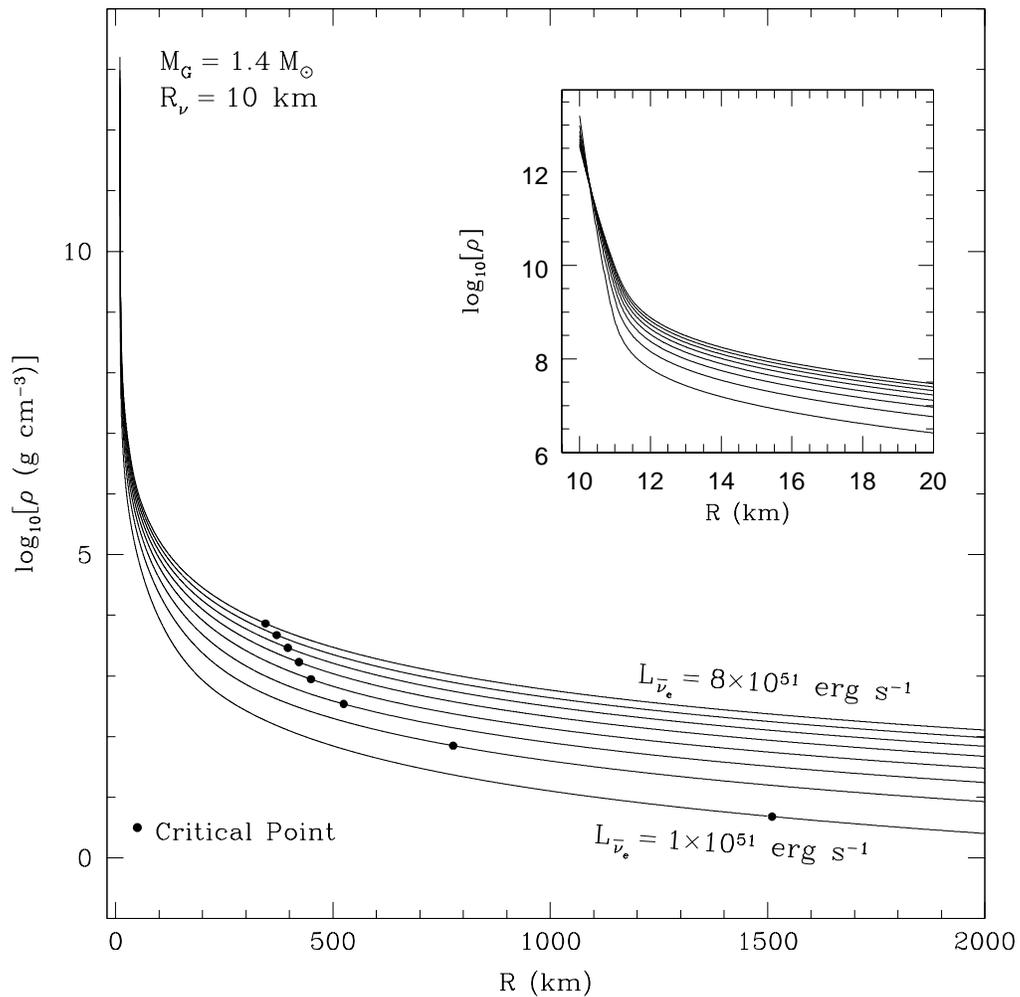}\kern+6in\hfill}
\caption{Log$_{10}$ of the mass density ($\rho$) in g cm$^{-3}$ versus radius ($R$) in km for the same range of neutrino 
luminosities as in Fig.~\ref{fig:vp} and for the same protoneutron star characteristics.  Dots mark the critical point.  
The inset shows
$\log_{10}\rho$ versus $R$ for the region close to the protoneutron star.  
Note the steep density gradient, which drops precipitously over as much as five orders of magnitude
in just over a kilometer.  As the neutrino luminosity decreases, $\rho(R_\nu)$ increases
in order to maintain our inner integral boundary condition on the $\nu_e$ neutrino optical depth (eq.~\ref{optical}).\label{fig:rp}}
\end{figure}

\begin{figure} 
\vspace*{6.0in}
\hbox to\hsize{\hfill\includegraphics{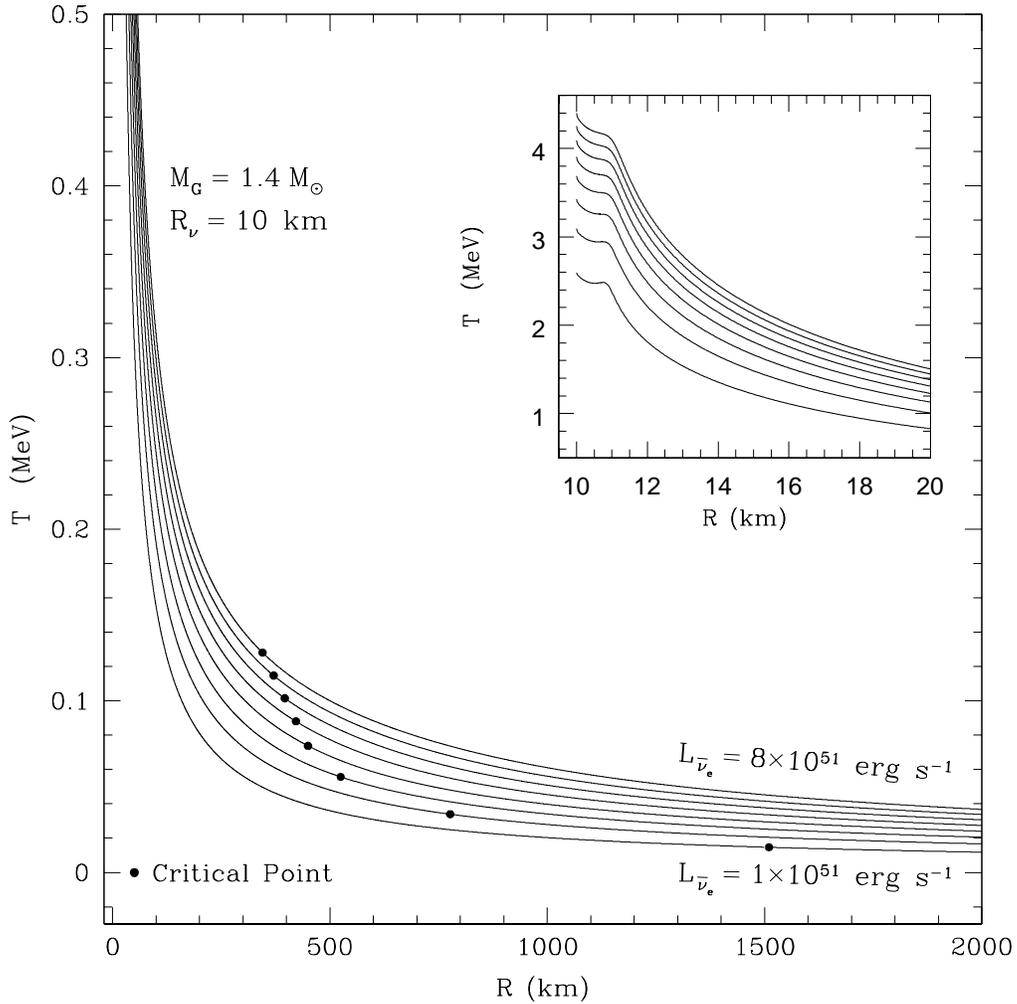}\kern+6in\hfill}
\caption{Matter temperature (${\rm T}$) in MeV versus radius in km for the same profiles as in Fig.~\ref{fig:vp}.
Dots mark the critical point.  Note that the important regime of possible $r$-processing lies 
between
0.5 MeV and approximately 0.08 MeV, a region extending out to 700 km for the highest luminosities and 
to less than 200 km for the lowest luminosities shown.  Comparing the temperature in this range of radii to
those in Fig.~\ref{fig:vp}, it's clear that the assumption of constant outflow velocity (Meyer and Brown 1997) is a 
poor approximation during nucleosynthesis in neutrino-driven winds. The inset shows the structure of the temperature 
as a function of radius very close to 
the neutrinosphere.  The bump in ${\rm T}$ is caused by the onset of heating (compare with Fig.~\ref{fig:qp}).\label{fig:tp}}
\end{figure}

Figures \ref{fig:vp}, \ref{fig:rp}, and \ref{fig:tp} show the velocity, mass density, and
temperature as a function of radius for eight different neutrino luminosities.  
For these models, $R_\nu$ is 10 km and the neutron star gravitational mass is 1.4 M$_\odot$. 
The average neutrino energies were set at $\avenue=11$ MeV, $\aveanue=14$ MeV, and $\aveunu=23$ MeV for 
the highest luminosity.  For each subsequent luminosity, the average energies were decreased 
according to $\langle\varepsilon_\nu\rangle\propto L_\nu^{1/4}$.  The luminosities 
were held in the ratios $\lumanue/\lumnue=1.3$ and $\lumanue/\lumunu=1.4$.  In these figures, critical points ($v=c_s$)
are shown as dots. Table \ref{tab:fiducial} lists the global properties of each of these models, including the
integrated energy deposition rate, $Q$ (eq.~\ref{bigQtot}),
and the mechanical luminosity or hydrodynamical power, $P_{\rm mech}=\dot{M}v_a^2/2$, where $v_a$ is the asymptotic velocity.
In addition, we include the asymptotic entropy $s_a$ and dynamical timescale defined as the $e$-folding time of the 
density at $T=0.5$ MeV:
\beq
\tau_\rho=\frac{1}{vy}\left|\frac{1}{\rho}\frac{\p\rho}{\p r}\right|^{-1}.
\label{taurho}
\eeq

\begin{figure} 
\vspace*{6.0in}
\hbox to\hsize{\hfill\includegraphics{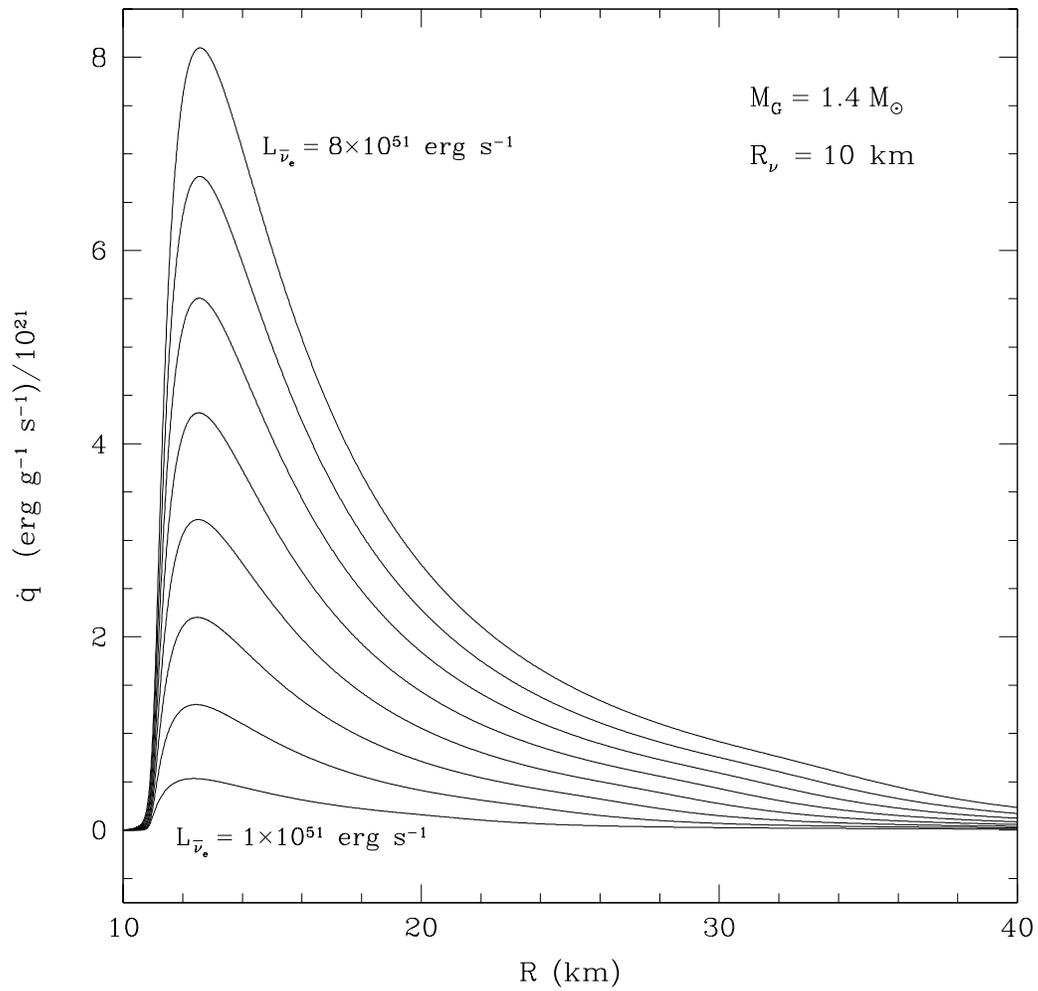}\kern+6in\hfill}
\caption{Energy deposition rate ($\dot{\rm q}$) in units of $10^{21}$ erg g$^{-1}$ s$^{-1}$ as a function of radius (in km) 
for the wind models in Figs.~\ref{fig:vp}$-$\ref{fig:tp}.
$\dot{q}$ profiles for $\bar{\nu}_e$ neutrino luminosities from 
$8\times10^{51}$ erg s$^{-1}$ to $1\times10^{51}$ erg s$^{-1}$ are depicted.
The total heating rates ($Q$; eq.~\ref{bigQtot}) for each of these models are given in Table \ref{tab:fiducial}.\label{fig:qp}}
\end{figure}

Figure \ref{fig:vp} shows clearly that even relatively close to the neutron star ($r\sim400$ km)
the flow is not homologous (i.e., $v$ is not proportional to $r$).  At much smaller radii, however, in the heating 
region ($r\lesssim60$ km, compare with Fig.~\ref{fig:qp}) the flow is homologous with $v\propto L_\nu r$.  We have 
found that simple parameterizations of $v(r)$, particularly for low neutrino luminosities, are not straightforward. 
Although not readily apparent, the asymptotic velocity ($v_a$) is a power law in $L_\nu$.  For 
high $L_\nu^{\rm tot}$, $v_a\propto L_\nu^{0.3}$. For low $L_\nu$, the index is $\sim0.46$.
The critical radius ($R_c$) also increases as a power of luminosity for low $L_\nu$.
For $M=1.4$ M$_\odot$, the index is $\sim0.95$, while for $M=2.0$ M$_\odot$ $R_c\propto L_\nu^{-0.85}$.
Note that if the protoneutron star cools as $L_\nu\propto t^{-0.9}$ then we should expect 
$R_c$ to grow linearly with time, implying that at late times the velocity of
the critical point away from the neutron star is approximately constant.  

\begin{figure} 
\vspace*{6.0in}
\hbox to\hsize{\hfill\includegraphics{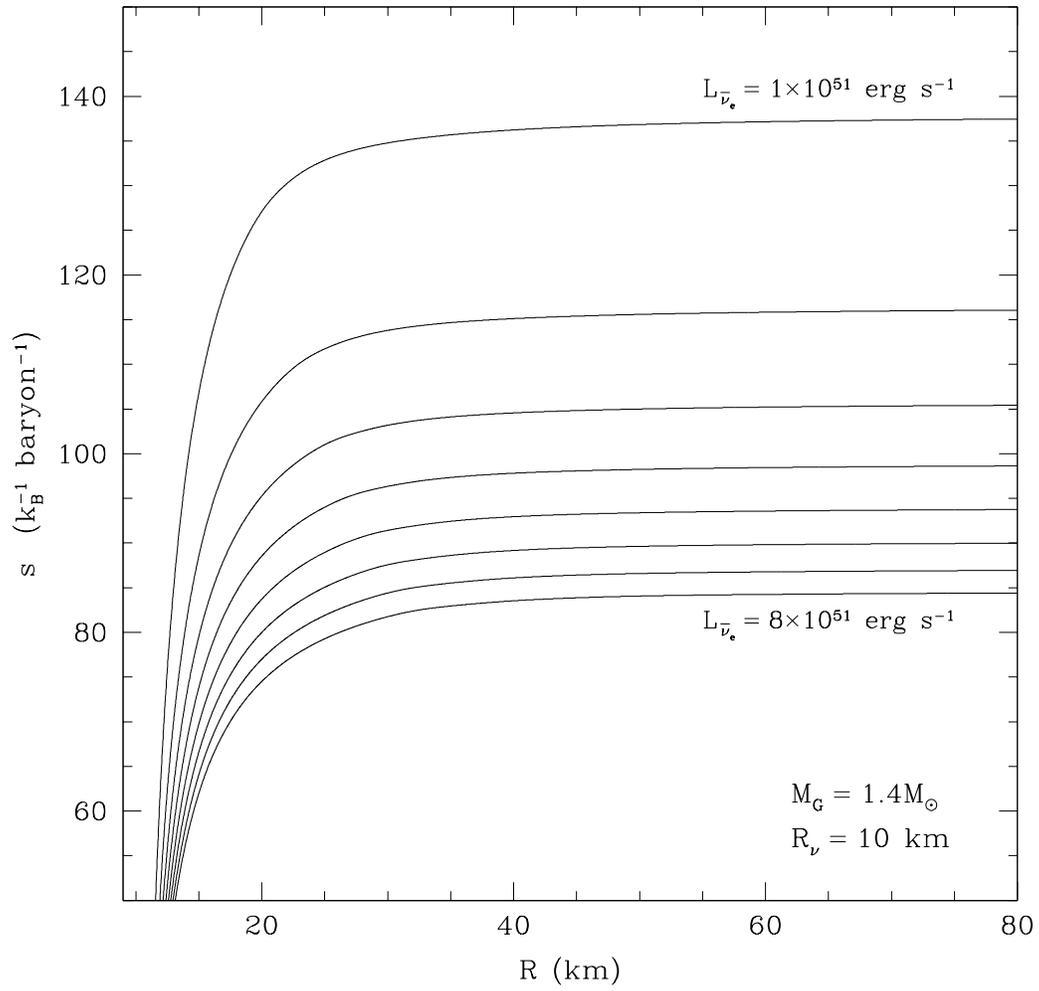}\kern+6in\hfill}
\caption{Entropy (${\rm s}$, per baryon per k$_{\rm B}$) versus radius in km for the same protoneutron star wind models
as in Figs.~\ref{fig:vp}-\ref{fig:qp}.  
Note that the entropy asymptotes quickly;
comparing this figure to Fig.~\ref{fig:tp}, we can see that in most cases $s$ increases less than 10 units for $T<0.5$ MeV. \label{fig:sp} }
\end{figure}

The stiff nature of the wind equations 
is manifest in Fig.~\ref{fig:rp}.  The inset shows that $\rho$ drops by more than four orders of magnitude in just 1-3 km, before the
neutrino heating rate reaches a maximum.  
Figure \ref{fig:qp} shows the total specific heating rate for each of the eight fiducial wind models.  Note that the
maximum in $\dot{q}$ occurs very close to the protoneutron star surface and that the position of this 
maximum is not a function of $L_\nu$.
At $r\sim12$ km the density gradient changes dramatically,
as the slow moving material in this inner atmosphere accelerates to infinity.  
Concomitant with this change in
$\rho(r)$ and the peak in $\dot{q}$ is a rise in $T$ and most of the entropy production. Figure \ref{fig:sp}
shows the entropy as a function of radius.  Note that
the entropy quickly comes to within a few percent of its asymptotic value ($s_a$).  
We find that $s_a$ is proportional to $L_\nu^{-0.24}$ for all luminosities, for 1.4 M$_\odot$.
For $M=2.0$ M$_\odot$, $s_a$ is proportional to $L_\nu^{-0.25}$.
As $L_\nu$ decreases, the gradients in $\rho$, $T$, and the entropy near $R_\nu$ become larger. 
Inspection of Fig.~\ref{fig:tp} shows that the $0.1$ MeV $\lesssim T\lesssim 0.5$ MeV region where 
the $r$-process (and preceding $\alpha$-process) might take place, lies between
$50$ km and  $500$ km for $\lumanue^{51}=8.0$ and between $25$ km and $150$ km for $\lumanue^{51}=1.0$.  
Importantly, Fig.~\ref{fig:vp} shows that the wind is still 
accelerating significantly at these radii.  This calls into question the assumption
of a constant velocity outflow used by many $r$-process modelers (e.g., Meyer and Brown 1997).

\begin{figure} 
\vspace*{6.0in}
\hbox to\hsize{\hfill\includegraphics{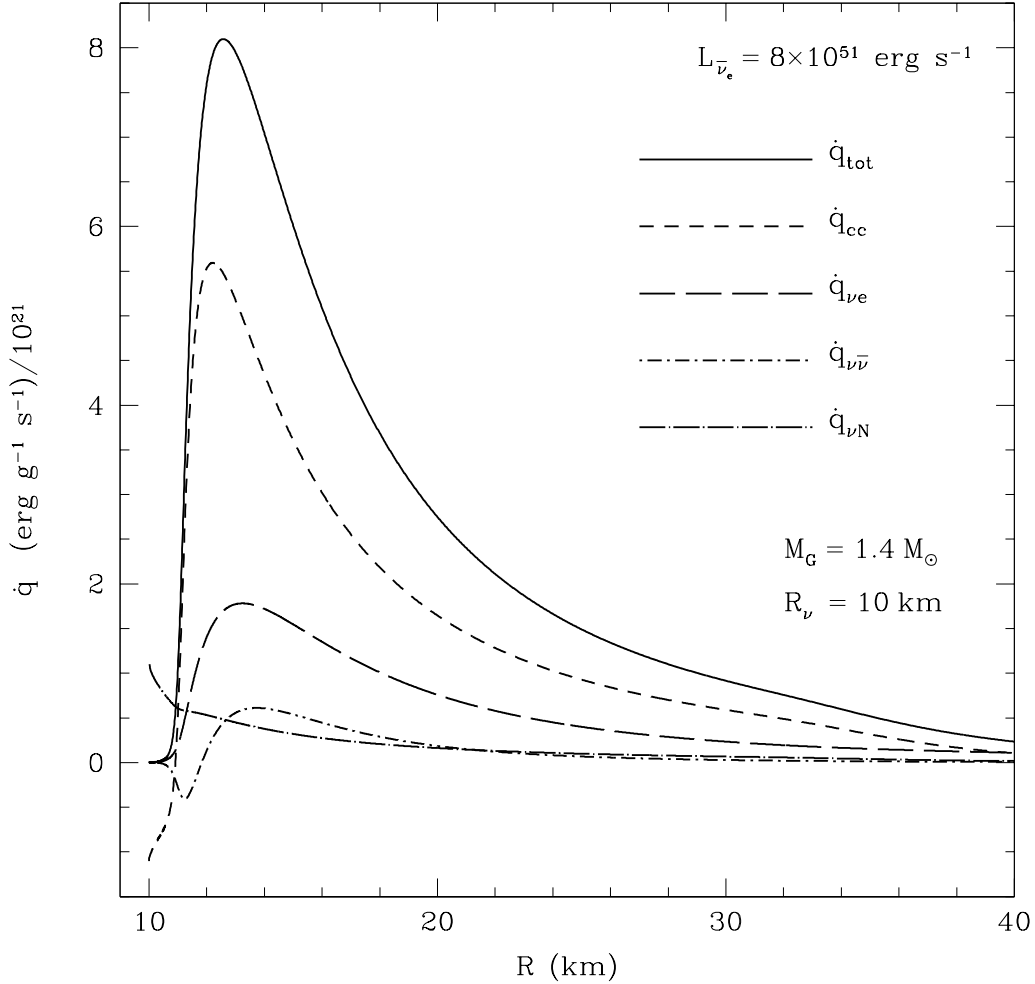}\kern+6in\hfill}
\caption{Contributions of specific neutrino processes to the energy deposition rate for 
$\lumanue=8\times10^{51}$ erg s$^{-1}$, $R_\nu=10$ km, and $M=1.4$ M$_\odot$.  The solid line shows the total heating rate
due to all processes, the short-dashed line is the net contribution from the charged-current reactions ($\dot{q}_{cc}$)
(eqs.~\ref{ccc} and \ref{hcc}), the long dashed line is neutrino-electron/positron scattering ($\dot{q}_{\nu e}$; eq.~\ref{qdotes}), 
the short dot-dashed line is the net energy deposition rate due to $\nu_i\bar{\nu}_i\leftrightarrow e^+e^-$  
($\dot{q}_{\nu\bar{\nu}}$; eqs.~\ref{coldpr} and \ref{heatpr}), 
and the long dot-dashed line is for neutrino-nucleon scattering ($\dot{q}_{\nu N}$; eq.~\ref{nunscatt}).  
Note the fairly rapid decrease in 
$\dot{q}_{cc}$ at $r\sim35$ km is due to the recombination of free neutrons and protons into $\alpha$ particles.  
As the neutrino luminosity decreases in these models, this transition moves inward in radius so that, for the lowest
luminosities, the charged current processes end abruptly at $\sim25$ km.  For models with larger $M/R_\nu$
$\dot{q}_{\nu e}$ and $\dot{q}_{\nu\bar{\nu}}$ become more important relative to $\dot{q}_{cc}$.  
However, even for $M=2.0$ M$_\odot$ and $R_\nu=10$ km, $\dot{q}_{cc}$ dominates at the peak in $\dot{q}_{\rm tot}$.\label{fig:qip}}
\end{figure}

\begin{figure} 
\vspace*{6.0in}
\hbox to\hsize{\hfill\includegraphics{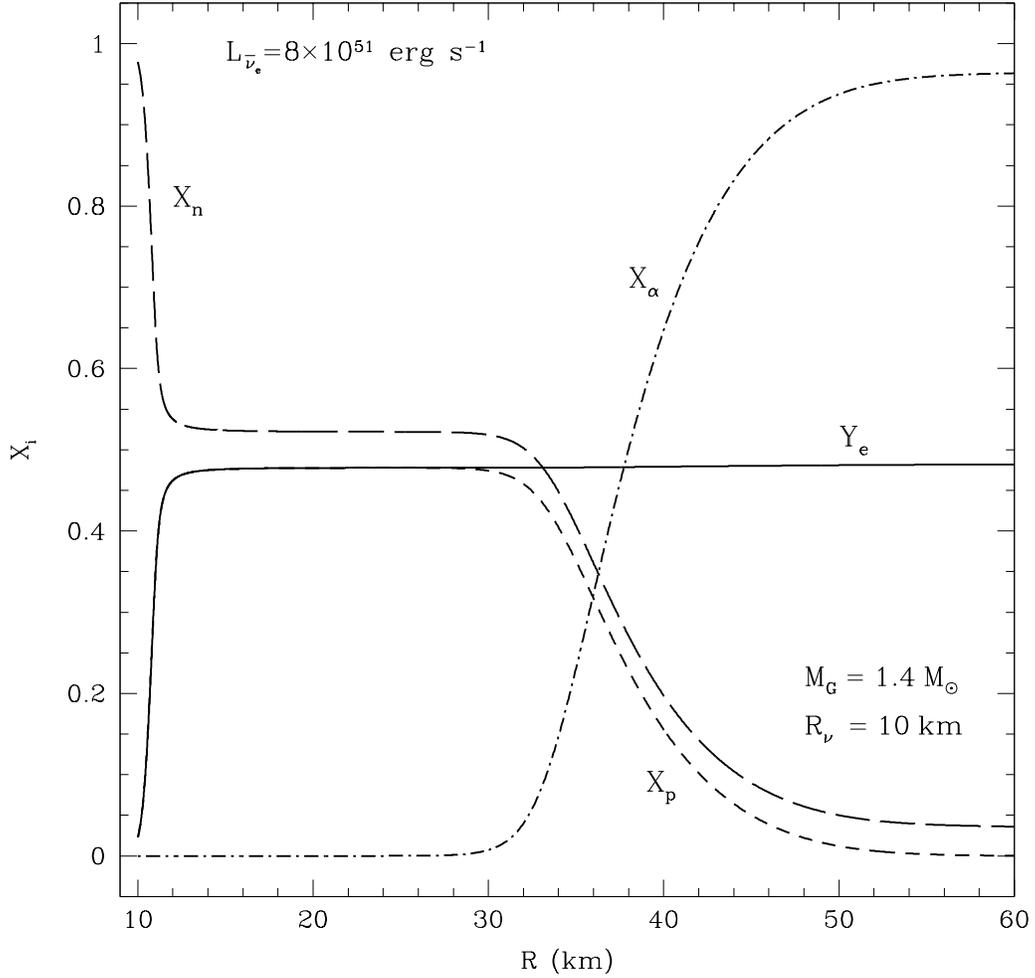}\kern+6in\hfill}
\caption{The electron fraction ($Y_e$, solid line), proton fraction ($X_p$, short dashed line),
neutron fraction ($X_n$, long dashed line), and alpha fraction ($X_\alpha$, dot-dashed line) for
the highest luminosity wind model ($\lumanue=8\times10^{51}$ erg s$^{-1}$) in Figs.~\ref{fig:vp}-\ref{fig:sp},
corresponding to the heating profile in Fig.~\ref{fig:qip}.  Note the transition at $r\sim35$ km
where $\alpha$ particles form and effectively shut off the charged-current heating rate.  The asymptotic $Y_e$ is set
very close to the neutron star ($r\sim15$ km) and only undergoes small subsequent change as a result of the
$\alpha$-effect (Fuller and Meyer 1995; 
McLaughlin, Fuller, and Wilson 1996)).  In this case, $Y_e$ changes by $\sim$1\% beyond $r=20$ km. 
The asymptotic electron fraction, $Y_e^a$, is well-approximated by eq.~(\ref{yeqw}) in the text.\label{fig:x}}
\end{figure}

Figure \ref{fig:qip} shows a breakdown of the total energy deposition into its separate components 
for the $L_{\bar{\nu}_e}^{51}=8.0$ fiducial model.  All of the heating processes conspire to give $\dot{q}(R_\nu)=0$,
in accordance with our boundary condition.  At the surface, the charged-current heating rate ($\dot{q}_{cc}$) 
provides net cooling, which balances heating
from neutrino-nucleon scattering ($\dot{q}_{\nu N}$).  $\dot{q}_{cc}$ generally dominates the other heating
processes in the models we consider here.  Note that heating due to neutrino-electron/positron scattering 
($\dot{q}_{\nu e}$) does not contribute as significantly as is indicated by \cite{qw1996} and \cite{otsuki},
because of their simplifying assumption that the energy transfer per scattering is $\sim\varepsilon_\nu/2$
instead of $\sim(\varepsilon_\nu-4T)/2$ (\S\ref{sec:nuescatt}).  
Comparing Fig.~\ref{fig:qip} with Fig.~\ref{fig:x}, we 
can see that the drop in $\dot{q}_{cc}$ at $r\simeq35$ km is due to the formation of $\alpha$ particles.  As the luminosity 
decreases this transition region moves in, so that for $\lumanue^{51}=0.60$, $\dot{q}_{cc}\rightarrow0$ near $r\sim22$ km.
As the ratio of $M/R_\nu$ increases, the heating rate due to $\nu_i\bar{\nu}_i\rightarrow e^+ e^-$ ($\dot{q}_{\nu\bar{\nu}}$)
increases substantially relative to both $\dot{q}_{\nu e}$ and $\dot{q}_{cc}$. 
In fact, for $M/R_\nu=2.0\,{\rm M}_\odot/10$ km, the peak in $\dot{q}_{\nu \bar{\nu}}$ is actually slightly 
higher than that for $\dot{q}_{\nu e}$.  However, $\dot{q}_{cc}$ still dominates heating at the
peak in $\dot{q}_{\rm tot}$ by a factor of $\sim1.5$.
Note that an increase in $L_{\nu_\mu}$ or $\aveunu$ increases the importance
of $\dot{q}_{\nu e}$ and $\dot{q}_{\nu \bar{\nu}}$ relative to $\dot{q}_{cc}$.  In particular, this increases the
energy deposition at larger radii.  We explore some of these effects on the asymptotic character of the wind 
in \S\ref{sec:energy_source}.

\cite{qw1996} found that $\dot{M} \propto L_\nu^{2.5}M^{-2}$ for Newtonian gravity.  We find that the index
of this power law in $L_\nu$ is slightly decreased in general relativity to 2.4$-$2.5,  but that
the dependence on $M$ is stronger; $\dot{M}$ is approximately proportional to $M^{-3.1}$ for high luminosities. The
index decreases to $-2.7$ for low luminosities.  For the full range of luminosities and masses considered
here, the mass outflow rate in our general-relativistic calculations
is approximately a factor of three smaller than in our Newtonian wind models.

Computing the volumetric integral of $\dot{q}$ as in eq.~(\ref{bigQtot}) yields 
the net energy deposition rate, $Q$ (see Table \ref{tab:fiducial}).
For the heating function we employ in this paper, $Q$ is roughly proportional to $L_\nu^{2.4}$ for all protoneutron
star masses and luminosities.  The small variations in this power law index compliment  
the variation in $\dot{M}$ with $L_\nu$ so that
the ratio $Q/\dot{M}$ varies by less than 2\% over the whole range of $L_\nu$ for each mass.
In all cases $Q/\dot{M}\simeq GM/R_\nu$ to within 10\%.
Whereas \cite{qw1996} found that $Q\propto M^{-1}$, we find 
a more stiff dependence on mass; for high $L_\nu$, $Q$ is proportional to $M^{-1.9}$ for 1.4 M$_\odot$ $\leq M\leq 1.6$ M$_\odot$.  
In addition, we find that their analytic
expression for $Q$ consistently underestimates the net energy deposition in our Newtonian models by as 
much as $\sim50$\%.

The wind mechanical power $P_{\rm mech}$ is proportional to $L_\nu^{3.4}$ for low $L_\nu$ and $M=1.4$ M$_\odot$.
This index decreases to 3.2 for $M=2.0$ M$_\odot$.  At high luminosity for all masses, $P_{\rm mech}\propto L_\nu^{3.2}$. 
Although the asymptotic velocity is set by the escape velocity from the protoneutron star and therefore
increases with increasing $M$, it does not increase as a power law in general relativity.  In fact, it increases more rapidly.
Therefore, even though $\dot{M}$ is approximately proportional to $M^{-2.5}$ at low luminosities, the increase in $v_a$ as
$M$ gets large forces $P_{\rm mech}$ to increase.  

The efficiency of energy deposition, $Q/L_\nu^{\rm tot}$,
ranges from $10^{-3}$ to $10^{-5}$ as $L_\nu$ decreases.  The efficiency of
conversion of neutrino energy to hydrodynamical power, $P_{\rm mech}/L_\nu^{\rm tot}$, ranges from 
only $2\times10^{-5}$ to less than $6\times10^{-8}$ for the models considered here.  
The efficiency for the conversion of net energy deposition to hydrodynamical power, $P_{\rm mech}/Q$,
decreases with luminosity as $L_\nu^{1.4}$ for 1.4 M$_\odot$ with $P_{\rm mech}/Q\simeq2.4\times10^{-3}$ for 
$\lumanue^{51}=0.6$.  This means that almost all of the net energy deposition goes into
overcoming the gravitational potential.  The excess energy, manifest at infinity as the mechanical
power, is very small in comparison with $Q$.  These quantities may
be potentially important if the wind is to emerge and escape to infinity in the expanding supernova
envelope.

\subsection{The Effects of General Relativity}

Over the range of masses presented here, we find significant enhancements in the entropy per baryon
using the full general-relativistic framework.  Over a broad range of luminosities for the 1.4 M$_\odot$ 
protoneutron star we find that $s_a$ is 25$-$30 units less in our Newtonian calculations than in our analogous 
general-relativistic calculations. Typical reductions in $\dot{M}$ and $Q$ are of order a factor of three 
and two, respectively. 
These differences were anticipated by \cite{qw1996}
and \cite{cardall} and more recently realized in the wind calculations of \cite{otsuki}.
As the latter showed, the general-relativistic  effects on $s_a$ are much more
the result of using the general-relativistic hydrodynamic formulation than of incorporating the general-relativistic 
corrections to $\dot{q}$ expressed in eqs.~(\ref{red}) and (\ref{null}).  
The inclusion of general relativity in the hydrodynamics
makes the structure of the protoneutron star more compact than in a Newtonian description.  This
makes the temperature and density gradients steeper just exterior to $R_\nu$ and particularly in the 
heating region.  Although $dT/dr$ decreases rapidly, $d\rho/dr$ drops much faster.  This effectively increases 
the specific energy deposition per unit mass and the entropy is enhanced significantly.

For comparison, we calculated several wind models with eqs.~(\ref{grv})$-$(\ref{grt}) 
that included the enhancement to $\dot{q}$ due to the bending of null geodesics, but did not include any 
redshift factors on $L_\nu$ or $\langle\varepsilon_\nu\rangle$.  For the 1.4 M$_\odot$ model with $\lumanue^{51}=8.0$
(see Table \ref{tab:fiducial}) this change nearly doubled both $Q$ and $\dot{M}$, while decreasing
$\tau_\rho$ from 3.68 to 1.95 ms.
Although more heating occurred at larger radii,
the peak in $\dot{q}$ also increased so as to offset any potential gain in $s_a$.  The net enhancement was just
6 units in entropy.  Conversely, keeping all the redshift
terms and eliminating $\Phi$ from $\Xi(r)$ in eq.~(\ref{null}) increased $\tau_\rho$ by more than 30\% and
left $s_a$ virtually unchanged.

Because the bending of null geodesics increases the net energy deposition close to the neutron star and
the redshift terms act to decrease energy deposition over the whole profile, it is clear from
this comparison that only an increase in $\dot{q}(r)$ that does not significantly increase $Q$ can have large effects on 
$s_a$ (see \S\ref{sec:energy_source} and Qian and Woosley 1996).

\subsection{The Electron Fraction}
\label{sec:yefiducial}

Figure \ref{fig:x} shows the evolution of the neutron ($X_n$), proton ($X_p$), alpha ($X_\alpha$), and electron fraction ($Y_e$) 
as a function of radius for the $\lumanue^{51}=8.0$ fiducial model (Table \ref{tab:fiducial}
and Figs.~\ref{fig:vp}-\ref{fig:qip}).  The electron fraction profile is computed using 
eq.~(\ref{yeeq}), solved simultaneously with the wind equations.  
In computing the wind solutions we assume nuclear statistical equilibrium between free nucleons and alpha particles.
$Y_e$ comes to within
a few percent of its asymptotic value ($Y_e^a$) in just the first five kilometers.  
This quick evolution is due primarily to the low matter velocities in the inner region.
For $\lumanue/\lumnue=8/6.15$ and $\aveanue/\avenue=14/11$, eq.~(\ref{yeqw}) predicts that
$Y_e^a\simeq0.478$.  Solving the differential equation we find remarkable agreement:
$Y_e\simeq0.477$ at $r=20$ km.  Just beyond this, for $r\gtrsim30$ km,
free nucleons form $\alpha$ particles and $Y_e$ rises slightly as $X_\alpha$ increases until
$Y_e=0.485$ at $r\simeq150$ km.  This is the $\alpha$ effect, whose import in this context was first 
noted by \cite{fuller_meyer} and \cite{mclaughlin_fuller}.  The magnitude here is only of order 1\%.  
For $Y_e^a<0.5$, we find generally that  the magnitude of the $\alpha$ effect increases as the luminosity decreases
for a given $R_\nu$.  In addition, for models in which $\lumanue/\lumnue$ and $\aveanue/\avenue$ are larger,
and, hence, $Y_e^a$ is naively lower (\`{a} la eq.~\ref{yeqw}), the magnitude of the $\alpha$ effect
is also enhanced.  However, for a reasonable range of $\lumanue/\lumnue$ and $\aveanue/\avenue$ as well
as total neutrino luminosities, the $\alpha$ effect never increases $Y_e^a$ by more than $\sim$10\%.
That is, if $Y_e=0.40$ before $\alpha$ particle formation, we find that the $\alpha$ effect increases $Y_e$ 
to no more than approximately $0.44$.

Since in these fiducial models we decrease the average neutrino energies with luminosity,
the threshold effect in the charged-current reactions, manifest in eq.~(\ref{yeqw}) 
by the neutron/proton mass difference ($\Delta$),
becomes important. Despite the fact that $\lumanue/\lumnue$ and $\aveanue/\avenue$ are maintained
as above, $Y_e^a$ eventually becomes greater than 0.5.  As $Y_e^a$ becomes much larger ($\sim0.52$),
it experiences what might also be termed an $\alpha$ effect: because an $\alpha$ particle has $Y_e=Z/A=0.50$,
the onset of $X_\alpha$ formation {\it decreases} $Y_e^a$.

\subsection{The Dynamical Timescale}

\begin{figure} 
\vspace*{6.0in}
\hbox to\hsize{\hfill\includegraphics{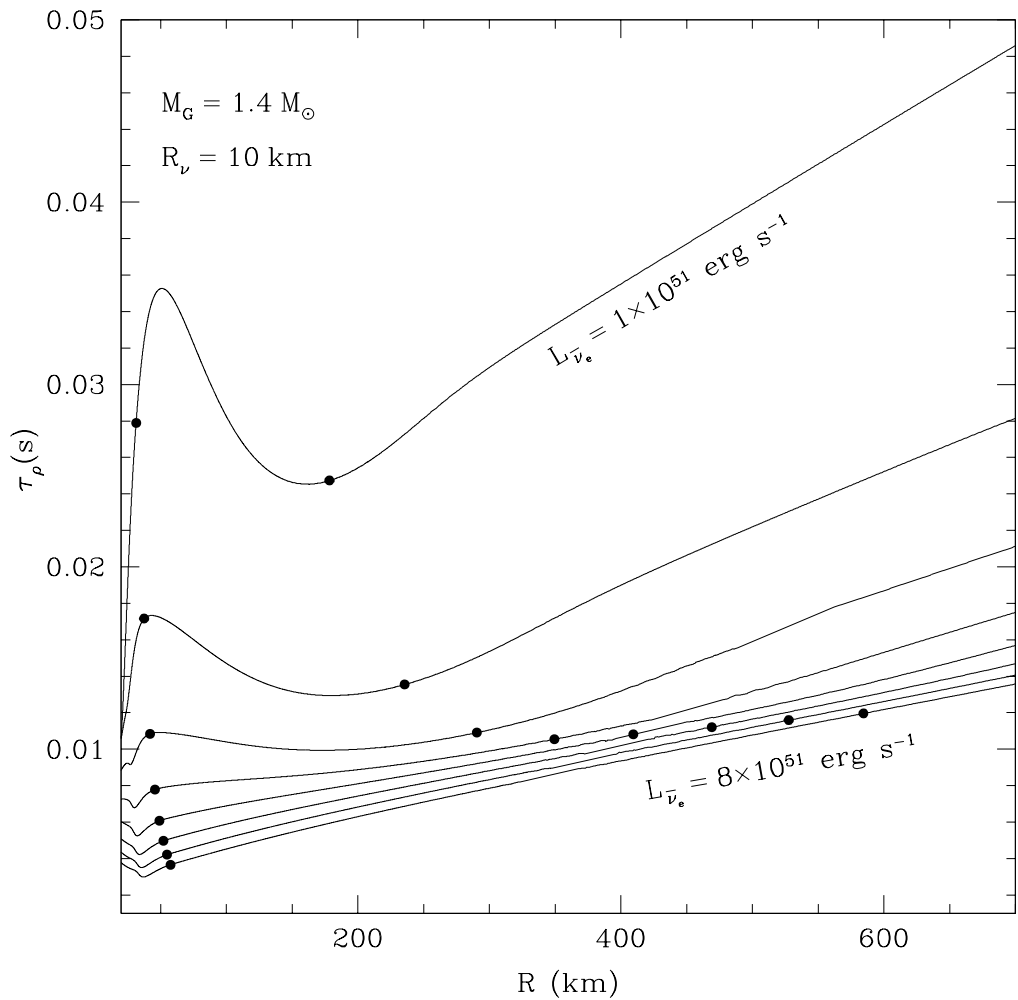}\kern+6in\hfill}
\caption{Dynamical timescale ($\tau_\rho$) defined in eq.~(\ref{taurho}) versus radius for the eight models shown in 
Figs.~\ref{fig:vp}-\ref{fig:sp}.  On each line, the first point marks the radius at which 
$T(r)=0.5$ MeV.  The second point marks where $T(r)=0.1$ MeV.  Note that for high luminosities, $\tau_\rho$ increases
by more than a factor of three over the range of radii and temperatures relevant for $r$-process nucleosynthesis.  For
the lowest luminosities $d\tau_\rho/dr$ changes sign and $\tau_\rho$ actually decreases by $\sim$30\% between the two points on
the $L_{\bar{\nu}_e}=1\times10^{51}$ erg s$^{-1}$ curve.\label{fig:timescale}}
\end{figure}

In discussing our results, in order to make an apposite  comparison with previous studies, we quote $\tau_\rho$ (eq.~\ref{taurho})
at $T=0.5$ MeV. However, using such a scale to characterize the nature of the resulting nucleosynthesis is suspect.
Figure \ref{fig:timescale} shows $\tau_\rho$ as a function of radius for the wind models presented in 
Figs.~\ref{fig:vp}-\ref{fig:sp}. The dots on each line of constant neutrino luminosity mark the
range $0.5$ MeV $\gtrsim T(r)\gtrsim0.1$ MeV, the temperature range relevant for neutron-capture nucleosynthesis.
Although the $r$-process may continue at temperatures well below 0.1 MeV, we include these dots to guide the eye.
The dynamical timescale of the wind, or the expansion timescale, has been defined in several different
ways by many researchers.  No clear consensus exists.  \cite{cardall} defined their dynamical timescale
as the $e$-folding time of the temperature at $T=0.5$ MeV.  \cite{hwq} defined their dynamical timescale in 
the same way, but then used an expansion timescale (1.28 times the dynamical time) to discuss their results.  
\cite{qw1996} and \cite{freiburghaus} used the ratio $r/v$ at $T=0.5$ MeV to characterize the expansion.
Finally, \cite{meyer_brown} connect the $e$-folding time of the density with their expansion timescale,
$r/2v$, by using the equation for $\dot{M}$ and dropping the acceleration term.  With these assumptions,
they obtain
\beq
\tau_{\rho}^{\rm \,MB}(t)=\tau_0\left(1+\frac{t}{2\tau_0}\right),
\label{mbtau}
\eeq
where $t$ is the time on a Lagrangean mass element in the flow. Figure \ref{fig:timescale}
shows clearly that any simple parameterization of the dynamical time, using any definition, is an oversimplification.
For $\lumanue^{51}=8$, $\tau_\rho$ increases by a factor of three over this range 
of temperatures.  
Equation (\ref{mbtau}) captures the increase of $\tau_\rho$ with radius for high luminosities, but 
overestimates the slope by about a factor of two.  
At low luminosities, of course, eq.~(\ref{mbtau}) does not capture $\tau_\rho(t)$ at all.  
At these luminosities, the dynamical timescale actually decreases over this range of temperatures.  This arises
because a region of positive curvature in the $v(r)$ profile of Fig.~\ref{fig:vp} develops between
$40$ km and $200$ km at low luminosities.  This observation simply underscores the danger in considering
a single dynamical time that is meant to characterize an actual wind profile.  

Figure \ref{fig:ffig} shows tracks of constant mass in the plane of $s_a$ versus $\tau_\rho$ for 
luminosities from $\lumanue^{51}=8.0$  to $\lumanue^{51}=0.70$,
for $R_\nu=10$ km.  Although the indices vary slightly for different masses, $s_a\propto L_\nu^{-0.25}$ and
$\tau_\rho\propto L_\nu^{-1.4}$, so that $s_a$ is approximately proportional to $\tau_\rho^{0.2}$.  
One can imagine that these curves represent evolutionary cooling tracks in time in the space of
$s_a$ and $\tau_\rho$ for constant $R_\nu$ and $M$. Simple extrapolation of this power law allows one to estimate at 
what $\tau_\rho$ a given entropy might obtain, for a given mass and neutron star radius.  
For example, the 1.4 M$_\odot$ trajectory in Fig.~\ref{fig:ffig}
will not reach $s_a\simeq200$ until $\tau_\rho\simeq0.4$ seconds. The corresponding neutrino luminosity at this point is
$\lumanue^{51}\simeq0.22$. Knowing that $\dot{M}\propto L_\nu^{2.4}$
allows one to estimate the mass outflow rate as $\dot{M}\simeq1.8\times10^{-8}$ M$_\odot$ s$^{-1}$.
These simple power laws and a knowledge of how $L_\nu$ might behave in time
allow us to put powerful constraints on the likely wind epoch of $r$-process 
nucleosynthesis, as we demonstrate in \S\ref{sec:evolution}.

\cite{wanajo} found that their dynamical timescale saturated at high neutrino luminosities near 
$2-3\times10^{52}$ erg s$^{-1}$.  This conclusion is an artifact of their definition for the dynamical timescale
and their numerical approach to the wind problem.
Our solution shows that for constant average neutrino energies, even up to 
$L_\nu^{\rm tot}\simeq8\times10^{52}$ erg s$^{-1}$, 
$\tau_\rho$ continues to decrease, roughly as $\tau_\rho\propto L_\nu^{-0.75}$.  The entropy also
decreases as the luminosity increases; $s_a\propto L_\nu^{-0.15}$.  Since these power laws are for 
constant average neutrino energy,
we deduce from the fiducial models that $\tau_\rho\propto L_\nu^{-0.75}\langle\varepsilon_\nu\rangle^{-2.6}$
and $s_a\propto L_\nu^{-0.15}\langle\varepsilon_\nu\rangle^{-0.4}$.  These are to be compared with the
analytic results of \cite{qw1996} who did not include general-relativistic effects:
$\tau_{\rm dyn}\propto L_\nu^{-1}\langle\varepsilon_\nu\rangle^{-2}$ 
and $s_a\propto L_\nu^{-1/6}\langle\varepsilon_\nu\rangle^{-1/3}$. 
In addition, although we find that $\tau_\rho$ decreases as $R_\nu$ decreases, for constant 
protoneutron star mass, neutrino luminosity, and average neutrino energy, using either the 
Newtonian or general-relativistic wind equations, we find that $\tau_\rho$ is generally 
10\%$-$15\% shorter in the Newtonian case.
This result is owed in part to the fact that the increase in $\dot{q}$ due to the bending of neutrino 
trajectories is insufficient to counter the decrease in $\dot{q}$ caused by gravitational redshift
of the neutrino luminosity and energy (Cardall \& Fuller 1997).  (See Table \ref{tab:fiducial}
for a comparison between the Newtonian and general-relativistic calculations for the 1.4 M$_\odot$
and $\lumanue^{51}=8.0$ fiducial model.)

\subsection{The Steady-State Approximation}

To conclude this section, we include a few words about the degree to which the protoneutron star wind 
can be considered quasi-stationary.  There are several timescales of importance.  The first is $\tau_d$, the timescale for decay
of the neutrino luminosity, set by the power-law index $\delta$ in the relation $L_\nu\propto t^{-\delta}$: $\tau_d=t/\delta$.  
The second is
the time $\tau_m$ for matter to move from $R_\nu$ to the critical point $R_c$, where it loses
sonic contact with the rest of the flow.  The third relevant timescale is the sound crossing time, $\tau_s$, between
$R_\nu$ and $R_c$.  $\tau_s$ varies from just $\sim10$ ms to more than 250 ms over the range of $L_\nu$s presented here. 
For low luminosities, $\tau_s\propto L_\nu^{-1.4}$.  $\tau_m$ is proportional to $L_\nu^{-2.8}$ for all luminosities and
varies from 4 seconds to more than 5000 seconds for the same range of $L_\nu$s.
However, these numbers for $\tau_m$ are quite deceiving. In our models, due to the exponential density gradient just exterior
to $R_\nu$, the matter is effectively trapped for $r\lesssim12$ km.  In fact, for the lowest luminosity cases presented here,
the matter velocity for $r\lesssim10.5$ km can be of order 10 cm s$^{-1}$.  In effect, then, the region shown in the inset 
in Fig.~\ref{fig:rp} is an atmosphere in hydrostatic equilibrium {\it from} which the wind emerges.  
From this region the matter would escape on timescales much longer than the total protoneutron star cooling time.  
If, instead, we redefine $\tau_m$ as the time necessary for a Lagrangean mass element to go from the peak of the 
heating profile (see Fig.~\ref{fig:qp}) at $r\sim 12$ km to $R_c$, we find that $\tau_m$ is of order $\sim 10$ ms 
for $\lumanue^{51}=8.0$ and $\tau_m\sim 1$ second for $\lumanue^{51}=0.5$. 
The steady-state approximation is only valid if $\tau_m,\tau_s\ll\tau_d$.  For example,
taking $L_\nu(t)\propto t^{-0.9}$ and high neutrino luminosities, $\dot{M}$ drops 10\% in roughly $\tau_d$.
At these luminosities, $\tau_d$ is approximately $20-30$ ms.  Although both $\tau_m$ and $\tau_s$
are less than $\tau_d$, they are not significantly so.  We conclude that the steady-state assumption might be reasonably
employed, but that caution is warranted.


\section{THE EVOLUTION OF PROTONEUTRON STAR WINDS}
\label{sec:evolution}

With the eigenvalue problem solved and some of the systematics in hand, in this section we explore
possible evolutionary trajectories using the steady-state solutions. 
Beyond surveying the entire relevant parameter space, we endeavor to model the 
whole of the Kelvin-Helmholtz cooling phase, including radial contraction
and the simultaneous evolution of the luminosity and average neutrino energy.  

Perhaps a second after core bounce,
as the wind emerges, the protoneutron star atmosphere will be extended ($R\sim30-50$ km) and perhaps highly luminous 
($L_\nu^{\rm tot}\sim 5\times10^{52}$ erg s$^{-1})$.  As the neutron star cools it will contract  
quasi-hydrostatically.  This may take as many as several seconds, depending upon
the nuclear equation of state.  The average neutrino energies during contraction may increase, peak 
near the time at which $R_\nu$ settles, and then decrease roughly linearly in time (Pons et al. $\!$1999).
The luminosity may decay quasi-exponentially or as a power law in time (Burrows and Lattimer 1986; Burrows 1988;
Pons et al. $\!$1999).

Figure \ref{fig:evolution} shows the luminosity, radius, and average energy as a function of time for our evolutionary models.
This picture is merely schematic, but illustrates a representative scenario.  In the following discussion
we take $L_\nu^{\rm tot}\propto t^{-0.9}$.  A simple rescaling in time can be performed for other power-law indices
or exponential luminosity decay.  Two possible tracks for the time evolution of $R_\nu(t)$ are shown.
The short dashed line is linear contraction such that $R_\nu(t=0.4\,{\rm s})=20.3$ km and $R_\nu(t=1\,{\rm s})=10$ km.
This is the evolutionary model, which we label as `$R_\nu(t)\propto1-at$'.  For comparison, the dot-dashed
line has $R_\nu(t)$ proportional to $t^{-1/3}$.  This model also has $R_\nu(t=0.4\,{\rm s})=20.3$ km.
At $t\simeq3.2$ seconds, $R_\nu=10$ km.  The power $1/3$ was obtained from a rough fit over approximately
one second of post-explosion evolution in a supernova model of S.~Bruenn.  In this calculation the supernova
was  simulated in one-dimension artificially for a Woosley and Weaver (1995) 15 M$_\odot$ progenitor starting from 
the collapse calculations of \cite{bruenn2001}.
We focus on the model with $R_\nu(t)\propto1-at$ instead of the model with
$R_\nu(t)\propto t^{-1/3}$ because it reaches a more compact configuration (i.e., maximum $M/R_\nu$)
at earlier times, that is, with higher luminosity.  As we explore in the next section,
large $M/R_\nu$, coupled with high luminosity and/or average neutrino energy gives short dynamical timescales
and relatively high entropies, both potentially important for $r$-process nucleosynthesis in the protoneutron star context.

In order to determine appropriate numbers for the ratios $\aveunu/\aveanue$, $\aveanue/\avenue$, $\lumanue/\lumunu$,
and $\lumanue/\lumnue$ we surveyed supernova simulations
(e.g., Mayle, Wilson, and Schramm 1987; Burrows, Hayes, and Fryxell 1995; Mezzacappa et al.~2001; 
Liebend\"{o}rfer et al.~2001;
Bruenn, De Nisco, and Mezzacappa 2001; Rampp and Janka 2000; S.~Bruenn 2001, private communication) and
protoneutron star cooling calculations (Burrows and Lattimer 1986; Pons et al. $\!$1999; J.~Pons 2000, private communication).
The common assumption of equipartition in luminosity between the three neutrino species 
is generally not realized in these calculations.  In fact,  
$L_{\nu_\mu}+L_{\bar{\nu}_\mu}+L_{\nu_\tau}+L_{\bar{\nu}_\tau}$ is usually of order 50$-$60\% of $L_\nu^{\rm tot}$.
In addition, the ratio $\aveanue/\avenue$ ranges from 1.1 to 1.4 and $\lumanue/\lumnue$ from 1.0 to 1.4.
The ratio of $\aveanue$ to $\aveunu$ also varies significantly.  Like the fiducial models
presented in \S\ref{sec:fiducial}, we use $\lumanue$ to index our evolutionary models in this section
and set $\aveunu/\aveanue=1.6$, $\aveanue/\avenue=1.3$, $\lumanue/\lumunu=1.4$,
and $\lumanue/\lumnue=1.3$.  At any time, all luminosities and average energies can then be computed
from Fig.~\ref{fig:evolution}.  We chose $\aveanue/\avenue$ and $\lumanue/\lumnue$ so as to accord with
the literature while also minimizing $Y_e^a$ (eq.~\ref{yeqw}), this being potentially favorable for
$r$-process nucleosynthesis (but see \S\ref{sec:high_ye}).  Note that with $\aveanue/\avenue=1.3$ and $\lumanue/\lumnue=1.3$
and the magnitude of the average energies set by Fig.~\ref{fig:evolution}, $Y_e^a\sim0.46$ at early times.
We feel these numbers are merely representative. We explore potentially important 
modifications to our prescription in \S\ref{sec:energy_source}.

For each point along the evolutionary models represented in Fig.~\ref{fig:evolution} we calculate
the steady-state wind solution. We do this for a range of protoneutron star masses from 1.4$-$2.0 M$_\odot$.  
Neutrino luminosity is always quoted as the local neutrino luminosity at the surface of the protoneutron star, 
not the luminosity at infinity.

\subsection{Results: Evolutionary Models}
\label{sec:baseline}

\begin{figure} 
\vspace*{6.0in}
\hbox to\hsize{\hfill\includegraphics{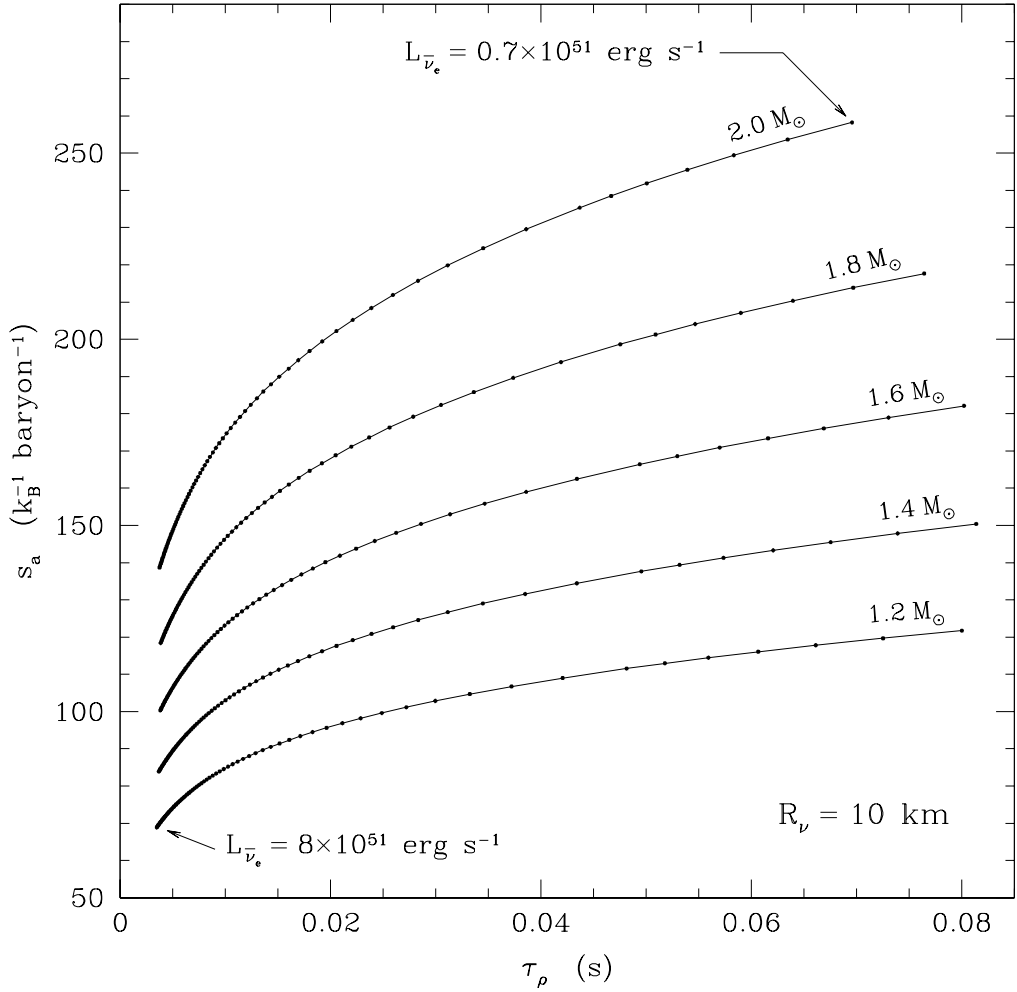}\kern+6in\hfill}
\caption{Tracks for constant protoneutron star mass in the plane of asymptotic entropy ($s_a$, per k$_{\rm B}$ per baryon) and 
dynamical timescale ($\tau_\rho$, eq.~\ref{taurho}) in seconds. $R_\nu$ is held constant at 10 kilometers. 
Each track covers a range in luminosity from $\lumanue=8.0\times10^{51}$ erg s$^{-1}$ to $\lumanue=0.7\times10^{51}$ erg s$^{-1}$.
Small dots are the points for which a model with a given luminosity was calculated.  All masses are gravitational masses.  
In these models, we take $\avenue=11$ MeV, $\aveanue=14$ MeV, and $\aveunu=23$ MeV for $\lumanue=8.0\times10^{51}$ erg s$^{-1}$.  
For each subsequent luminosity, the average energies were decreased according to $\langle\varepsilon_\nu\rangle\propto L_\nu^{1/4}$.
In Table \ref{tab:fiducial}, we summarize the global properties at several luminosities along the 1.4 M$_\odot$ 
trajectory in this figure.\label{fig:ffig}}
\end{figure}

\begin{figure} 
\vspace*{6.0in}
\hbox to\hsize{\hfill\includegraphics{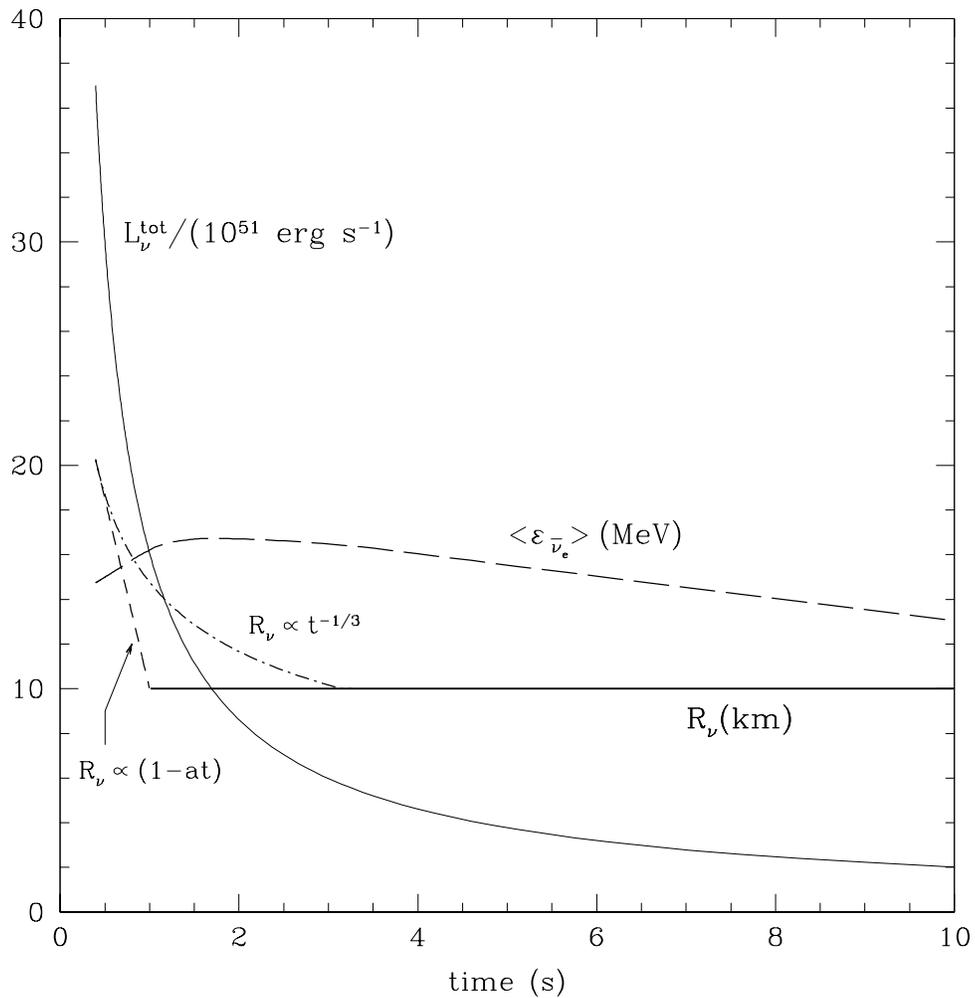}\kern+6in\hfill}
\caption{A schematic showing total neutrino luminosity in 10$^{51}$ erg s$^{-1}$ (thin solid line), 
average $\bar{\nu}_e$ neutrino energy in MeV (long dashed line), and neutrinosphere radius ($R_\nu$) in km.
Two possible evolutions for $R_\nu(t)$ are shown.  Linear contraction such that $R_\nu(t=0.4\,{\rm s})=20.3$ km
and $R_\nu(t=1\,{\rm s})=10.0$ km, which we label as `$R_\nu(t)\propto1-at$', is shown as a short dashed line.
Contraction with $R_\nu(t)\propto t^{-1/3}$ is shown as a dot-dashed line.  
The thick solid line denotes 10 km, the final $R_\nu$ for the protoneutron star.  
$L_\nu^{\rm tot}$ is proportional to $t^{-0.9}$. We set $\aveunu/\aveanue=1.6$, $\aveanue/\avenue=1.3$, 
$\lumanue/\lumunu=1.4$, and $\lumanue/\lumnue=1.3$. \label{fig:evolution} }
\end{figure}

\begin{figure} 
\vspace*{6.0in}
\hbox to\hsize{\hfill\includegraphics{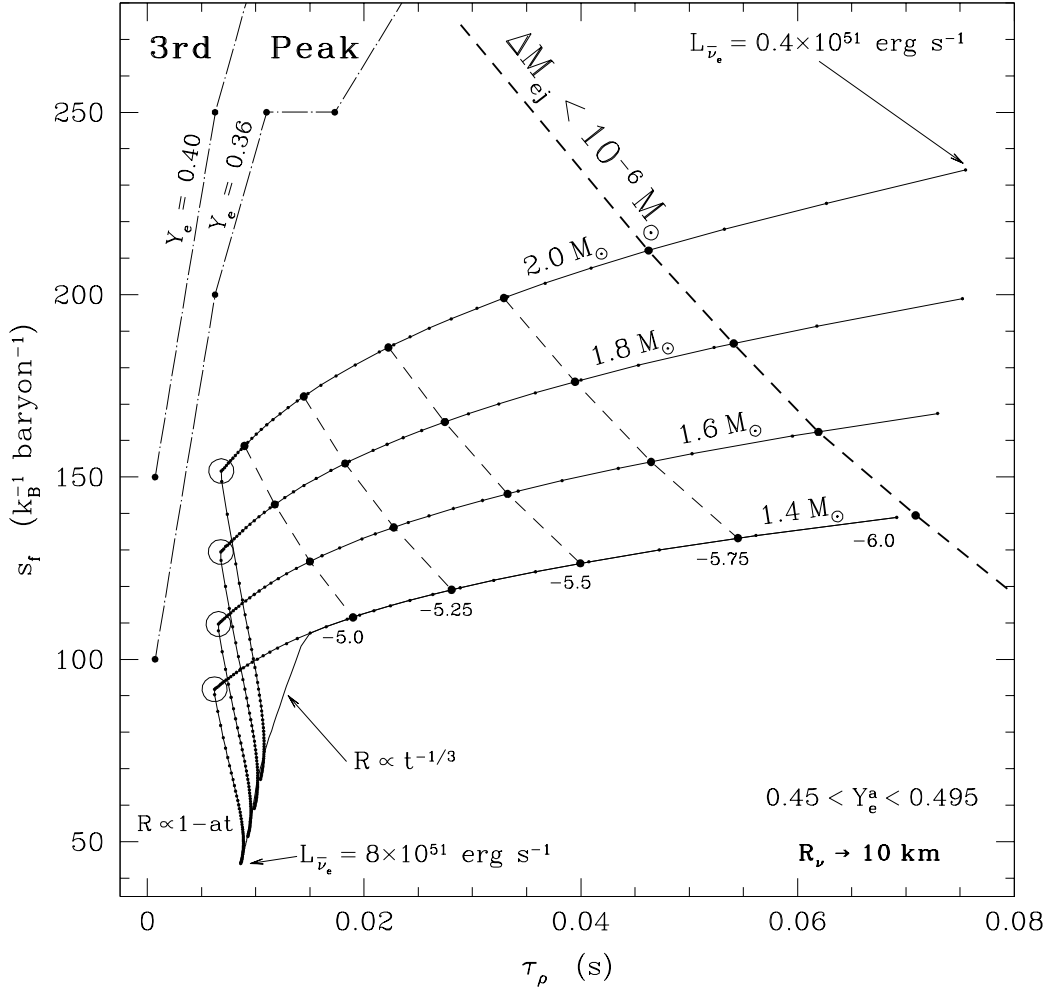}\kern+6in\hfill}
\caption{Evolutionary wind models in the plane of asymptotic entropy ($s_a$ per k$_{\rm B}$ per baryon) and 
dynamical timescale ($\tau_\rho$, eq.~\ref{taurho}) in seconds.  Tracks with $R_\nu(t)\propto(1-at)$ 
(see Fig.~\ref{fig:evolution}) for $M=1.4$, 1.6, 1.8, and 2.0 M$_\odot$ are shown as solid lines with small dots.
The solid line without dots is the track with $R_\nu(t)\propto t^{-1/3}$ for 1.4 M$_\odot$.  
All tracks start at small $s_a$ and $\tau_\rho$ with $R_\nu\simeq20.3$ km and $\lumanue^{51}=8.0$.
The tracks end at high $s_a$ and long $\tau_\rho$ with $\lumanue^{51}=0.4$.
Taking $L_\nu\propto t^{-0.9}$, this range spans the first $\sim12$ seconds of protoneutron star wind evolution.
The large open circles mark $\lumanue^{51}=3.4$ for each mass trajectory.
Dashed lines connecting all mass tracks with large dots are lines of constant $\log_{10}[\Delta M_{\rm ej}]$
(see eq.~\ref{mej}, Fig.~\ref{fig:deltam}, and text for details).  The dashed line connecting all mass tracks on the right,
marked with $-6.0$, delineates at what point in the evolution only 10$^{-6}$ M$_\odot$ of material is yet to be ejected. 
In all of the more than 350 models presented here the asymptotic electron fraction is always in the range 
$0.45\leq Y_e^a\leq 0.495$. Generally, $Y_e^a$ increases as the luminosity and average neutrino energies decrease.
Finally, the dash-dotted lines in the upper left corner indicate, for constant values of $Y^a_e$, in what range
of $s_a$ and $\tau_\rho$ production of the third $r$-process abundance peak is possible in the calculations 
of \cite{meyer_brown}.  See \S\ref{sec:evolution} for discussion and details.\label{fig:fig}}
\end{figure}

Shown in Fig.~\ref{fig:fig} are evolutionary trajectories for $M=1.4$, 1.6, 1.8, and 2.0 M$_\odot$
in the plane of $\tau_\rho$ versus asymptotic entropy, $s_a$.
Note that $s_a$ does not include contributions from $\beta$-decays during nucleosynthesis.  
During $r$-process nucleosynthesis these processes may increase $s_a$ by $\gtrsim10$ units, depending
on the dynamical timescale (Meyer and Brown 1997).
However, because we post-process our wind models to obtain the nucleosynthetic yield 
and include only alpha particles and free nucleons in the equation of state we employ in solving the eigenvalue problem 
(see \S\ref{sec:equations}), such an entropy increase is not included in $s_a$.
Figure \ref{fig:fig} is analogous to Fig.~\ref{fig:ffig}, but for changing $R_\nu(t)$, $\langle\varepsilon_\nu\rangle(t)$,
and $L^{\rm tot}_\nu(t)$ using the evolution depicted in Fig.~\ref{fig:evolution}.  
The evolutionary trajectories labelled with `$R_\nu\propto1-at$' are solid lines with small dots.  
For comparison, we also show the evolution for $M=1.4$ M$_\odot$ with $R_\nu(t)\propto t^{-1/3}$ as a thin solid line without dots.   
The small dots on each evolutionary track are separate $L_\nu$s.  All tracks start with $\lumanue^{51}=8.0$.
The lowest luminosity shown on this plot is  $\lumanue^{51}=0.4$ for each track.
The time evolution for any mass begins with high luminosity,
large $R_\nu$, and, hence, low $s_a$ ($\sim50-70$) and moderate $\tau_\rho$ ($\sim9$ ms).  
As $R_\nu$ gets smaller in the first second of
evolution, the trajectories with $R_\nu\propto1-at$ move to much higher $s_a$ and slightly smaller $\tau_\rho$ before 
they cease contraction at $R_\nu=10$ km.
The $s_a$ reached at this luminosity is set by $M/R_\nu$, with the 2.0 M$_\odot$ model reaching $s_a\simeq150$.
Our evolutionary models contract from $\sim 20$ km to $R_\nu=10$ km in approximately 1 second.    At this point in the evolution, 
$R_\nu$ is fixed and each track makes a sharp turn toward much longer $\tau_\rho$ and only moderately higher $s_a$.  
This turnoff point 
is marked with a large open circle on each track and has $\lumanue^{51}=3.4$ and $L_\nu^{\rm tot}=1.57\times10^{52}$ erg s$^{-1}$.  
At this point the trajectories join lines of constant $R_\nu$, like those in Fig.~\ref{fig:ffig}.
Due to the relatively slow contraction,
the model with $R_\nu(t)\propto t^{-1/3}$ never exhibits such a sharp turn in the $s_a-\tau_\rho$ plane and eventually
joins the other 1.4 M$_\odot$ evolutionary track at $\tau_\rho\sim0.015$ seconds, corresponding to $t\simeq3.15$ seconds
and $\lumanue^{51}=1.3$.  Note that the turnoff point at $\lumanue^{51}=3.4$ marks the point of minimum $\tau_\rho$ for each
model with $R_\nu\propto1-at$. Table \ref{tab:evol} gives the global properties of our neutrino-driven wind models at 
$\lumanue^{51}=8.0$, $\lumanue^{51}=3.4$, and $\lumanue^{51}=0.4$, including the asymptotic electron fraction, $Y_e^a$.  
For comparison, we also include in Table \ref{tab:evol} the model with $R_\nu(t)\propto t^{-1/3}$ at  $\lumanue^{51}=3.4$.

\begin{figure} 
\vspace*{6.0in}
\hbox to\hsize{\hfill\includegraphics{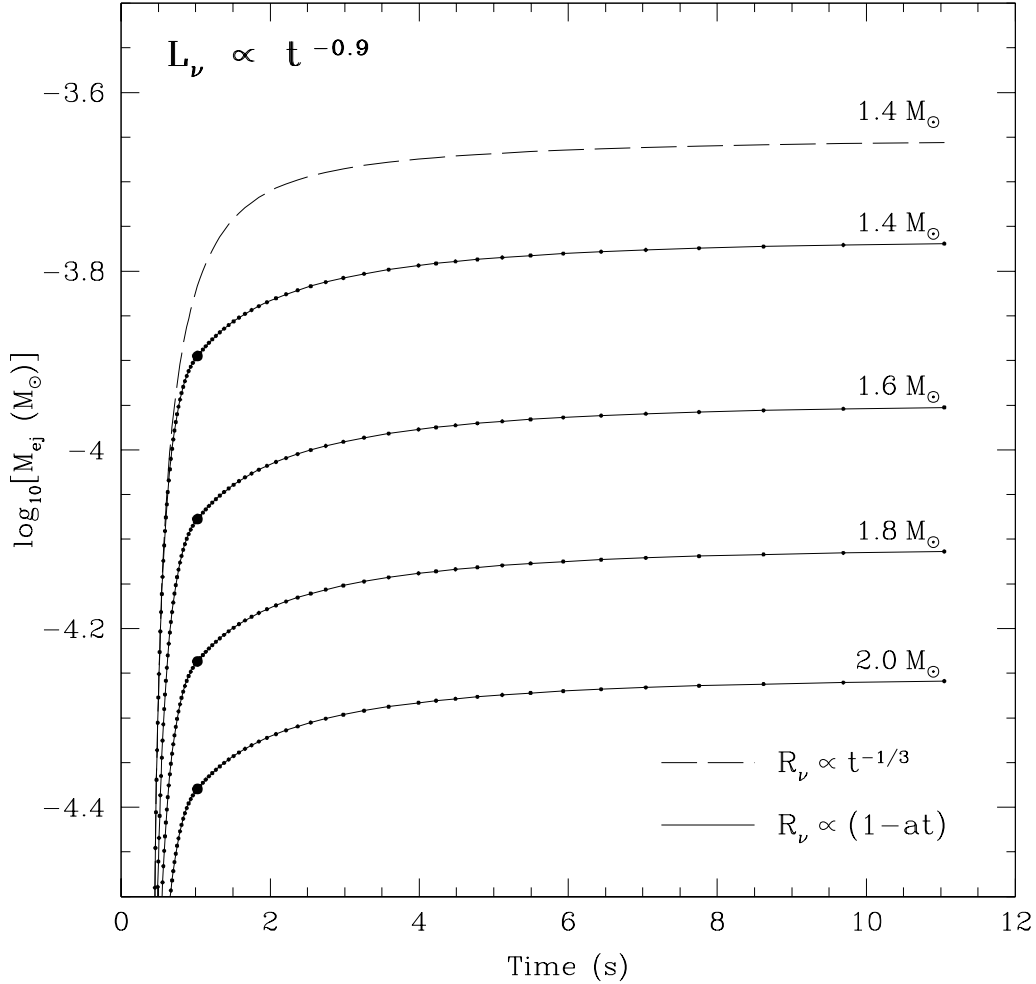}\kern+6in\hfill}
\caption{The log of the integrated total mass ejected $M_{\rm ej}$ (eq.~\ref{mej}) in units of M$_\odot$ 
as a function of time in seconds, assuming $L_\nu(t)^{\rm tot}\propto t^{-0.9}$ for the tracks of constant mass 
shown in Fig.~\ref{fig:fig}. 
Evolutionary models with $R_\nu(t)\propto(1-at)$ (see Fig.~\ref{fig:evolution}) are shown as solid lines with small dots,
which correspond to the dots in Fig.~\ref{fig:fig}, indicating luminosity (time).
$M_{\rm ej}$ for the track in Fig.~\ref{fig:fig} with $R_\nu(t)\propto t^{-1/3}$ and $M=1.4$ M$_\odot$ is 
shown here as a long dashed line without dots.
The large dot on each of the solid curves at $t\sim1$ second shows the point at which the model has contracted to $R_\nu=10$ km.  
This corresponds to the point on Fig.~\ref{fig:fig} where a given track of constant mass with $R_\nu(t)\propto(1-at)$ takes a sharp 
turn at $\tau_\rho\sim0.006$ seconds. \label{fig:mt}}
\end{figure}

\begin{figure} 
\vspace*{6.0in}
\hbox to\hsize{\hfill\includegraphics{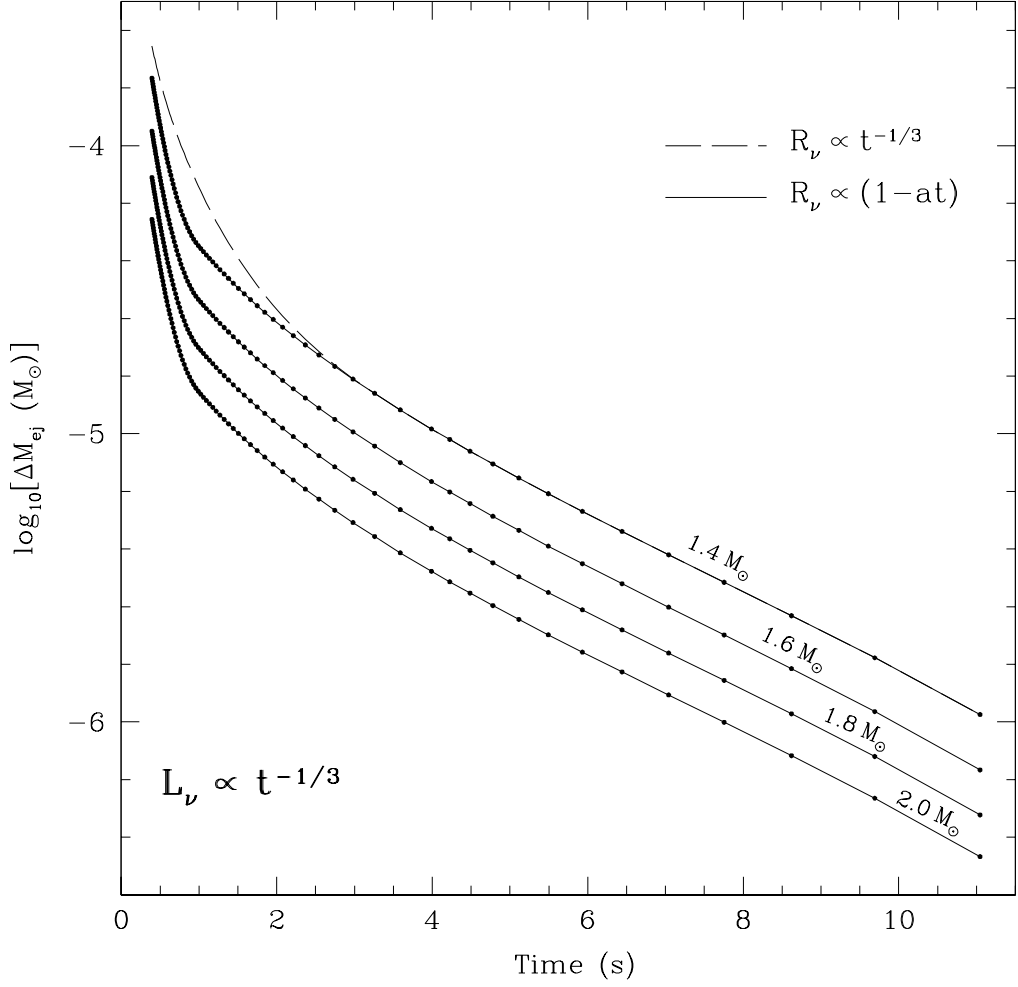}\kern+6in\hfill}
\caption{Log$_{10}[\Delta M_{\rm ej}(=M_{\rm ej}^{\rm tot}-M_{\rm ej}(t))]$, in units of M$_\odot$, versus time in seconds
for the models in Fig.~\ref{fig:mt}.  The long dashed line is for the model with $R_\nu(t)\propto t^{-1/3}$.
The four solid lines with dots correspond to those in Figs.~\ref{fig:fig} and \ref{fig:mt} for the 1.4, 1.6, 1.8, and
2.0 M$_\odot$ evolutionary models, which we label with `$R_\nu(t)\propto1-at$'.  Lines of constant $\log_{10}[\Delta M_{\rm ej}]$
are shown in Fig.~\ref{fig:fig} as dashed lines connecting big dots on the evolutionary models with $R_\nu(t)\propto(1-at)$.\label{fig:deltam}}
\end{figure}

Having solved for $\dot{M}$ at every point along these evolutionary tracks and assuming that
$L_\nu(t)\propto t^{-0.9}$, we calculate the total mass ejected in the wind as a function of time:
\beq
M_{\rm ej}(t)=\int_0^t\,\dot{M}(t\pr)\,dt\pr.
\label{mej}
\eeq
Figure \ref{fig:mt} shows this integral for all five of the models presented in Fig.~\ref{fig:fig}.  The dashed
line shows $M_{\rm ej}(t)$ for the model with $R_\nu(t)\propto t^{-1/3}$.  As one would expect, because of the slower 
radial contraction of the protoneutron
star, this model ejects more matter than the corresponding trajectory with $R_\nu\propto(1-at)$.  In this case,
the difference is about 30\%.  The small dots on each of the solid lines mark $L_\nu(t)$ for each model and correspond to
the luminosity points on each track in Fig.~\ref{fig:fig}.  The large dots on each line mark $\lumanue^{51}=3.4$, the luminosity at
which each track in Fig.~\ref{fig:fig} reaches $R_\nu=10$ km and turns sharply.
Note that the 2.0 M$_\odot$ model ejects only $\sim6\times10^{-5}$ M$_\odot$ of
material, about three times less than the 1.4 M$_\odot$ model.
Extrapolating the results of Fig.~\ref{fig:mt} we can compute the total $M^{\rm tot}_{\rm ej}$ for $t\rightarrow\infty$.
We can then compute, at any time, the mass yet to be ejected by the wind, $\Delta M_{\rm ej}(t)=M^{\rm tot}_{\rm ej}-M_{\rm ej}(t)$.  
Figure \ref{fig:deltam} shows $\Delta M_{\rm ej}(t)$ versus time for each track in Fig.~\ref{fig:fig}.
The lines and dots correspond with those in Fig.~\ref{fig:mt}. 
In Fig.~\ref{fig:fig}, we plot lines of constant $\log_{10}[\Delta M_{\rm ej}(t)]$ in units of M$_\odot$ as dashed 
lines connecting big dots on each of the four evolutionary trajectories with $R_\nu\propto(1-at)$.  
The thick dashed line on the far right side of the plot, labelled $-6.0$,
is the line beyond which, only 10$^{-6}$ M$_\odot$ will be ejected.  

In the calculations presented here, we have arbitrarily defined
the point in time when the wind begins.  The absolute magnitude of $M_{\rm ej}$ in Fig.~\ref{fig:mt} 
for each model is therefore also arbitrary.  Only the ratios of these ejected masses or $\Delta M_{\rm ej}(t)$
for an individual trajectory are of real import.  The $-6.0$ line in Fig.~\ref{fig:fig} is of
particular significance because if all (or most) supernovae produce $r$-process elements then the total 
yield per supernova must be $10^{-5}-10^{-6}$ M$_\odot$ (e.g., Qian 2000).  
Therefore, if an $r$-process epoch is to exist along any of the trajectories shown in Fig.~\ref{fig:fig},
then it must begin at or before the line labelled $-6.0$ in order to eject sufficient mass.
If $r$-processing begins to the right of this line, less than 10$^{-6}$ M$_\odot$ will be ejected.
For any $L_\nu(t)$ and $R_\nu(t)$, such a bound must exist. 
We have explored the position
of this boundary for a variety of relationships for $L_\nu(t)$.  Taking reasonable $e$-folding timescales ($\tau$) and 
$L_\nu(t)\propto e^{\,-t/\tau}$ the $-$6.0 line moves to even shorter $\tau_\rho$.  For slower power law decay,
the boundary moves to longer $\tau_\rho$.  For $M=1.4$ M$_\odot$ and $L_\nu(t)\propto t^{-0.8}$ it moves from $\tau_\rho\simeq0.07$
to $\tau_\rho\simeq0.085$ seconds.   
Although the wind may eventually evolve to arbitrarily long dynamical timescales,
we conclude that the range of $\tau_\rho$ relevant for $r$-process nucleosynthesis is significantly constrained
by the $\log_{10}[\Delta M_{\rm ej}]=-6$ line in Fig.~\ref{fig:fig}.
In fact, this range is smaller than previous calculations suggest.
In addition, note that even if a given wind model produces $r$-process elements, only a fraction of the 
total mass ejected during that $r$-processing epoch will be in $r$-process elements; much of
the mass will remain in alpha particles.
Conservatively, then, if transonic protoneutron star winds are the primary site for the $r$-process,
this constraint on the amount of mass ejected per supernova implies that the epoch of $r$-process nucleosynthesis 
must occur for $\tau_\rho$ less than $\sim0.085$ seconds.
 
For this range in $\tau_\rho$, there is also only a relatively small range of $s_a$ available 
to the transonic protoneutron star wind.
As evidenced by the calculations of \cite{twj1994} and \cite{qw1996} and borne out in Fig.~\ref{fig:fig}, 
$s_a$ as large as $\simeq400$ is simply outside what can be obtained for
reasonable dynamical timescales, even including the effects of general relativity. 
If we extrapolate the curves shown here to later times (lower luminosities), even the 2.0 M$_\odot$ trajectory 
does not reach 400 until $\tau_\rho\simeq0.5$ seconds.  At this point 
$\dot{M}\simeq1.5\times10^{-9}$ M$_\odot$ s$^{-1}$.  If  $r$-processing could
occur during these late stages, it would need to persist for many thousands of seconds to yield 
even 10$^{-6}$ M$_\odot$ of ejecta.  Moreover, the survey calculations of \cite{hwq} and \cite{meyer_brown}
show that with $s_a\sim400$  and $\tau_\rho\sim0.5$ seconds, one requires $Y_e^a\lesssim0.3$
to achieve an appreciable third peak $r$-process.  A $Y_e^a$ this low is extremely unlikely.
As the protoneutron star cools, the neutrino luminosity and average neutrino energy will be correlated.
Protoneutron star cooling calculations (Burrows and Lattimer 1986; Pons et al. $\!$1999) 
indicate that the average neutrino energies will fall throughout these late evolutionary stages as the luminosity decreases.
As the magnitude of $\aveanue$ and $\avenue$ decrease, the asymptotic electron fraction (eq.~\ref{yeqw})
must increase on account of the energy threshold for the reaction $\bar{\nu}_e p\rightarrow n e^+$ (the `threshold effect').
For the 1.4 M$_\odot$ model, these constraints are even more severe.  This 
track reaches $s_a\sim400$ only when $\tau_\rho$ is several {\it seconds} and $\dot{M}$ 
is of order 10$^{-11}$ M$_\odot$ s$^{-1}$.

Note that for a given mass and $\tau_\rho$, $s_a$ is 10-30 units higher in Fig.~\ref{fig:ffig} than in Fig.~\ref{fig:fig}
owing to the lower average neutrino energies for a given luminosity in our fiducial models (\S\ref{sec:fiducial}) than in the
evolutionary models we consider in this section.  One might argue that by quickly decreasing $\langle\varepsilon_\nu\rangle$ for all 
neutrino species with respect to the luminosity that the trajectory would move more quickly to 
higher $s_a$, and, hence, be more likely to yield $r$-process ejecta.  While such a change would certainly drive $s_a$ higher,
it would also make $\tau_\rho$ increase faster, decrease $\dot{M}$ significantly, and the  
threshold effect would drive $Y_e^a$ higher.  Thus, such a modification can only make the constraints tighter.
 
We conclude that the late-time $r$-process as obtained in \cite{woosley1994} is
extremely unlikely in the context of a transonic wind.
In essence, because $s_a$ is initially set by $M/R_\nu$ for a given model,
once each trajectory reaches $R_\nu(t)=10$ km, the wind evolves quickly to much larger $\tau_\rho$
and only modestly higher asymptotic entropy. 
That is, for constant $R_\nu$, $s_a$ is roughly 
proportional to $\tau_\rho^{0.2}$. By the time a transonic wind evolves to high entropy,
the dynamical timescale is too long and $Y_e^a$ is too high to allow for a robust $r$-process.
The slope of the trajectories in the $s_a-\tau_\rho$
plane shown in Fig.~\ref{fig:fig} guarantee that if the wind enters a regime of very high entropy 
it does so with very large $\tau_\rho$ and minute $\dot{M}$, so as to preclude any significant $r$-process yield.

Instead, we propose an $r$-process epoch just a second or two after explosion, coinciding with
the end of the protoneutron star contraction phase.
In this scenario, the wind moves into an early-time $r$-processing regime 
in the $s_a-\tau_\rho$ plane and then out of this regime at later times so that the constraint on $M_{\rm ej}$ is satisfied.
With this in mind, the behavior of the wind trajectories during contraction, and particularly
the point in each track where $\tau_\rho$ is at a minimum ($\lumanue^{51}=3.4$), is suggestive and tantalizing.  As 
\cite{hwq} noted, a small dynamical timescale, even for only moderate entropies, can yield a successful
$r$-process.  More recently, \cite{otsuki} have shown that a successful $r$-process can be realized in this context.
For these reasons we turn our attention to an early-time, high luminosity, short-$\tau_\rho$,
and modest entropy $r$-process.  For reference, in Fig.~\ref{fig:fig}, we include long dashed lines of
constant $Y_e$, taken from the $r$-process survey calculations of \cite{meyer_brown}, above and to the left of
which, for a given $Y^a_e$, production of the third $r$-process peak at $A\sim195$ is assured.  Caution is encouraged in
taking these lines too seriously.  They were computed along specific $T$ and $\rho$ trajectories for fixed
$Y^a_e$ and dynamical timescale, $\tau^{\rm MB}$, given by eq.~(\ref{mbtau}), which overestimates $d\tau/dt$ by roughly a factor
of two at these high luminosities (see Fig.~\ref{fig:timescale} and \S\ref{sec:structure}).  
These lines are only suggestive. 


\section{Nucleosynthesis in the Evolutionary Protoneutron Star Wind Models}
\label{sec:nucleosynthesis}

Any successful $r$-process site must do more than simply produce nuclei with $A\gtrsim195$.
Observations of $r$-nuclei in ultra metal-poor halo stars (notably, CS22892-052, HD115444, and CS31082-001) 
show that the abundance pattern for $A\gtrsim135$ is 
nearly identical to the scaled solar $r$-process abundance
(Sneden et al. $\!$1996; Burris et al. $\!$2000; Westin et al. $\!$2000; Hill et al. $\!$2001).  
Simply producing the platinum peak is no guarantee that the solar abundance distribution 
is reliably reproduced (Meyer and Brown 1997).  In the wind scenario, particularly, one must construct 
the time-integrated yield as the neutrino luminosity decays and the global wind structure evolves.

For each luminosity point on the 1.4 M$_\odot$ evolutionary trajectory with $R_\nu\propto(1-at)$ in Fig.~\ref{fig:fig}
we obtained a unique $\dot{M}$ and velocity, temperature, and density profile.  For
each individual profile we compute the nucleosynthetic yield ($Y$) as a function of the atomic mass, $A$.
With $Y(A)$ and $\dot{M}$ at every point, assuming $L_\nu(t)\propto t^{-0.9}$, we compute the weighted sum
to get the total amount of ejected material at each $A$, $M_{\rm ej}(A)$. 
For the 1.4 M$_\odot$ model we find no significant nucleosynthesis beyond $A\sim100$.
In fact, most of the mass is concentrated at a peak in Sr, Y, and Zr.
Inspecting the yield at each luminosity (time) reveals that when the protoneutron star has contracted 
to $R_\nu=10$ km, the point of minimum $\tau_\rho$ in Fig.~\ref{fig:fig} denoted by a large open circle, 
the nucleosynthetic flow reaches a maximum in $A$.  That is, all the points with $\tau_\rho\gtrsim0.006$ seconds 
on the 1.4 M$_\odot$ evolutionary track,
even though they have higher entropy, produce lower average $A$ ejecta.  
This can be understood simply: as the dynamical timescale of the wind gets longer, more seed nuclei are formed.  
Hence, for a given $Y_e^a$ and $s_a$, the neutron-to-seed ratio decreases (Hoffman, Woosley, and Qian 1996; 
Meyer and Brown 1997; Freiburghaus et al. $\!$1999).  
Therefore, the point of minimum $\tau_\rho$, when $M/R_\nu$ reaches a maximum, 
affords the best possibility for a robust $r$-process.  
For all protoneutron star masses, the evolutionary models with $R_\nu\propto(1-at)$ turn sharply at 
$\lumanue^{51}=3.4$ and for $\tau_\rho$ between 6 and 7 ms.  Unfortunately, although the $\lumanue^{51}=3.4$ 
point produced the highest average $A$ ejecta of any other luminosity 
along the 1.4 M$_\odot$ track in Fig.~\ref{fig:fig}, the nucleosynthesis did not even proceed to the second abundance peak.

We also calculated the nucleosynthetic yield for the 1.6, 1.8, and 2.0 M$_\odot$ trajectories at $\lumanue^{51}=3.4$,
assuming that these  points of minimum $\tau_\rho$ would also yield the highest average $A$ ejecta
of any of the points in a given mass trajectory.
None successfully generated nucleosynthesis beyond the second $r$-process abundance peak.
Even the 2.0 M$_\odot$ model, which has $s_a\simeq151$ and $\tau_\rho\simeq0.0068$ seconds for $\lumanue^{51}=3.4$,
did not proceed beyond $A\sim135$.  Hence, for the $Y_e^a$s derived and the time evolution we have assumed, we fail 
to produce viable $r$-process nucleosynthesis in any of our evolutionary transonic protoneutron star wind models.


\section{What is To Be Done?}
\label{sec:what}

We have already ruled out the possibility of a late-time $r$-processing epoch at long
dynamical timescales and high entropies ($s_a\gtrsim400$) using constraints on the amount of material ejected, the slope
of the evolutionary tracks in the $s_a-\tau_\rho$ plane, and the inexorable rise in $Y_e^a$ 
as the protoneutron star cools. 

We are left wondering what reasonable modifications might generically yield third-peak $r$-process 
nucleosynthesis for a canonical protoneutron star with 1.4 M$_\odot$ and $R_\nu=10$ km.
In this section we further explore the viability of the early-time $r$-process.
We consider the likely range of $Y_e^a$, $s_a$, and $\tau_\rho$ accessible to transonic neutrino-driven winds,
and present the physical conditions we require for production of the both the second and third $r$-process abundance
peaks.

\subsection{The Asymptotic Electron Fraction: $Y_e^a$}
\label{sec:high_ye}

One might argue that $Y_e^a$ is simply too high in these winds to yield successful
nucleosynthesis.  There are several important points in this regard.
First, our evolutionary models, which failed to produce nucleosynthesis beyond $A\sim135$, all had
$Y_e^a\sim0.46$ at $\lumanue^{51}=3.4$.  This $Y_e^a$ favors the formation of $A\sim90$ nuclei and produces many seed nuclei,
thus decreasing the neutron-to-seed ratio for a given entropy and dynamical timescale.
Second, both \cite{woosley1994} and \cite{wanajo} obtained unacceptably large over-productions of nuclei near $A\sim90$ 
in the early phase of their wind calculations, at high luminosity and low entropy.
Third, \cite{hoffman_1996a} find that this overproduction problem is solved if $Y_e^a\gtrsim0.485$.  
These three points together imply that if the $r$-process occurs generically in protoneutron star winds 
then $Y_e^a$ must be either less than 0.40 or greater than 0.48 to avoid the overproduction problem
at $A\sim90$.  Naively, it might seem that $Y_e^a\lesssim0.40$ is favored because this would
naturally increase the neutron-to-seed ratio by increasing the number fraction of neutrons.  
However, there are several reasons why $Y_e^a\gtrsim0.48$ might actually be viable.
First, for $Y_e^a\gtrsim0.485$ \cite{hoffman_1996a} found that some interesting $p$-process elements were produced,
which were  previously unaccounted for (e.g., $^{92}$Mo).
Second, the most detailed transport studies done to date
(Mezzacappa et al. $\!$2001; Liebend\"{o}rfer et al. $\!$2001;
Rampp and Janka 2000)
indicate that $\aveanue/\avenue\simeq1.1-1.2$ and $\lumanue/\lumnue \simeq1.1$.
Depending on the magnitude of $\aveanue$ and $\avenue$, these numbers put $Y_e^a\gtrsim0.48$, as per eq.~(\ref{yeqw}).
Third, \cite{hwq} find that as $Y_e^a$ increases from 0.48 to $\sim0.495$ the requisite
entropy for third-peak production actually {\it decreases} for fixed dynamical timescale.
The last point implies that having a high $Y_e^a$ might slightly relieve the constraints on $s_a$
set by an early-time, high-luminosity $r$-process.  The fact that some $p$-process elements
might also be produced in a high $Y_e^a$ environment is attractive.
Together, we feel that the above points make it plausible that 
$Y_e^a\gtrsim0.48$ in protoneutron star winds.  Such a conclusion, constrains the three-dimensional 
space $s_a-\tau_\rho-Y_e^a$ significantly.

Of course, having $Y_e^a\sim0.30-0.35$ might also cure the overproduction problem at $A\sim90$
while increasing the neutron-to-seed ratio dramatically, so as to allow for third-peak production at
the entropies and timescales obtained for the 1.4 M$_\odot$ evolutionary model in Fig.~\ref{fig:fig}.
However, to attain $Y_e^a\lesssim0.35$ one requires $\lumanue/\lumnue\gtrsim1.55$ for $\aveanue/\avenue=20$ MeV/12 MeV.
Such conditions would be extreme in light of the detailed transport calculations carried out to date.
However, if $Y_e^a$ does reach values this low in profiles like those in Fig.~\ref{fig:fig}, 
there are two constraints worth pointing out.  The first is that $Y_e^a$ must evolve with the
neutrino luminosities and average energies so that very little mass is left to be ejected
by the time $Y_e^a$ increases to 0.40.  Otherwise, the same overproduction problems at $A\sim90$ may occur.
The second constraint is that if $Y_e^a$ is sufficiently low to guarantee third-peak production,
it must eject not more than $\sim10^{-5}$ M$_\odot$ of $r$-process material per supernova.  

We conclude that $Y_e^a$ may be $\gtrsim$0.47 in protoneutron star winds. This follows from the
fact that $Y_e^a$ below 0.40 is very unlikely and if $Y_e^a$ is in the range 0.40$-$0.46, models suffer 
from overproduction of $A\sim90$ nuclei.  With this in mind, in the next section we consider modifications
to our transonic wind models that might increase the entropy or decrease the dynamical timescale.

\subsection{Entropy and Dynamical Timescale}

One might choose to increase $s_a$ by changing the bulk protoneutron star characteristics.
Increasing the ratio $M/R_\nu$ increases $s_a$ significantly with only modest decreases in $\tau_\rho$.
However, this ratio cannot be increased arbitrarily.  $M\gtrsim1.5$ M$_\odot$ may be disfavored in light 
of neutron star binary observations (Arzoumanian 1995) and 
$R_\nu<9$ km seems unlikely due to constraints on the high density nuclear equation of state 
(e.g., Lattimer and Prakash 2001).  In order to explore this,
however, we varied $M$ and $R_\nu$ at $\lumanue^{51}=3.4$ in our evolutionary models with $R_\nu\propto1-at$.
In Table \ref{tab:evol2}, we summarize these results for $M=$1.8 M$_\odot$,
1.6 M$_\odot$, and 1.4 M$_\odot$.  These models should be compared with the
models with $\lumanue^{51}=3.4$  in Table \ref{tab:evol}.  Unfortunately,
for $Y_e^a\simeq0.46$ for each model, we did not obtain third-peak $r$-process nucleosynthesis.
Increasing $Y_e^a$ artificially in our nucleosynthesis calculations to 0.48, the neutron-to-seed ratio
stays low and we fail to generate $r$-process elements beyond the second peak.  

Although reasonable increases in $M/R_\nu$ are favorable for the $r$-process, they are insufficient
for strong third-peak nucleosynthesis.
Therefore, in an effort to obtain a robust $r$-process in a canonical 1.4 M$_\odot$, $R_\nu=$10 km protoneutron star,
we are only left with the option of modifications to the energy deposition profile.

\subsection{Possible Alterations to $\dot{q}$}
\label{sec:energy_source}

\cite{qw1996} showed that an artificial energy source 
at radii between 20 km and 30 km, beyond the peak in $\dot{q}$
could substantially increase the entropy and decrease the dynamical timescale.
In fact, any extra energy source that broadens the energy deposition profile, thus increasing $\dot{q}$
in a region of low mass density, increases $s_a$ and decreases $\tau_\rho$.

We noted in \S\ref{sec:evolution} the difference in entropy, for a given $\tau_\rho$, between
the fiducial tracks in Fig.~\ref{fig:ffig} and the evolutionary tracks in Fig.~\ref{fig:fig}.  
Comparing the $M=1.6$ M$_\odot$ tracks on both plots, the difference in $s_a$ between the two at 
$\tau_\rho=0.02$ seconds is $\sim$12 units.  The fiducial model (with higher $s_a$) 
has a total neutrino luminosity almost twice that of the evolutionary model, but its average neutrino
energies are more than 35\% lower.  The increase in entropy is caused by an interplay between the charged-current heating
rate ($\dot{q}_{cc}$) and the neutrino-electron scattering heating rate ($\dot{q}_{\nu e}$).  The
former is proportional to $\lumanue\langle\varepsilon_{\bar{\nu}_e}^2\rangle+\lumnue\langle\varepsilon_{{\nu}_e}^2\rangle$,
while $\dot{q}_{\nu_i e}$ is proportional to $\sum_i L_{\nu_i}\langle\varepsilon_{\nu_i}\rangle$.  
Clearly, for fixed neutrino luminosities, as the average neutrino energy drops, $\dot{q}_{\nu_i e}$
becomes more important relative to $\dot{q}_{cc}$.  
Because $\dot{q}_{cc}\rightarrow 0$ as the alpha fraction increases (see Figs.~$\!$\ref{fig:qip} and \ref{fig:x}), 
the fact that $\dot{q}_{\nu_i e}$ increases in importance
effectively broadens the energy deposition profile, thus increasing the entropy.  
Although the heating rate due to $\nu_i\bar{\nu}_i$ annihilation peaks close to the protoneutron star surface, it
also contributes to the total energy deposition rate at radii larger than where $\dot{q}_{cc}\rightarrow 0$.
Because $q_{\nu_i\bar{\nu}_i}\propto\sum_i L^2_{\nu_i}\langle\varepsilon_{\nu_i}\rangle$, for fixed neutrino
luminosity and decreasing average neutrino energy, this process also becomes more important relative to $q_{cc}$, 
thereby enhancing the effect on $s_a$. 
Although the total effect here is relatively small, at short $\tau_\rho$ any increases in $s_a$ are of potential significance.
Small average energies coupled with higher luminosities are one way to achieve moderately higher $s_a$ and shorter $\tau_\rho$. 
Note that due to the threshold effect (eq.~\ref{yeqw}) in the charged-current reactions, if one decreases 
$\aveanue$ and $\avenue$, $Y_e^a$ will increase, and any potential gains in $s_a$ might be mitigated.
However, as we discussed in \S\ref{sec:high_ye}, Fig.~10 of \cite{hwq} shows that the 
entropy required for third-peak nucleosynthesis actually decreases for high $Y_e^a$.  

Similarly, one might also increase $\aveunu$ and $L_{\nu_\mu}$
relative to the same quantities for the electron and anti-electron types.  
Since the $\nu_\mu$- and $\nu_\tau$-type neutrinos do not participate in the
charged-current reactions, any increase in their luminosity or average energy effectively increases the importance 
of $\dot{q}_{\nu_i e}$ and $\dot{q}_{\nu_i\bar{\nu}_i}$ with respect to $\dot{q}_{cc}$.  As we noted in \S\ref{sec:evolution}, 
the evolution of luminosity and energy shown in Fig.~\ref{fig:evolution} is only suggestive.  
For this reason we explored modifications to the ratios $\aveunu/\aveanue=1.6$, $\aveanue/\avenue=1.3$, $\lumanue/\lumunu=1.4$,
and $\lumanue/\lumnue=1.3$. For our extreme 2.0 M$_\odot$ evolutionary model with $\lumanue^{51}=3.4$
(see Table \ref{tab:evol} for comparison), we set $\lumnue=\lumanue=\lumunu$ and $\avenue=15$ MeV, $\aveanue=22$ MeV, 
and $\aveunu=34$ MeV.   This increased $Y_e^a$ from 0.469
to 0.484, increased $Q$ by more than a factor of three, decreased $s_a$ by 17 units to $\sim$134, and
decreased $\tau_\rho$ by 40\% to 4.1 ms.  These modifications were insufficient to produce
third-peak nucleosynthesis.

Following \cite{qw1996},
we artificially increased $\dot{q}(r)$ in the region $20$ km $\leq r \leq 50$ km for our 1.4 M$_\odot$ evolutionary model 
at $\lumanue^{51}=3.4$  so that $Q$ (eq.~\ref{bigQtot}) went from $1.21\times10^{49}$ erg s$^{-1}$
to $1.33\times10^{49}$ erg s$^{-1}$, an increase of 10\%.  Because $Q$ is the volume integral of $\rho\dot{q}$ and
$\rho$ is small at these radii ($10^{6}-10^{7}$ g cm$^{-3}$), $\dot{q}$ must be enhanced substantially in 
order to affect a 10\% change in $Q$.
With this extra energy deposition we found that maximum increases in $s_a$ and decreases in $\tau_\rho$,
depending on the degree of augmentation of $\dot{q}$ as a function of $r$, were 17 units and 50\%, respectively.
$\dot{M}$ increased by just 8\%.  
We made the same sort of modification to the 2.0 M$_\odot$ model with $\lumnue=\lumanue=\lumunu$ and 
$\avenue=15$ MeV, $\aveanue=22$ MeV, and $\aveunu=34$ MeV.  In this case, we increased $Q$ by 6\% and
found that $s_a$ was increased from 134 to 150 and that  $\tau_\rho$ decreased from 4.1 ms to 2.6 ms.
That $\tau_\rho$ can decrease so significantly as a result of $\lesssim10$\% changes in $Q$ demonstrates 
the importance of conducting a full transport study in this context.  We save such
an investigation for a future work, but emphasize that the shape of the energy deposition profile
may be the final arbiter in determining the true potential of this site as the seat of $r$-process nucleosynthesis.

\subsection{The Early-Time $r$-Process}
\label{sec:early}

Although we have described the general physics of protoneutron star winds, the resulting nucleosynthesis,
and modifications to our models that might enhance the wind's entropy and decrease its dynamical timescale,
none of the models we have presented so far produces a robust, third-peak $r$-process.  
However, setting $M=2.0$ M$_\odot$, $R_\nu=9$ km, $\lumanue=\lumnue=\lumunu=8.0\times10^{51}$ erg s$^{-1}$, and 
$\avenue=14.5$ MeV, $\aveanue=22$ MeV, $\aveunu=34$ MeV, and adding an artificial heating source between
20 km and 50 km that increases $Q$ by 4\%, we derive a wind with $s_a\simeq150$, $\tau_\rho\simeq1.3$ ms,
and $Y_e^a\sim0.477$.  This extremely compact and luminous protoneutron star yields a wind profile that 
produces third-peak $r$-process nucleosynthesis.
We have found that for $s_a\sim150$ and $0.47\lesssim Y_e^a\lesssim0.495$, we require
$\tau_\rho\lesssim1.3$ ms in order to generate a significant $A\sim195$ yield. 
These are the necessary conditions we derive from our general-relativistic wind models for
an early-time protoneutron star $r$-process epoch.  If a wind trajectory, like those in Fig.~\ref{fig:fig} 
were to pass into the region $s_a\sim150$ and $\tau_\rho\lesssim1.3$ ms with $Y_e^a$ less than 0.50, 
some third-peak material would be produced.
Note that \cite{otsuki} attained third-peak nucleosynthesis for similar wind conditions.
Artificially setting $Y_e^a$ equal to $0.40$, their model with $s_a\sim140$ and $\tau_\rho\sim1.2$ ms 
successfully produced abundance peaks at $A\sim130$ and $A\sim195$.

Although a full nucleosynthesis survey, using real wind profiles, is required to map the space 
$100\lesssim s_a\lesssim200$, $\tau_\rho\lesssim1.5$ ms, and $0.46\lesssim Y_e\lesssim0.50$,
we note some features of potential importance for the short $\tau_\rho$, early-time $r$-process.
First, as we decrease $Y_e^a$ from 0.495 to 0.47, for a given $\tau_\rho$ and $s_a$, the neutron-to-seed
ratio stays roughly constant.  
Hence, the ratio of the abundance yield at $A\sim130$ to that at $A\sim195$ is relatively insensitive to $Y_e^a$.
Second, $r$-process nucleosynthesis at very short timescales and high electron fractions is
possible because the number of seed nuclei formed is very small.
As a consequence, we expect the nucleosynthetic yield in this regime to be 
sensitive to changes in the input nuclear physics and, in particular, the three-body reactions
important in seed nuclei formation (e.g., $^4$He$(\alpha n, \gamma)^9$Be; Kajino et al.~2001).

Our requisite conditions for third-peak nucleosynthesis, $\tau_\rho\lesssim1.3$ ms, $s_a\simeq150$, and high $Y_e^a$, 
disfavor $r$-process nucleosynthesis generically in neutrino-driven winds from neutron stars with $M=$1.4 M$_\odot$.  
Our results in Table 3 indicate that 
even for $R_\nu=8$ km, $s_a$ is 35 units too small and $\tau_\rho$ is a factor of about three too long
for the $r$-process to proceed to the third abundance peak.
We find these gaps in entropy and timescale very difficult to bridge.
For $M=1.4$ M$_\odot$ we require $R_\nu\lesssim6.5$ km to reach this $s_a$ and $\tau_\rho$.
It is unlikely that any high-density nuclear equation of state could accommodate
such a small radius.
Even taking $M=1.6$ M$_\odot$ and $R_\nu=9$ km, without invoking an artificial heating source,
we fail to reach $\tau_\rho\lesssim1.3$ ms and $s_a\simeq150$.
Although the importance of the distribution of energy deposition and extra heating sources should 
be borne in mind, our unmodified wind models require a 
very massive and highly luminous protoneutron star with small radius. 
Indeed, considering the fact that our successful wind models originate from neutron stars with 
$M\gtrsim2.0$ M$_\odot$ and $R_\nu\lesssim9$ km,
we are forced to consider the possibility that the primary site for the $r$-process is not
a protoneutron star at all.   A neutrino-driven outflow generated near the event horizon
of a black hole might bear many of the characteristics of our successful protoneutron
star models. Perhaps very short timescale outflows or jets originating from the
compact inner accretion disk created in the collapsar models of \cite{macfadyen} attain the
necessary conditions for $r$-process nucleosynthesis. 
Such outflow models would benefit by being generated in a region with  high $M/R_\nu$, 
without being subject to the constraints imposed on neutron stars by the nuclear equation of state.


\section{DISCUSSION}
\label{sec:discussion}

We have constructed a robust and efficient algorithm for solving the neutrino-driven protoneutron star wind
problem using both general relativity and Newtonian gravity.  We employed physical boundary conditions
for the transonic wind, a well-motivated neutrino energy deposition function, and an equation of state suited 
to this problem.  For the first time, we included the differential equation for the evolution of $Y_e$ in
radius and the proper sonic point boundary condition.  Using this computational tool, we studied the structure 
and systematics of neutrino-driven winds with an eye toward assessing the suitability of this site for $r$-process 
nucleosynthesis.  We have examined a wide range of protoneutron
star radii, masses, and neutrino spectral characteristics.  By positing an expression for $L_\nu^{\rm tot}(t)$,
we have modeled potential contraction and cooling scenarios that might exist in Nature and calculated
the total mass ejected for the corresponding evolutionary trajectories.  Employing general relativistic hydrodynamics, 
we find significant enhancements in the asymptotic entropy 
of the wind, in good agreement with the post-Newtonian models of \cite{qw1996},
the analytic approximations of \cite{cardall}, and the work of \cite{otsuki}.
In addition, we find that modest modifications to the net energy deposition rate can 
markedly improve the conditions for $r$-process nucleosynthesis.
Indeed, we feel that changes in the profile of energy deposition represent the most
viable alterations to our models, which might lead to robust $r$-process nucleosynthesis
in the protoneutron star context.

Our results indicate that only an early-time epoch of $r$-process nucleosynthesis at high $L_\nu$, small $\tau_\rho$, 
and modest entropy is possible.  A late-time $r$-process, at very high entropy ($\gtrsim300$), 
long $\tau_\rho$, and low $L_\nu$ is not viable.  There are several components to this argument.  
As the luminosity of the protoneutron star decays, both the asymptotic entropy and dynamical timescale of the wind
increase.  The former is conducive to $r$-process nucleosynthesis.  The latter is not.  Hence, the inexorable
rise in both compete.  Fundamentally, for the transonic wind, we find that the asymptotic 
entropy does not increase fast enough to compensate for the deleterious rise in $\tau_\rho$.  
In addition, as the luminosity decays and $\tau_\rho$ increases, $\dot{M}$ decreases. For example, our
1.4 M$_\odot$ evolutionary model only reaches $s_a\sim300$ when $\dot{M}\sim5\times10^{-10}$ M$_\odot$ s$^{-1}$ and 
$\tau_\rho\sim5$ seconds. 
Clearly, even if the $r$-process could exist under these conditions, such an epoch would have to 
continue for many thousands of seconds to produce even 10$^{-6}$ M$_\odot$ of $r$-process ejecta.
Finally, the continued rise in $Y_e^a$ at late times 
as $\avenue$ and $\aveanue$ decrease (due to the energy threshold for the reaction $\bar{\nu}_e p \rightarrow n e^+$,
the {\it threshold effect}) also argues against a high-$s_a$, long-$\tau_\rho$ $r$-process.  

For these reasons, we conclude that if $r$-process nucleosynthesis occurs in protoneutron star winds, 
it must occur at early times, at or just after the moment when $R_\nu$ reaches a minimum.  
Our 1.4 M$_\odot$ evolutionary trajectory with $R_\nu\propto1-at$ in Fig.~\ref{fig:fig} does not attain 
sufficiently high entropies and short timescales
for successful third peak $r$-process nucleosynthesis.
We have calculated the nucleosynthetic yield at every luminosity (time) in this trajectory and at 
no point does the resulting nucleosynthesis go beyond $A\sim100$.  Interestingly, however, we find that
the luminosity point that yields the highest average $A$ ejecta corresponds to the point in time
where $R_\nu$ reaches a minimum.  This point is also a minimum in $\tau_\rho$ for the trajectory.
We calculated the nucleosynthesis at this same luminosity point for each mass in Fig.~\ref{fig:fig}.
Even for the 2.0 M$_\odot$ model with $s_a\sim150$, we did not obtain nucleosynthesis beyond the second abundance
peak.  As evidenced by the survey calculations of \cite{hwq} and \cite{meyer_brown} and the wind
calculations of \cite{otsuki}, these models are outside a regime of
successful third-peak nucleosynthesis.  
However, in \S\ref{sec:energy_source} we have shown that reasonable modifications
to the spectral character of the neutrinos and the energy deposition function might conceivably
shorten $\tau_\rho$ sufficiently for the $r$-process to proceed in some of these models.  In \S\ref{sec:early},
we found that winds with $s_a\simeq150$, $\tau_\rho\lesssim1.3$ ms, and $0.47\lesssim Y_e^a\lesssim0.495$ can
generate third-peak $r$-process elements.

If transonic protoneutron star winds are the primary site for $r$-process nucleosynthesis,
then a successful $r$-process wind model must enter this $s_a-\tau_\rho-Y_e^a$ regime.
The wind, starting just after re-ignition of the supernova shock,
begins with large $R_\nu$ and $L^{\rm tot}_\nu$ and, hence, low entropy ($s_a\sim50$) and short dynamical 
timescales ($\tau_\rho$ is several ms).  In order to avoid overproduction of $A\sim90$ nuclei,
$Y_e^a$ is high ($\gtrsim0.48$) during this low entropy contraction phase (see \S\ref{sec:high_ye}; Hoffman et al.~1996).  
As the protoneutron star contracts, it moves to much higher
entropy and shorter $\tau_\rho$.  Just as $R_\nu$ reaches a minimum and the protoneutron star is
at its most compact, $s_a$ is sufficiently high ($\sim150$), and $\tau_\rho$ is sufficiently short ($\lesssim1.3$ ms) to 
guarantee successful third peak $r$-process nucleosynthesis.  
This epoch does not persist.  Because $R_\nu$ is now constant in time the wind evolves quickly along trajectories like
those in Figs.~\ref{fig:ffig} and \ref{fig:fig} with $s_a\propto\tau_\rho^{0.2}$ to much longer timescales and
only moderately larger $s_a$.  This, coupled with the rise in $Y_e^a$ due to the threshold effect, 
effectively shuts off the $r$-process so that no more than 
$\sim10^{-5}$ M$_\odot$ is ejected.  We emphasize that this scenario requires the wind
to move into and then out of an $r$-processing regime in the space of $s_a-\tau_\rho-Y_e^a$.

Such a picture is provocative, but not yet convincing.
Simply obtaining a wind solution that has proper $s_a$, $\tau_\rho$, and $Y_e^a$ to guarantee
production of the third peak is hardly sufficient to explain the remarkable agreement
between the $r$-element abundances with atomic masses at and beyond barium in 
ultra-metal-poor halo stars and the observed solar $r$-process inventory.
It is difficult to understand how such a scenario might consistently reproduce the barium abundance, all the lanthanides,
the platinum peak, and the actinides (Cayrel et al.~2001; Hill et al.~2001).  
In addition, while the data for these halo stars show remarkable consistency
with the solar abundances above $A\sim135$, below this mass they are markedly 
inconsistent and there is significant star-to-star scatter.  Perhaps some subset of all 
supernovae account for the region below $A\sim135$ and never undergo a vigorous $r$-process.
Perhaps others do obtain the required $s_a$, $\tau_\rho$, and $Y_e^a$ and account for the 
full range of nuclides, including the first abundance peak and proceeding to uranium 
(Wasserburg and Qian 2000; Qian and Wasserburg 2000; Sneden et al. $\!$1996; Burris et al. $\!$2000).  
As we demonstrate in \S\ref{sec:evolution} with our $R_\nu\propto t^{-1/3}$ model, slow radial 
contraction of the protoneutron star may preclude any significant $r$-process yield, as the 
wind would then never enter a regime of short dynamical timescale with $s_a\sim150$.  

The supernova progenitor structure might be important in this regard.
The two-dimensional calculations of \cite{bhf1995} and \cite{janka_muller} 
indicate that a transonic protoneutron star wind can form just tenths
of seconds after the successful re-ignition of the supernova shock.  The pressure of the wind is sufficient for
it to emerge into the expanding supernova ejecta.  However, \cite{janka_muller} found that
as the supernova shock passes through the Si-O interface in their one-dimensional 15 M$_\odot$
progenitor it causes a strong reverse shock that slows the wind expansion from a 
$v\sim2\times10^{9}$ cm s$^{-1}$ to a few times 10$^{8}$ cm s$^{-1}$. 
It is possible that a termination or reverse shock might generally disrupt the transonic wind as it
propagates toward the protoneutron star.
Exactly how far in radius the reverse shock propagates will be a function of the hydrodynamical power
of both the wind and the reverse shock as the neutrino luminosity decays, each being 
functions of the progenitor structure.  With sufficient power, the reverse shock may continue to 
the sonic point. This would put the whole region between the protoneutron 
star surface and the supernova shock in sonic contact, thereby converting a transonic wind into a subsonic breeze.  
We have conducted preliminary hydrodynamical calculations, which suggest this could occur in certain circumstances.  
Steady-state solutions to the wind problem can also be formulated in this context,
but with an outer boundary pressure set by conditions at the supernova shock.
\cite{qw1996} explored the effects of an external boundary pressure on their wind models.
This increased the wind entropy by just 11 units, but increased the dynamical timescale by more than 60\%.
Such a change would be detrimental to an early-time, short-$\tau_\rho$ $r$-process.
However, it may be that the reverse shock does not have sufficient power to disrupt the 
wind interior to the sonic point.  
In the steady-state, across the shock boundary, the velocity will decrease, the density will increase
so as to maintain $\dot{M}$, and both the temperature and entropy will increase.

In order to test the effects of a termination or reverse shock on the 
nucleosynthesis, we inserted a shock by hand at a radius ($r_{\rm sh}$) of 4000 km in our wind model 
with $M=2.0$ M$_\odot$ and $R_\nu=9$ km (see \S\ref{sec:early}), far outside the sonic point ($r\simeq180$ km).  
At $r_{\rm sh}$ the matter velocity was $5.1\times10^{9}$ cm s$^{-1}$, $\rho$ was $17$ g cm$^{-3}$, and $T$ 
was approximately $0.023$ MeV.  Using the Rankine-Hugoniot shock jump conditions, we estimate that 
$v\simeq7v^{\prime}$, $\rho\simeq\rho^{\prime}/7$, and $T\sim T^{\prime}/2$, 
where unprimed quantities are for the wind just before the shock ($r<r_{\rm sh}$), and primed quantities are for 
the flow just after the shock ($r>r_{\rm sh}$).  These conditions increase $s_a$ by about 10 units in the post-shock region
and increase $\tau_\rho$ significantly, due to the sudden decrease in $v$.
This had subtle, but potentially significant effects on the resulting $r$-process yield.
To appreciate this, one must understand that without the slowing of the fluid trajectory 
by passage through the reverse shock, the $r$-process freeze-out in our wind models 
occurs for temperatures below 0.01 MeV, because only then are beta decays along the $r$-process 
path fast enough to compete with the rapid material expansion.  By contrast, if the material
slows (and reheats) by passage through the shock, the $r$-process freeze-out happens at higher 
temperatures, typically near 0.05 MeV.  Although the average number of neutrons captured 
per seed nucleus is the same for the shocked
and unshocked (but otherwise identical) trajectories, the distribution
of those neutron captures is different.  In particular, for the case
considered here, the shocked trajectory had a factor of three larger yield at A$\sim$195.  The reason is that, when
the trajectory slowed and reheated by shock passage, the nuclear flow
changed and allowed more nuclei to leak out of the N=82 closed shell
and proceed up to N=126 (A$\sim$195) at the expense of flow from the
N=50 closed shell (A$\sim$80) to N=130.  An additional interesting effect was
that, unlike the unshocked trajectory, the shocked trajectory showed evidence
of formation of a rare-earth element peak at A$\sim$165.  Surman et al.~(1997) 
argued that this peak forms during freeze-out as the r-process path
rapidly moves through the Z$\sim$60, N$\sim$104 region of somewhat
enhanced nuclear stability in the nuclide
chart.  In the winds we study here, the shocked trajectories favor such
a freeze-out while the unshocked trajectories do not.  We conclude that
the finer details of the $r$-process abundance curve may depend in interesting
ways on the location and strength of a termination or reverse shock.

These hydrodynamical issues are part of the larger question of fallback in Type-II supernovae.
It is possible that the most massive supernova progenitors, with their extended hydrogen envelopes
and dense core structures, experience significant fallback over timescales much longer than the cooling time 
(Chevalier 1989; Woosley and Weaver 1995).  Even if the wind were able to emerge from the neutron star for 10-20 
seconds after explosion, it might not have sufficient power to overcome extended fallback over
minutes and hours.  
Even without a large overlying hydrogen envelope (Type Ib, Ic), the neutrino-driven wind
may be hindered by any progenitor with a large inner core and outer core binding energy.
Therefore, we speculate that an early-time $r$-process,  unencumbered by fallback or reverse shocks, 
is most likely in less massive Type-II, -Ib, or -Ic supernova progenitors.
Accretion-induced collapse may offer even more potential in this context for a fully developed, 
early-time transonic wind as, in this case, there is no overlying mantle to impede
the wind's emergence (Fryer et al.~1999).

\section{Conclusions}

Our major conclusions are the following:  
\begin{itemize}
\item For a given mass outflow rate, we find a significantly shorter dynamical timescale ($\tau_\rho$) 
than indicated by many previous investigations.
\item Because the temperature, density, and velocity gradients often used in defining
a dynamical timescale evolve on a mass element as it moves away from the protoneutron star,
employing a unique dynamical timescale to characterize the wind is not recommended.
\item For a given protoneutron star radius, the asymptotic entropy ($s_a$) is
proportional to $\tau_\rho^{0.2}$.
\item A late-time, high entropy ($s_a\gtrsim300$), long timescale neutrino-driven wind, 
is not a viable astrophysical site for $r$-process nucleosynthesis.  Although the wind may eventually
realize very high entropy, the mass outflow rate will be too small and the dynamical timescale
too long at such an epoch to account for the galactic $r$-process abundance.
\item The third-peak $r$-process elements can be produced in significant abundance
in protoneutron star winds only at early times, at modest entropy ($s_a\sim150$),
and very short dynamical timescale ($\tau_\rho\sim1$ ms).
\item Winds originating from protoneutron stars with mass 1.4 M$_\odot$ and radius $R_\nu=10$ km do 
not produce elements beyond $A\sim100$ at any time during wind evolution.
\item We derive that third-peak $r$-process arises naturally in the context of spherical,
transonic protoneutron star winds only in the unlikely case of protoneutron stars with 
$M\gtrsim2.0$ M$_\odot$ and $R_\nu\lesssim9$ km.
\item Shocks in the protoneutron star wind, exterior to the sonic point, caused by the wind's
interaction with the inner supernova can significantly influence the third-peak and rare-earth
$r$-process element abundances.
\end{itemize}

All of these conclusions hold for the generally high $Y_e$'s (0.46-0.49) derived in this work. 
Only in the unlikely case that $Y_e$ in the wind is $\lesssim0.35$ are our conclusions dramatically
altered.  In this case, third-peak $r$-process nucleosynthesis might be obtained in our transonic
models for protoneutron star masses less than 2.0 M$_\odot$ and radii greater than 9 km,
at early times in the wind evolution.  The uncertainty in the spectral characteristics
of the electron and anti-electron neutrinos in determining the asymptotic electron fraction
during wind formation and evolution is primary on our list of unresolved issues.
Others include:
\begin{itemize}
\item the character of the energy deposition profile as obtained from detailed neutrino transport
in a self-consistent calculation,
\item the high-density nuclear equation of state, which would elucidate the range of protoneutron star masses
and radii relevant,
\item the hydrodynamical interaction of the wind and shocks in the expanding post-supernova environment,
\item the effects of rotation and magnetic fields, 
\item exactly how, even given successful third-peak nucleosynthesis for a given model,
the universality of the observed $r$-process distribution for $A\gtrsim135$ can be accounted for
generically by the progenitor-dependent parameter space of neutrino-driven winds.
\end{itemize}

These last points 
leave the prospect of the $r$-process in this context an open question.
We conclude from this study that if the $r$-process occurs in protoneutron star winds, it most likely 
occurs at early times after the preceding supernova, in winds with very short dynamical timescales ($\lesssim1.3$ ms), 
moderate entropies ($\sim150$), and, possibly, high electron fractions (0.47$\lesssim Y_e^a\lesssim0.495$).
In contrast to the early-time scenario,
because protoneutron star winds only enter a high entropy ($s_a\gtrsim$300) regime with very low $\dot{M}$ 
($\lesssim10^{-9}$ M$_\odot$ s$^{-1}$) and extremely long timescales ($\tau_\rho\gtrsim$ seconds),
a late-time $r$-process is simply not viable.
Conditions suitable for an early-time $r$-process are realized in our models only by very 
compact protoneutron stars with $M=2.0$ M$_\odot$ and $R_\nu\lesssim9$ km. 
The conditions necessary for third-peak $r$-process nucleosynthesis are not realized in neutrino-driven 
transonic winds from canonical neutron stars with $M=1.4$ M$_\odot$ and $R_\nu=10$ km.
In fact, for this neutron star mass, we require $R_\nu\lesssim6.5$ km.  Although such neutron star masses and
radii are not entirely excluded by current high density equations of state, such radii seem unlikely
to obtain in the early post-supernova phase.  The short-timescale jet outflows from the dense 
inner accretion disk around a black hole formed in the collapsar models of \cite{macfadyen}
might attain the necessary entropies and timescales for the $r$-process,
since in that context $M/R_\nu$ can be significantly larger than in the protoneutron star context.
Importantly, it should be noted that we consistently produce $r$-process nucleosynthesis $below$ 
$A\sim135$.  Perhaps protoneutron stars of canonical mass and radius (1.4 M$_\odot$, 10 km) 
produce elements in this mass range generically, thus accounting (due to progenitor structure
and temporal characteristics of the neutrino spectrum) for the variations in abundance observed in 
these elements in ultra-metal-poor halo stars.


\section{Acknowledgements}
T.~A.~T. is supported by a NASA GSRP fellowship.  B.~S.~M. is supported
by NASA grant NAG5-4703 and NSF grant AST 98-19877.  The authors would like to thank
Philip Pinto for helpful discussions and both Jos\'{e} Pons and Steven Bruenn for providing
useful information from their protoneutron star cooling and supernova calculations, respectively.
In addition, we thank George Fuller, Christian Cardall, and Jason Pruet for their comments and 
critical reading of the text.  Wind profiles are available upon request from T.~A.~T. 



\pagebreak

\begin{table}
\begin{scriptsize}
\begin{center}
\caption{Fiducial Wind Models: 1.4 M$_\odot$
\label{tab:fiducial}}
\begin{tabular}{ccccccc}
\tableline
\tableline
\vspace*{-.15cm}
& & & & & & \\ 
$\lumanue^{51}$  & $L_\nu^{\,\rm tot}$ (10$^{51}$ erg s$^{-1}$) & $\dot{\rm M}$ (M$_\odot$ s$^{-1}$) 
& $Q$ (10$^{48}$ erg s$^{-1}$)  & $P_{\rm mech}$ (10$^{48}$ erg s$^{-1}$) & $\tau_\rho$ (ms) & $s_{\rm a}$  \\
\vspace*{-.15cm}
& & & & & & \\ 
\tableline
\tableline
\vspace*{-.15cm}
& & & & & & \\ 
8.0  & 37.0 & $9.05\times10^{-5}$ & 35.1 & 0.848 & 3.68 & 83.9 \\
\vspace*{-.15cm}
& & & & & & \\ 
8.0\tablenotemark{{\rm \,a}} & 37.0 & $2.70\times10^{-4}$ & 75.1 & 2.54 & 3.28 & 68.2 \\
\vspace*{-.15cm}
& & & & & & \\ 
7.0  & 32.4 & $6.56\times10^{-5}$ & 25.1 & 0.561 & 4.24 & 86.4 \\
\vspace*{-.15cm}
& & & & & & \\ 
6.0  & 27.6 & $4.47\times10^{-5}$ & 17.1 & 0.346 & 5.01 & 89.5 \\
\vspace*{-.15cm}
& & & & & & \\ 
5.0  & 23.1 & $2.84\times10^{-5}$ & 10.9 & 0.194 & 6.14 & 93.4  \\
\vspace*{-.15cm}
& & & & & & \\ 
4.0  & 18.5 & $1.63\times10^{-5}$ & 6.28 & 0.0929 & 7.93 & 98.4  \\
\vspace*{-.15cm}
& & & & & & \\ 
3.0  & 13.9 & $8.03\times10^{-6}$ & 3.09 & 0.0346 & 11.29 & 105.4 \\
\vspace*{-.15cm}
& & & & & & \\ 
2.0  & 9.25 & $3.74\times10^{-6}$ & 1.14 & $8.59\times10^{-3}$ & 19.19 & 116.2  \\
\vspace*{-.15cm}
& & & & & & \\ 
1.0  & 4.63 & $5.44\times10^{-7}$ & 0.211 & $8.24\times10^{-4}$ & 49.54 & 137.6  \\
\vspace*{-.15cm}
& & & & & & \\ 
0.6  & 2.78 & $1.58\times10^{-7}$ & 0.062 & $1.50\times10^{-4}$ & 100.9 & 156.3  \\
\vspace*{-.15cm}
& & & & & & \\ 
\tableline 
\tableline
\tablenotetext{{\rm a}}{Newtonian calculation}
\end{tabular}
\end{center}
\end{scriptsize}
\end{table}

\begin{table}
\begin{scriptsize}
\begin{center}
\caption{Evolutionary Wind Models: $\lumanue^{51}=8.0$, 3.4, and 0.4 \label{tab:evol}} 
\begin{tabular}{cccccccccc}
\tableline
\tableline
\vspace*{-.15cm}
& & & & & & & & & \\ 
Mass (M$_\odot$) & $\lumanue^{51}$ & $L_\nu^{\rm tot}$ (10$^{51}$ erg s$^{-1}$) & 
$R_\nu$ (km) & $\dot{\rm M}$ (M$_\odot$ s$^{-1}$) 
& $Q$ (10$^{48}$ erg s$^{-1}$) & $P_{\rm mech}$ (10$^{48}$ erg s$^{-1}$) & $\tau_\rho$ (ms) & $s_{\rm a}$ 
& $Y_e^a$ \\
\vspace*{-.15cm}
& & & & & & & & \\ 
\tableline
\tableline
\vspace*{-.15cm}
& & & & & & & & & \\ 
2.0 & 8.0 & 37.0 &  20.3 & $2.19\times10^{-4}$ & 53.3 & 1.27 & 10.45 & 67.08 & 0.467\\
\vspace*{-.15cm}
& & & & & & & & & \\ 
1.8 & 8.0 & 37.0 & 20.3 & $3.03\times10^{-4}$ & 67.2  & 1.67 & 9.87 & 59.05  & 0.465\\
\vspace*{-.15cm}
& & & & & & & & & \\ 
1.6 & 8.0 & 37.0 &  20.3 & $4.38\times10^{-4}$ & 83.8 & 2.27 & 9.27 & 51.39  & 0.463 \\
\vspace*{-.15cm}
& & & & & & & & & \\ 
1.4 & 8.0 & 37.0 & 20.3 & $6.69\times10^{-4}$ & 108.2 & 3.25 & 8.63 & 44.02 & 0.460 \\
& & & & & & & & \\ 
\tableline
& & & & & & & & & \\ 
2.0 & 3.4 & 15.7 & 10.0 & $9.80\times10^{-6}$ & 5.94 & 0.0826 & 6.83 & 151.61 &  0.469\\
\vspace*{-.15cm}
& & & & & & & & & \\ 
1.8 & 3.4 & 15.7 & 10.0 & $1.39\times10^{-5}$ & 7.34 & 0.108 & 6.78 & 129.43  & 0.467 \\
\vspace*{-.15cm}
& & & & & & & & & \\ 
1.6 & 3.4 & 15.7  &10.0 & $2.04\times10^{-5}$ & 9.29 & 0.148 & 6.56 & 109.70  &  0.465\\
\vspace*{-.15cm}
& & & & & & & & & \\ 
1.4 & 3.4 & 15.7 &10.0 & $3.14\times10^{-5}$ & 12.1 & 0.213 & 6.21 & 91.78 & 0.462 \\
\vspace*{-.15cm}
& & & & & & & & & \\ 
1.4\tablenotemark{\,a}& 3.4 & 15.7 & 14.7 & $9.03\times10^{-4}$ & 21.8 & 0.424 & 9.98 & 64.01 & 0.457\\
& & & & & & & & & \\ 
\tableline
& & & & & & & & & \\ 
2.0 & 0.4 & 1.85 &10.0 & $1.24\times10^{-7}$ & 0.0757 & $1.97\times10^{-4}$ & 75.49 & 234.20 &  0.492\\
\vspace*{-.15cm}
& & & & & & & & & \\ 
1.8 & 0.4 & 1.85 &10.0 & $1.72\times10^{-7}$ & 0.0923 & $2.48\times10^{-4}$ & 75.21 & 198.88  & 0.491 \\
\vspace*{-.15cm}
& & & & & & & & & \\ 
1.6 & 0.4 & 1.85 & 10.0 & $2.46\times10^{-7}$ & 0.114 & $3.23\times10^{-4}$ & 72.92 & 167.40  &  0.490\\
\vspace*{-.15cm}
& & & & & & & & & \\ 
1.4 & 0.4 & 1.85 &10.0 & $3.67\times10^{-7}$ & 0.144 & $4.44\times10^{-4}$ & 69.14  & 138.92  & 0.489\\
& & & & & & & & & \\ 
\tableline 
\tableline 
\end{tabular}
\end{center}
\tablenotetext{a}{The 1.4 M$_\odot$ trajectory in Fig.~\ref{fig:fig} with $R_\nu\propto t^{-1/3}$, for $\lumanue^{51}=3.4$. }
\end{scriptsize}
\end{table}

\begin{table}
\begin{scriptsize}
\begin{center}
\caption{Modified Evolutionary Wind Models:
$\lumanue^{51}=3.4$ and $Y_e^a\simeq0.46$ \label{tab:evol2}} 
\begin{tabular}{cccccccc}
\tableline
\tableline
\vspace*{-.15cm}
& &  & & & & \\ 
Mass (M$_\odot$) &  $L_\nu^{\rm tot}$ (10$^{51}$ erg s$^{-1}$) & $R_\nu$ (km) & $\dot{\rm M}$ (M$_\odot$ s$^{-1}$) 
& $Q$ (10$^{48}$ erg s$^{-1}$) & $P_{\rm mech}$ (10$^{48}$ erg s$^{-1}$) & $\tau_\rho$ (ms) & $s_{\rm a}$ \\
\vspace*{-.15cm}
& &  & & & \\ 
\tableline
\tableline
\vspace*{-.15cm}
& &  & & & & \\ 
1.8& 15.7 & 9.0 & $1.01\times10^{-5}$ & 6.18 & 0.092 & 5.63 & 146.44  \\
\vspace*{-.15cm}
& &  & & & & \\ 
1.6& 15.7 & 9.0 & $1.48\times10^{-5}$ & 7.79 & 0.122 & 5.61 & 122.98  \\
\vspace*{-.15cm}
& &  & & & & \\ 
1.4& 15.7 & 9.0 & $2.29\times10^{-5}$ & 10.2 & 0.174 & 5.39 & 102.15 \\
\vspace*{-.15cm}
& &  & & & & \\ 
1.4& 15.7 & 8.5 & $1.94\times10^{-5}$ & 9.25 & 0.157 & 4.97 & 108.48 \\
\vspace*{-.2cm}
& &  & & & & \\ 
1.4& 15.7 & 8.0 & $1.62\times10^{-5}$ & 8.35 & 0.141 & 4.54 & 115.86 \\
\vspace*{-.15cm}
& &  & & & & \\ 
\tableline 
\tableline 
\end{tabular}
\end{center}
\end{scriptsize}
\end{table}

\begin{thebibliography}{cccccccccccc}

\bibitem[Arzoumanian (1995)]{arzoumanian} Arzoumanian, Z. 1995, PhD Thesis, Princeton University

\bibitem[Bahcall (1964)]{bahcall} Bahcall, J. N. 1964, Physical Review, 136, B1164

\bibitem[Bruenn, De Nisco, \& Mezzacappa (2001)]{bruenn2001} Bruenn, S. W., De Nisco, K. R., \& Mezzacappa, A. 2001, preprint (astro-ph/0101400)

\bibitem[Burbidge et al.~(1957)]{burbidge} Burbidge, E. M., Burbidge, G. R., Fowler, W. A., \& Hoyle, F. 1957, Rev. Mod. Phys., 29, 547

\bibitem[Burris et al.~(2000)]{burris}Burris, D. L., Pilachowski, C. A., Armandroff, T. E., Sneden, C., Cowan, J. J., \& Roe, H. 2000, ApJ, 544, 302

\bibitem[Burrows \& Lattimer (1986)]{bl1986}Burrows, A. \& Lattimer, J. M. 1986, ApJ, 307, 178

\bibitem[Burrows (1988)]{burrows_sn_neutrinos}Burrows, A. 1988, ApJ, 334, 891

\bibitem[Burrows, Hayes, \& Fryxell (1995)]{bhf1995}Burrows, A., Hayes, J., \& Fryxell, B. A. 1995, ApJ, 450, 830 

\bibitem[Cardall \& Fuller (1997)]{cardall} Cardall, C. Y. \& Fuller, G. M. 1997, ApJL, 486, 111 

\bibitem[Chevalier (1989)]{chevalier} Chevalier, R. 1989, ApJ, 346, 847

\bibitem[Cayrel et al.~(2001)]{cayrel} Cayrel, R., Hill, V., Beers, T.~C., Barbuy, B., Spite, M., Spite, F., Plez, B., Andersen, J., Bonifacio, P., Francois, P., Molaro, P., Nordstr\"{o}m, B., \& Primas, F.~2001, Nature, 409, 691

\bibitem[Cowan et al.~(1996)]{cowan}Cowan, J. J., Sneden, C., Truran, J. W., \& Burris, D. L. 1996, ApJL, 460, 115 

\bibitem[Duncan, Shapiro, \& Wasserman (1986)]{dsw}Duncan, R. C., Shapiro, S. L., \& Wasserman, I. 1986, ApJ, 309, 141 

\bibitem[Eggleton (1971)]{eggleton1971}Eggleton, P. P. 1971, MNRAS, 151, 351

\bibitem[Flammang (1982)]{flammang}Flammang, R. A. 1982, MNRAS, 199, 833

\bibitem[Freiburghaus, Rosswog, \& Thielemann (1999)]{frt} Freiburghaus, C., Rosswog, S., \& Thielemann, F.-K. 1999, ApJL, 525, 121

\bibitem[Freiburghaus et al.~(1999)]{freiburghaus} Freiburghaus, C., Rembges, J.-F., Rauscher, T., Kolbe, E., Thielemann, F.-K., Kratz, K.-L., Pfeiffer, B., \& Cowan, J. J. 1999, ApJ, 516, 381

\bibitem[Fryer et al.~(1999)]{fryer} Fryer, C. L., Benz, W., Herant, M., \& Colgate, S. 1999, ApJ, 516, 892

\bibitem[Fuller \& Meyer (1995)]{fuller_meyer}Fuller, G. M. \& Meyer, B. S. 1995, ApJ, 453, 792

\bibitem[Fuller \& Qian (1996)]{qfredshift} Fuller, G. M. \& Qian, Y.-Z. 1996, Nuc. Phys. A, 606, 167

\bibitem[Hill et al.~(2001)]{hill}Hill, V., Plez, B., Cayrel, R., \& Beers, T.~C. 2001, proceedings of Astrophysicaql Ages and Timescales, ASP Conference Series, preprint (astro-ph/0104172)

\bibitem[Hoffman et al.~(1996)]{hoffman_1996a}Hoffman,  R. D., Woosely, S. E., Fuller, G. M., \& Meyer, B. S.  1996, ApJ, 460, 478

\bibitem[Hoffman, Woosley, \& Qian (1997)]{hwq} Hoffman,  R. D., Woosely, S. E., \& Qian,Y.-Z. 1997, ApJ, 482, 951

\bibitem[Horowitz \& Li (2000)]{cp_violation} Horowitz, C. J. \& Li, G. 2000, Proceedings of Intersections Conference, Qubec, preprint (astro-ph/0010042)

\bibitem[Itoh et al.~(1996)]{itoh1996}Itoh, N., Hayashi, H., Nishikawa, A., \& Kohyama, Y. 1996, ApJS, 102, 411

\bibitem[Janka \& M\"{u}ller (1995)]{janka_muller} Janka, H.-Th. \& M\"{u}ller, E. 1995, ApJL, 448, 109

\bibitem[Janka (1991)]{janka1991}Janka, H.-T. 1991, A\&A, 244, 378 

\bibitem[Janka \& Hillebrandt (1989)]{janka_hillebrandt} Janka, H.-Th. \& Hillebrandt, W. 1989, AA, 224, 49

\bibitem[Kajino et al.~(2001)]{kajino}Kajino, T., Otsuki, K., Wanajo, S., Orito, M., \& Mathews, G. 2001, to appear in Few-Body Systems Suppl., preprint (astro-ph/0006079)

\bibitem[Kippenhahn, Weigert, \& Hoffmeister (1968)]{kippenhahn1968}Kippenhahn, R., Weigert, A., \& Hoffmeister, E. 1968, Meth. Comput. Phys., 7, 129

\bibitem[Lattimer \& Prakash (2001)]{lattimer_prakash} Lattimer, J. M. \& Prakash, M. 2001, ApJ, 550, 426 

\bibitem[Liebend\"{o}rfer et al.~(2001)]{lieben2001} Liebend\"{o}rfer, M., Mezzacappa, A., Thielemann, F.-K., Messer, O. E. B., Hix, W. R., \& Bruenn, S. W. 2000, PRD, submitted, preprint (astro-ph/0006418)

\bibitem[London \& Flannery (1982)]{london_flannery} London, R. A. \& Flannery, B. P. 1982, ApJ, 258, 260

\bibitem[MacFadyen \& Woosley (1999)]{macfadyen} MacFadyen, A.~I. \& Woosley, S.~E. 1999, ApJ, 524, 262

\bibitem[Mayle, Wislon, \& Schramm (1987)]{mayle_wilson} Mayle, R., Wilson, J. R., \& Schramm, D. N. 1987, ApJ, 318, 288

\bibitem[McLaughlin, Fuller, \& Wilson (1996)]{mclaughlin_fuller} Mclaughlin, G., Fuller, G. M., \& Wilson, J. R. 1996, ApJ, 472, 440

\bibitem[McWilliam et al.~(1995a)]{mcwilliam_a}McWilliam, A., Preston, G. W., Sneden, C., \& Searle, L. 1995a, AJ, 109, 2757

\bibitem[McWilliam et al.~(1995b)]{mcwilliam_b}McWilliam, A., Preston, G. W., Sneden, C., \& Shectman, S. 1995b, AJ, 109, 2757

\bibitem[Meyer \& Brown (1997)]{meyer_brown} Meyer, B. S. \& Brown, J. S. 1997, ApJS, 112, 199

\bibitem[Meyer et al.~(1992)]{meyer_1992} Meyer, B. S.,  Howard, W. M.,  Mathews, G. J.,  Woosley, S. E., \&  Hoffman, R. D. 1992, ApJ, 399, 656 

\bibitem[Mezzacappa et al.~(2001)]{mezz2001} Mezzacappa, A., Liebend\"{o}rfer, M., Messer, O. E. B., Hix, W. R., Thielemann, F.-K., \& Bruenn, S. W. 2001, PRL, 86, 1935

\bibitem[Myra \& Burrows (1990)]{myra} Myra, E.~S. and Burrows, A. 1990, ApJ, 364, 222

\bibitem[Nobili, Turolla, \& Zampieri (1991)]{nobili1991}Nobili, L., Turolla, R.,  \& Zampieri, L. 1991, ApJ, 383 250

\bibitem[Otsuki et al.~(2000)]{otsuki}Otsuki, K., Tagoshi, H., Kajino, T., \& Wanajo, S.-Y. 2000, ApJ, 533, 424

\bibitem[Pons et al.~(1999)]{pons1999}Pons, J. A., Reddy, S., Prakash, M., Lattimer, J. M., \& Miralles, J. A. 1999, ApJ, 513, 780

\bibitem[Press et al.~(1992)]{press}Press, W. H., Teukolsky, S. A., Vetterling, W. T., \& Flannery, B. P. 1992, Numerical Recipes in Fortran 77 (2d Ed.; New York, NY: Cambridge University Press)

\bibitem[Qian et al.~(1993)]{qian93}Qian, Y.-Z., Fuller, G.~M., Mathews, G.~J., Mayle, R.~W., Wilson, J.~R., Woosley, S.~E. 1993, PRL, 71, 1965

\bibitem[Qian (2000)]{qian}Qian, Y.-Z. 2000, ApJL, 534, 67

\bibitem[Qian \& Fuller (1995A)]{qf1995A}Qian, Y.-Z. \& Fuller, G. M. 1995a, PRD, 51, 14, 1479

\bibitem[Qian \& Fuller (1995B)]{qf1995B}Qian, Y.-Z. \& Fuller, G. M. 1995b, PRD, 52, 2, 656

\bibitem[Qian \& Wasserburg (2000)]{qian_wasserburg}Qian, Y.-Z. \& Wasserburg, G. J. 2000, Phys. Reps., 333, 77

\bibitem[Qian \& Woosley (1996)]{qw1996}Qian, Y.-Z. \& Woosley, S. E. 1996, ApJ, 471, 331 

\bibitem[Rampp \& Janka (2000)]{rampp2000} Rampp, M. \& Janka, H.-Th. 2000, ApJL, 539, 33

\bibitem[Reddy et al.~(1998)]{reddy_1998}Reddy, S., Prakash, M., \& Lattimer, J. M. 1998, PRD, 58, 013009

\bibitem[Rosswog et al.~(1999)]{rosswog} Rosswog, S., Liebend\"{o}rfer, M., Thielemann, F.-K., Davies, M., Benz, W., \&Piran, T. 1999, A\&A, 341, 499

\bibitem[Salmonson \& Wilson (1999)]{salmonson_wilson} Salmonson, J. D. \& Wilson, J. R. 1999, ApJ, 517, 859

\bibitem[Sneden et al.~(1996)]{sneden} Sneden, C., McWilliam, A., Preston, G. W., Cowan, J. J., Burris, D. L., \& Armosky, B. J. 1996, ApJ, 467, 819

\bibitem[Sumiyoshi et al.~(2000)]{sumiyoshi} Sumiyoshi, K., Suzuki, H., Otsuki, K., Teresawa, M., \& Yamada, S. 2000, PASJ, 52, 601

\bibitem[Surman et al.~(1992)]{surman}Surman, R., Engel, J., Bennett, J.~R., \& Meyer, B.~S. 1992, PRL, 79, 1809

\bibitem[Takahashi, Witti, \& Janka (1994)]{twj1994} Takahashi, K., Witti, J., \& Janka, H.-T. 1994, A\&A, 286, 857 

\bibitem[Thompson, Burrows, \& Horvath (2000)]{thompson}Thompson, T. A., Burrows, A., \& Horvath, J. E.  2000, PRC, 62, 035802 

\bibitem[Tubbs (1979)]{tubbs_scatter}Tubbs, D. L. 1979, ApJ, 231, 846

\bibitem[Tubbs \& Schramm (1975)]{tubbs_schramm}Tubbs, D. L. \& Schramm, D. N. 1975, ApJ, 201, 467

\bibitem[Wallerstein (1997)]{wallerstein}Wallerstein,  G., Iben, I., Parker, P., Boesgaard, A. M., Hale, G. M.,  Champagne, A. E., Barnes, C. A., K\"{a}ppeler, F., Smith, V. V.,  Hoffman, R. D.,  Timmes, F. X., Sneden, C.,  Boyd, R. N.,  Meyer, B. S., \& Lambert, D. L. 1997, Rev. Mod. Phys., 69, 995 

\bibitem[Wanajo et al.~(2000)]{wanajo} Wanajo, S., Kajino, T., Mathews, G. J., \& Otsuki, K. 2001, accepted to ApJ

\bibitem[Wasserburg \& Qian~(2000)]{wasserburg_qian}Wasserburg, G. J. \& Qian, Y.-Z. 2000, ApJL, 529, 21

\bibitem[Westin et al.~(2000)]{westin} Westin, J., Sneden, C., Gustafsson, B., \& Cowan, J. J. 2000, ApJ, 530, 783

\bibitem[Woosley \& Hoffman (1992)]{wh1992}Woosley, S. E. and Hoffman, R. D. 1992, ApJ, 395, 202 

\bibitem[Woosley et al.~(1994)]{woosley1994}Woosley, S. E., Wilson, J. R., Mathews G. J., Hoffman, R. D., \& Meyer, B. S. 1994, ApJ, 433, 209

\bibitem[Woosley \& Weaver (1995)]{woosley_weaver} Woosley, S. E. \& Weaver, T. A. 1995, ApJS, 101, 181

\end{thebibliography}
\end{document}